\documentclass[oneside,range]{ar2e}

\usepackage{graphicx}
\usepackage[hypertex]{hyperref}

\input symbols.tex

\catcode`\@=11
\renewenvironment{thebibliography}[1]{%
   \small
   \section*{References}%
   \def\@tempa{#1}%
   \ifx\@tempa\@empty
   \list{}{\def\@biblabel##1{##1}\labelwidth0pt\leftmargin1em\itemindent-1em\itemsep0pt\labelsep0pt
            \usecounter{enumiv}%
            \let\p@enumiv\@empty}%
   \else
   \list{\@biblabel{\arabic{enumiv}}}%
           {\settowidth\labelwidth{\@biblabel{#1}}%
            \leftmargin\labelwidth\itemsep3pt
            \advance\leftmargin\labelsep
            \usecounter{enumiv}%
            \let\p@enumiv\@empty
            \renewcommand\theenumiv{\arabic{enumiv}}}%
   \fi
      \if@openbib
        \renewcommand\newblock{\par}
      \else
        \renewcommand\newblock{\hskip .11em \@plus.33em \@minus.07em}%
      \fi
      \sloppy\clubpenalty4000\widowpenalty4000%
      \sfcode`\.=\@m}
     {\def\@noitemerr
       {\@latex@warning{Empty `thebibliography' environment}}%
      \endlist}
\renewcommand\section{\@startsection {section}{1}{\z@}%
                                   {-3.5ex \@plus -1ex \@minus -.2ex}%
                                   {2.3ex \@plus.2ex}%
                                   {\reset@font\large\bfseries\boldmath}}
\renewcommand\subsection{\@startsection{subsection}{2}{\z@}%
                                     {-3.25ex\@plus -1ex \@minus -.2ex}%
                                     {1.5ex \@plus .2ex}%
                                     {\reset@font\large\bfseries\boldmath}}
\renewcommand\subsubsection{\@startsection{subsubsection}{3}{\z@}%
                                     {-3.25ex\@plus -1ex \@minus -.2ex}%
                                     {.5ex \@plus .2ex}%
                                     {\reset@font\normalsize\bfseries\boldmath}}
\renewcommand\l@subsubsection[3]{%
  \ifnum \c@tocdepth > 2
    \addpenalty{\@secpenalty}%
    \setlength\@tempdima{28pt}%
    \begingroup
      \parindent \z@ \rightskip \@pnumwidth
      \parfillskip -\@pnumwidth
      \leftskip\@tempdima\footnotesize
      \leavevmode{\itshape#1}\nobreak\leaders\hbox{$\m@th \mkern \@dotsep mu.\mkern \@dotsep
       mu$}\hfill \nobreak\hbox to\@pnumwidth{\hss #2}\par
    \endgroup
  \fi\fi}

\def\@cite#1#2{[{#1\if@tempswa , #2\fi}]}

\catcode`\@=12

\def\OMIT#1{{}}
\renewcommand{\lsim}{\mathrel{\!\mathpalette\vereq<}\!}
\renewcommand\gsim{\mathrel{\!\mathpalette\vereq>}\!}
\def\vereq#1#2{\lower3.5pt\vbox{\baselineskip1.5pt \lineskip1.5pt
\ialign{$#1\hfill##\hfil$\crcr#2\crcr\sim\crcr}}}

\newcommand{\ov}{\overline}
\newcommand{\TeV}{\tev}
\newcommand{\GeV}{\gev}
\newcommand{\MeV}{\mev}
\newcommand{\nubar}{\nub}

\jname{ARNPS}
\jyear{2006}
\jvol{56}
\ARinfo{}

\begin{document}

\title{\boldmath \CP violation and the CKM matrix}

\markboth{H\"ocker \& Ligeti}{\CP violation and the CKM matrix}

\author{Andreas H\"ocker
\affiliation{CERN, CH--1211 Geneva 23, Switzerland}
Zoltan Ligeti
\affiliation{Lawrence Berkeley National Laboratory,
  University of California, Berkeley, CA 94720 \\
Center for Theoretical Physics, Massachusetts Institute of Technology,
Cambridge, MA 02139}}


\begin{abstract}\\
Our knowledge of quark-flavor physics and \CP violation increased 
tremendously over the past five years. It is confirmed that the Standard 
Model correctly describes the dominant parts of the observed 
\CP-violating and flavor-changing phenomena. Not only does \CP 
violation provide some of the most precise constraints on the flavor 
sector, but several measurements performed at the \B-factories 
achieved much better precision than had been expected. 
We review the present status of the Cabibbo-Kobayashi-Maskawa matrix and \CP 
violation, recollect the relevant experimental and theoretical inputs, 
display the results from the global CKM fit, and discuss their 
implications for the Standard Model and some of its extensions. \\
\mbox{ }\hfill{\scriptsize CERN-PH-EP/2006-007, LBNL-59882, MIT-CTP 3729}
\end{abstract}

\maketitle

\section{Introduction}

\subsection{Quark Mixing and \CP Violation in the Standard Model}

In the Standard Model (SM), quark masses and their mixing and \CP violation have a
common origin.  Local gauge invariance forbids bare mass terms for fermions,
just as it forbids gauge boson masses.  Masses are generated after spontaneous
electroweak symmetry breaking, owing to the Yukawa couplings of the fermions to
the Higgs condensate.  Therefore \CP violation and flavor-changing interactions
also probe the electroweak scale, complementary to other studies at the energy
frontier.  Because the nature of the Higgs condensate is not known, it is
interesting to study as many different aspects of it as possible.

Historically, flavor physics and \CP violation were excellent probes of new
physics:  the absence of the decay $\KL\to \mu^+\mu^-$ predicted
the charm quark; \CP violation in $\Kz\Kzb$ mixing predicted 
the third generation; the $\Kz\Kzb$ mass difference predicted 
the charm mass; the mass difference of the heavy and light mass eigenstates 
of the $\Bz\Bzb$ mesons predicted the heavy top mass.  These
measurements put strong constraints on extensions of the SM, and imply that if
there is new physics at the TeV scale, it must have a very special flavor and \CP
structure to satisfy these constraints.

The SM contains three sources of \CP violation (if the quark masses are nonzero
and nondegenerate).  One of them occurs in the Cabibbo-Kobayashi-Maskawa (CKM) 
matrix~\cite{C,KM} that describes the mixing of the quark generations, and the 
corresponding \CP-violating
parameter is of order unity.  This source of \CP violation is discussed in
detail in this review.  With the SM amended to include  neutrino masses, \CP
violation is also possible in the mixing of leptons.  We have no experimental
information on the size of this effect yet.  The third source of \CP violation
occurs in flavor-conserving strong-interaction processes.  However, the
experimental upper bound on the neutron electric dipole moment implies that this
$(\theta_{\rm QCD}/16\pi^2)F_{\mu\nu} \widetilde F^{\mu\nu}$ term in the SM
Lagrangian is at best tiny, $\theta_{\rm QCD} \lsim 10^{-10}$ (for a review,
see~\cite{Dine:2000cj}). This is usually referred to as the {\em strong \CP
problem}.  Thus we know that neither \CP violation in the lepton sector nor that
corresponding to $\theta_{\rm QCD}$ can play any role in quark-flavor physics
experiments.

Interestingly, there already exists strong evidence for \CP violation beyond
the  SM. Assuming that the evolution of the universe began in a
matter-antimatter symmetric state or that inflation washed out any previously
existing asymmetry, the three sources of \CP violation in the SM mentioned above
cannot explain the observed baryon-to-entropy ratio, $n_B/s \approx 9 \times
10^{-11}$.  The SM contains all three Sakharov conditions~\cite{Sakharov:1967dj}
--- baryon number violation, $C$ and \CP violation, and deviation from thermal
equilibrium --- required to generate such an asymmetry.  However, the asymmetry
generated by the SM is many orders of magnitudes too
small~\cite{Farrar:1993hn}.  This implies that there must be \CP violation
beyond the SM, but it does not have to be in the quark sector, nor in
flavor-changing processes.\footnote
{
	The baryon asymmetry may be explained by leptogenesis~\cite{leptogen}.
	The discovery of neutrino masses makes the seesaw mechanism and the
	existence of very massive right-handed neutrinos likely.  These may
	decay out of equilibrium through \CP-violating interaction, generating 
	nonzero lepton number, which can subsequently be converted by part into a 
	baryon asymmetry
	by $(B+L)$-violating but $(B-L)$-conserving sphaleron processes present in
	the SM. 
}
Nevertheless, because most new physics scenarios predict or naturally have
observable effects in quark-flavor physics experiments, it is interesting to
thoroughly test this sector of the SM.

\subsection{The CKM Matrix}

The Yukawa interaction of the quarks is given by
\beq\label{Yukawa}
{\cal L}_Y = - Y_{ij}^d\, \ov{Q_{Li}^I}\, \phi\, d_{Rj}^I
  - Y_{ij}^u\, \ov{Q_{Li}^I}\, \varepsilon\, \phi^* u_{Rj}^I + {\rm h.c.} ,
\eeq
where $Y^{u,d}$ are $3\times 3$ complex matrices, $\phi$ is the Higgs field,
$i$ and $j$ are generation labels, and $\varepsilon$ is the $2\times 2$ antisymmetric
tensor.  The $Q_L^I$ are left-handed quark doublets, and $d_R^I$ and $u_R^I$ are
right-handed down- and up-type quark singlets, respectively, in the
weak-eigenstate basis.  When $\phi$ acquires a vacuum expectation
value, $\langle\phi\rangle = (0, v/\sqrt2)$, Eq.~(\ref{Yukawa}) yields Dirac
mass terms for quarks with $3\times3$ mass matrices
\beq
\label{eq:mumd}
   M^u = \frac{v Y^u}{\sqrt{2}}\,, \qquad
   M^d = \frac{v Y^d}{\sqrt{2}}\,.
\eeq
To move from the basis of the flavor eigenstates to the basis
of the mass eigenstates, one performs the transformation
\beq
        U_L^{u(d)} M^{u(d)} U_R^{u(d)\dag}
        = {\rm diag}\Big(m_{u(d)},
                          m_{c(s)},
                          m_{t(b)}
                     \Big)\,,
\eeq
where $U_L^{u,d}$ and $U_R^{u,d}$ are unitary matrices and the masses $m_q$ are
real.  The quark mass matrices are diagonalized by different transformations for
the left-handed up- and down-quarks, which are part of the same $\SU(2)_L$
doublet,
\beq\label{misalign}
Q_L^I = \pmatrix{u_{Li}^I\cr d_{Li}^I} = (U_L^{u\dagger})_{ij}
  \pmatrix{u_{Lj}\cr (U_L^u U_L^{d\dagger})_{jk}\, d_{Lk}} .
\eeq
By convention, we pulled out $(U_L^{u\dagger})_{ij}$, so that the 
``misalignment" between the two transformations operates on the down-type quark
mass eigenstates.  Thus the charged-current weak interaction is modified by the
product of the diagonalizing matrices of the up- and down-type quark mass
matrices, the so-called CKM matrix~\cite{C,KM},
\beq\label{eq:ckm1}
        \VCKM = U_L^u U_L^{d\dag}
= \pmatrix{
        V_{ud} & V_{us} & V_{ub} \cr
        V_{cd} & V_{cs} & V_{cb} \cr
        V_{td} & V_{ts} & V_{tb} \cr
        } .
\eeq
However, the neutral-current part of the Lagrangian in the mass
eigenstate basis remains unchanged, \ie, there are no flavor-changing neutral
currents (FCNC) at tree level.

Being the product of unitary matrices, $\VCKM$ itself is unitary, $\VCKM
\VCKM^\dag={\rm I}$. This requirement and the freedom to arbitrarily  choose the
global phases of the quark fields reduce the initial nine unknown  complex
elements of $\VCKM$ to three real numbers and one phase, where the latter
accounts for \CP  violation. Because these four numbers effectively govern the
rates of all tree- and loop-level electroweak transitions that involve the
charged current, it is a compelling exercise to  overconstrain $\VCKM$. If
inconsistencies among different measurements occurred, it would reveal the
existence of physics beyond the SM.

The charged-current couplings among left-handed quark fields
are proportional to the elements of $\VCKM$. The right-handed
quarks have no $W$-boson interaction in the SM, and
the $Z$, photon and gluon couplings are flavor-diagonal. For left-handed 
leptons, flavor-mixing can be introduced in a manner similar to that of 
quarks.

The first CKM element, the Cabibbo quark-mixing angle $\theta_C$, was
introduced in 1963~\cite{C} to explain the small
weak-interaction decay rates for particles carrying strangeness. When \CP
violation was discovered in 1964 by the  observation of the \CP-odd decay
$\KL\to\pip\pim$~\cite{cpviol}, researchers had not yet perceived 
the intimate relation  between the dynamical
rules of quark-flavor mixing and the phenomenon of \CP violation. 
Hence the terrain was  open for speculations. In 1970, 
Glashow, Iliopoulos \& Maiani (GIM)~\cite{GIM} used the unitary
quark-mixing ansatz to postulate the existence of a
fourth quark with quantum number charm to explain the observed
suppression of strangeness-changing neutral currents (\eg, $\KL\to\mu^+\mu^-$).
This mechanism yields the absence of tree-level FCNCs
in the SM. In 1973 the concept of quark-flavor mixing and \CP violation 
were unified when Kobayashi \& Maskawa
showed that for at least three generations of quarks, there would be enough
physical degrees of freedom left in the quark-flavor mixing matrix to allow for
a nonzero phase~\cite{KM}. This example demonstrates how,  throughout the
history of particle physics, discoveries and developments in  flavor physics
have led to spectacular progress. The subsequent  discovery of bottom and
top quarks, and even a third lepton generation, as  well as the observation of
direct \CP violation in the kaon system backed  the KM idea. The \B-factories
provided the proof that the KM mechanism is the dominant source of \CP 
violation at the electroweak scale with the measurement of $\stwob$ in 
agreement with the KM expectation~\cite{sin2b_dicovery}.

Beyond the description of \CP violation, the three-generation CKM matrix
provides a consistent framework for understanding a huge variety of observations.
This includes understanding the very different neutral-meson 
mixing frequencies governed by FCNC box diagrams (the GIM mechanism is not 
effective for top quark loops and hence large mixing amplitudes are
possible), to the precise predictions of rare decays of \kaon and \B
mesons. Using the measurements of the $\Kz\Kzb$ and $\Bz\Bzb$
mixing frequencies, the KM ansatz could be used to predict the scales of charm 
and top-quark masses, respectively. Consequently, the hierarchical structure 
of the CKM matrix guides experimentalists in their quest for observables
that can precisely determine its elements: $a)$ The quarks occurring in the box 
diagrams govern the mixing frequency, and hence determine whether or not 
the neutral meson mixing can be observed experimentally; 
$b)$ the lifetimes of the lightest mesons with a given flavor depend on
the CKM elements in the dominant weak decays; to measure
time-dependent \CP asymmetries, the lifetime and the mixing 
frequency should be comparable; $c)$ the CKM mixing hierarchy between 
the generations determines the relative size of the observable \CP-violating 
effects, and which mixing and decay amplitudes contribute to 
the process. The conjunction of these considerations have led to the 
understanding that asymmetric-energy \B-factories would allow
theoretically clean measurements of \CP-violating observables in the 
$B_d$ sector. As a result of the successful operation of the \B-factories,
for the first time, significantly overconstraining the CKM 
matrix with many dynamically different processes, and hence testing 
the KM theory, was possible. This---and not the measurement of the CKM 
phase's {\em value} 
per se---has been and is the primary motivation for investing important 
resources into the physics of quark flavors.
The overwhelming success of the KM theory hides however the problem that the 
{\em origin} of the observed flavor structure (mass hierarchy, mixing and \CP 
violation) is not understood within the SM.

\subsection{Unitarity Constraints}

We have seen in the previous section that quark-flavor (or weak interaction) 
eigenstates and mass eigenstates are not equivalent. Hence the quark flavors 
mix through their couplings to the charged weak current. The CKM matrix, which
describes the three-generation quark-flavor mixing in the SM,
would be greatly simplified if, for instance, two of the up- or down-type 
quarks had equal masses and were therefore indistinguishable 
(see Sec.~\ref{sec:jarlskog}). Because there is no empirical evidence
for such degeneracy, the full variable space for the CKM elements must 
be considered. Fortunately, the number of free parameters can be greatly 
reduced by very general considerations. Unitarity and the 
freedom to arbitrarily choose the global phase of a quark field, reduce the 
original $2n_g^2$ unknowns (where $n_g=3$ is the number of generations) to 
$(n_g-1)^2$ unknowns. Among these $n_g(n_g-1)/2$ are rotation angles 
and $(n_g-1)(n_g-2)/2$ phases describe \CP violation.
Three generations allow for only a single \CP-violating phase. 
(In the lepton sector and if neutrinos are Majorana fermions, there are $n_g-1$
additional \CP-violating phases, which are however not observable in oscillation
experiments.)

Among the infinite number of possibilities to parameterize the CKM matrix 
in terms of four independent parameters, we recall the two most 
popular ones for the purpose of this review.

\subsubsection{CKM Parameterizations}

Chau and Keung~\cite{chau} proposed the ``standard parameterization'' 
of $\VCKM$. It is obtained by the product of three (complex) 
rotation matrices, where the rotations are characterized by the Euler 
angles $\theta_{12}$, $\theta_{13}$ and $\theta_{23}$, which are the 
mixing angles between the generations, and one overall phase $\delta$,
\beq
\label{eq:ckmPdg}
\VCKM = \pmatrix{
        c_{12}c_{13}
                &    s_{12}c_{13}
                        &   s_{13}e^{-i\delta}  \cr
        -s_{12}c_{23}-c_{12}s_{23}s_{13}e^{i\delta}
                &  c_{12}c_{23}-s_{12}s_{23}s_{13}e^{i\delta}
                        & s_{23}c_{13} \cr
        s_{12}s_{23}-c_{12}c_{23}s_{13}e^{i\delta}
                &  -c_{12}s_{23}-s_{12}c_{23}s_{13}e^{i\delta}
                        & c_{23}c_{13}
        }\,,
\eeq
where $c_{ij}=\cos\!\theta_{ij}$ and $s_{ij}=\sin\!\theta_{ij}$ for
$i<j=1,2,3$. This parameterization satisfies exactly the unitarity
relation.\footnote
{
        This phase $\delta$ describes \CP violation. It should not be
        confused with the \CP-conserving hadronic phases that are
	introduced later.
}

Following the observation of a hierarchy between the mixing angles,
$s_{13} \ll s_{23} \ll s_{12} \ll 1$, Wolfenstein~\cite{wolfenstein} proposed an
expansion of the CKM matrix in terms of the four parameters $\lambda$,
$A$, $\rho$ and $\eta$ ($\lambda\simeq|V_{us}|\approx 0.23$ being the 
expansion parameter), which is widely used in the contemporary literature.
We use the definitions to {\em all orders}~\cite{buras}
\beqn
\label{eq:burasdef}
        s_{12}             &\equiv& \lambda\,,\nonumber \\
        s_{23}             &\equiv& A\lambda^2\,, \\
        s_{13}e^{-i\delta} &\equiv& A\lambda^3(\rho -i\eta)\,.\nonumber
\eeqn
Inserting the above definitions 
into Eq.~(\ref{eq:ckmPdg}) gives exact expressions for
all CKM elements. Although expanding Eq.~(\ref{eq:ckmPdg}) in $\lambda$ 
is illustrative,
\beq\label{Wolfenstein}
V = \pmatrix{ 
	1-\frac12\lambda^2-\frac18\lambda^4 
		& \lambda 
			& A\lambda^3(\rho-i\eta) \cr 
	-\lambda 
		& 1-\frac12\lambda^2 -\frac18\lambda^4(1+4A^2)
			& A\lambda^2 \cr 
	A\lambda^3(1-\rho-i\eta) 
		& -A\lambda^2 + \frac{1}{2}A\lambda^4 \left(1 - 2(\rho+i\eta)\right)
			& 1 - \frac{1}{2}A^2\lambda^4
	} + {\cal O}(\lambda^5)\,,
\eeq
it is not necessary, nor is it necessary to truncate the expansion at any 
order in $\lambda$.  In Ref.~\cite{ckmfitter2004} and in this review no 
such approximation is made.

\subsubsection{The Jarlskog Invariant}
\label{sec:jarlskog}

A phase-convention-independent measure of \CP violation in the SM is 
given by
\beqa
\label{eq:commu}
\Im\det\Big(\big[M^u M^{u\dagger},M^d M^{d\dagger}\big]\Big) &=& 2\, J\,
  (m_t^2-m_c^2)(m_t^2-m_u^2)(m_c^2-m_u^2) \nn\\
&&{} \ \times (m_b^2-m_s^2)(m_b^2-m_d^2)(m_s^2-m_d^2)\,.
\eeqa
The Jarlskog invariant~\cite{jarlskog}, $J$, contains the dependence 
on the CKM elements,
\beq
\label{eq:jarlskog}
	\Im\left(V_{ij}V_{kl}V_{il}^*V_{kj}^*\right)
        = J \!\sum_{m,n=1}^3\!\! \varepsilon_{ikm}\, \varepsilon_{jln}\,,
\eeq
where $V_{ij}$ are the CKM matrix elements and $\varepsilon_{ikm}$
is the total antisymmetric tensor. Hence owing to the unitarity of 
$\VCKM$, the nonvanishing imaginary parts of all quadri-products of 
CKM elements are equal up to their sign. 
One representation of Eq.~(\ref{eq:jarlskog}) 
reads, for instance, $J = \Im (V_{ud}V_{cs}V_{us}^*V_{cd}^*)$.
A nonvanishing KM phase and hence \CP violation requires 
$J\ne0$. The Jarlskog parameter expressed in the standard 
parameterization~(\ref{eq:ckmPdg}) reads
\beq
	J = c_{12}c_{23}c_{13}^2s_{12}s_{23}s_{13}{\rm sin}\delta\,,
\eeq
and, using the Wolfenstein parameterization, 
\beq
        J = A^2\lambda^6\eta\left(1-\lambda^2/2\right)
                        + \mathcal{O}(\lambda^{10})
         \approx 3 \times 10^{-5}\,.\nonumber
\eeq
The empirical value of $J$ is small compared with its mathematical maximum 
of $1/(6\sqrt{3}) \approx 0.1$, showing that \CP  violation is suppressed 
as a consequence of the strong hierarchy exhibited by the CKM matrix 
elements. Remarkably, to account for \CP violation (see Eq.~(\ref{eq:commu})) 
requires not only a nonzero $J$, but also nondegenerate masses of the
up-type and down-type quarks.

Physically meaningful quantities are phase-convention invariant. 
Such invariants are the moduli, $|V_{ij}|$, and the quadri-products 
$V_{ij}V_{kl}V_{il}^*V_{kj}^*$ (\eg, the Jarlskog invariant $J$). 
Higher-order invariants can be rewritten as functions of these~\cite{CPV-TheBook}.
The Wolfenstein parameters  $\lambda=\Vus/\sqrt{\Vud^2+\Vus^2}$ and 
$A=|V_{cb}/V_{us}|/\lambda$ are phase-convention invariant, 
whereas $\rho$ and $\eta$ with $\rho+i\eta=V_{ub}^*/(A\lambda^3)$ are not.
We use phase-invariant representations and formulas throughout 
this review.

\subsubsection{Unitarity Triangles}

The allowed region for the CKM phase can be elegantly
displayed by means of the Unitarity Triangle (UT) described
by the {\em rescaled} unitarity relation between the first column
of the CKM matrix and the complex conjugate of the third column 
(\ie, corresponding to the $B_{\!d}$ meson system\footnote
{
	The subscript in $B_{\!d}$ is omitted in the following.
})
\beq
\label{eq:utriangle}
   0 = \frac{V_{ud}V_{ub}^*}{V_{cd}V_{cb}^*}
        + \frac{V_{cd}V_{cb}^*}{V_{cd}V_{cb}^*}
        + \frac{V_{td}V_{tb}^*}{V_{cd}V_{cb}^*}
     = {\cal O}(1) + 1 + {\cal O}(1)\, ,
\eeq
Note that twice the area of the {\em non-rescaled} UT equals the 
Jarlskog parameter $J$ (this is true for any one of the six UTs, because
they all have the same area). This provides a geometrical interpretation 
of the phase-convention invariance of $J$: A phase redefinition of the CKM 
matrix rotates the UT while leaving its area, and hence $J$, invariant.
It is a remarkable property of the UT in the $B$ system
that its three sides are proportional to the same power of $\lambda$
(so all sides of the rescaled UT~(\ref{eq:utriangle})
are of order one), which predicts large observable \CP violation.
In comparison, the corresponding UT for the kaon sector is strongly
flattened
\beq
   0 = \frac{V_{ud}V_{us}^*}{V_{cd}V_{cs}^*}
        + \frac{V_{cd}V_{cs}^*}{V_{cd}V_{cs}^*}
        + \frac{V_{td}V_{ts}^*}{V_{cd}V_{cs}^*}
     = {\cal O}(1) + 1 + {\cal O}(\lambda^4)\, ,
\eeq
implying small observable \CP violation.

\begin{figure}[t!]
\centerline{\includegraphics*[width=7.5cm]{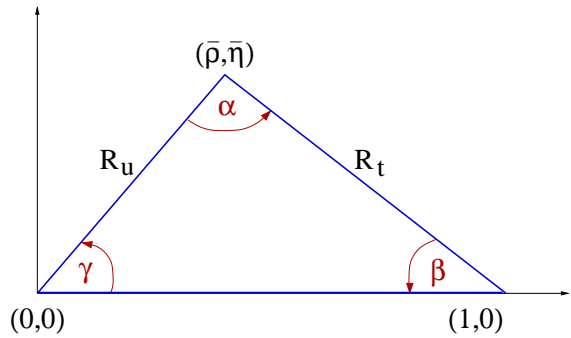}}
  \caption[.]{\label{fig:utriangle}\em
        The rescaled Unitarity Triangle.}
\end{figure}

The UT~(\ref{eq:utriangle}) is sketched in Fig.~\ref{fig:utriangle}
in the complex plane, where the apex is given by the following
phase-convention independent definition~\cite{Anikeev:2001rk},
\beqn
\label{eq:rhoetabar}
        \rhobar + i\etabar
        \equiv-\frac{V_{ud}V_{ub}^*}{V_{cd}V_{cb}^*}\,.
\eeqn
When the above equation is expressed in the Wolfenstein parameterization, 
one finds to all orders in $\lambda$~\cite{ckmfitter2004}\footnote
{
        Expanding Eq.~(\ref{eq:rhoetabar}) in $\lambda$ leads to the 
	well-known approximation
        \beqns
              \rhobar = \rho(1 - \lambda^2/2) + {\cal O}(\lambda^4)\,,
		\qquad
              \etabar = \eta(1 - \lambda^2/2) + {\cal O}(\lambda^4)\,.
	\eeqns
}
\beq
\label{eq:rhoetabarinv}
        \rho + i\eta
        \;=\; \frac{\sqrt{1-A^2\lambda^4}(\rhobar+i\etabar)}
                   {\sqrt{1-\lambda^2}\left[1 -
                   A^2\lambda^4(\rhobar+i\etabar)\right]}\,.
\eeq
The sides $R_u$ and $R_t$ of the UT (the third side along the real axis 
is normalized to unity) read to all orders,
\beq
\label{eq:rut}
R_u =
        \bigg|\frac{V_{ud}V_{ub}^*}{V_{cd}V_{cb}^*} \bigg|
                = \sqrt{\rhobar^2+\etabar^2}\,, \qquad
R_t =
        \bigg|\frac{V_{td}V_{tb}^*}{V_{cd}V_{cb}^*}\bigg|
                = \sqrt{(1-\rhobar)^2+\etabar^2}\,.
\eeq
The three angles of the UT, $\alpha$, $\beta$, $\gamma$, are defined by
\beq
\label{eq:utdefinitions}
\alpha = \arg\bigg(\! - \frac{V_{td}V_{tb}^*}{V_{ud}V_{ub}^*} \bigg)\,, \qquad
\beta  = \arg\bigg(\! - \frac{V_{cd}V_{cb}^*}{V_{td}V_{tb}^*} \bigg)\,, \qquad
\gamma = \arg\bigg(\! - \frac{V_{ud}V_{ub}^*}{V_{cd}V_{cb}^*} \bigg)\,,
\eeq
and the KM phase in the standard parameterization~(\ref{eq:ckmPdg})
is $\delta = \gamma+A^2\lambda^4\eta+\mathcal{O}(\lambda^6)$. 

At hadron machines there is copious production of $\Bs$ mesons. Here
the corresponding Unitarity Triangle, \UTs, is obtained by multiplying the 
second column of the CKM matrix by the complex conjugate of the third column, 
which is equivalent to replacing all $d$-quarks with $s$-quarks. Appropriately 
rescaled, one has
\beq
   0 =
   \frac{V_{us}V_{ub}^*}{V_{cs}V_{cb}^*} 
        + \frac{V_{cs}V_{cb}^*}{V_{cs}V_{cb}^*}
        + \frac{V_{ts}V_{tb}^*}{V_{cs}V_{cb}^*}	
     = {\cal O}(\lambda^2) + 1 + {\cal O}(1)\, ,
\eeq
The apex of the \UTs is defined accordingly by 
\beq
\label{eq:rhosetasbar}
        \rhobars + i\etabars
        \equiv-\frac{V_{us}V_{ub}^*}{V_{cs}V_{cb}^*}\,,	
\eeq
which is a short vector of order $\lambda^2$.
The small angle $\beta_s$ opposing the short side is given by 
\beq\label{betasdef}
   \beta_s  = \arg\bigg(\! - \frac{V_{ts}V_{tb}^*}{V_{cs}V_{cb}^*} \bigg)\,.
\eeq
The quantity $\stwobs$ can be determined from a time-dependent 
analysis of $\Bs\to\ J/\psi\phi$ decays.

\subsection{\CP Violation and Neutral-Meson Mixing}
\label{sec:CPviolation}

The most precise and among the theoretically cleanest information about 
the phase of the CKM matrix stems at present from the measurements of 
time-dependent \CP asymmetries in $B$ decays, so we need to review 
briefly the relevant formalism.  We use the \Bz system
as an example.  In general, all forms of \CP violation are related to
interference phenomena, because \CP violation is due to irreducible phases 
in the Lagrangian, which are observable only in interference experiments.

The conceptually simplest form of \CP violation, which can occur in both 
charged- and neutral-meson as well as baryon decays, is {\it \CP violation 
in decay}. If at least two amplitudes with nonzero relative weak $(\phi_k)$ 
and strong $(\delta_k)$ phases (that are odd and even under \CP, respectively) 
contribute to a decay,
\beq
A_f = \langle f | H |B\rangle = 
  \sum_k A_k\, e^{i\delta_k}\, e^{i\phi_k}\,, \qquad
\Abar_{\ov f} = \langle \ov f | H |\Bbar\rangle = 
  \sum_k A_k\, e^{i\delta_k}\, e^{-i\phi_k}\,.
\eeq
Then it is possible that $|{\ov A_{\ov f} / A_f}| \neq 1$, and thus \CP symmetry
is violated.  If there are two contributing amplitudes then $|{\ov A_{\ov f}|^2
- |A_f}|^2 \propto \sin(\phi_1-\phi_2) \sin(\delta_1-\delta_2)$. Because these
kinds of \CP asymmetries depend on strong phases, their interpretation is
usually model dependent (unless the strong phase can be eliminated by relating 
several decays to each other).  The best established observations of this type 
of \CP violation are $\Re\,\epsPr$ and the charge asymmetry in $\Bz\to
K^+\pi^-$. 

In neutral-meson decays there are other possibilities in which \CP violation
can occur. The two $B$-meson mass eigenstates are linear combinations of 
the flavor eigenstates,
\beq
\label{Bpq}
|B_{L,H}\rangle = p |B^0\rangle \pm q |\Bbar^0\rangle\,,
\eeq
with two complex parameters $p$ and $q$. The \CP symmetry
is violated if the mass eigenstates are not equal to the \CP eigenstates
({\it \CP violation in mixing}). This happens if $|q/p| \neq 1$, \ie, if the
physical states are not orthogonal, $\langle B_H | B_L\rangle \neq 0$, which
could not occur in a classical system.  The theoretical prediction of
$|q/p|$ requires the calculation of inclusive nonleptonic rates, which is
feasible in the heavy-quark limit, but in practice the uncertainties may be
sizable. This type of \CP violation is well established in the charge asymmetry
of semileptonic $\KL$ decay (given by $\delta_L \approx 2\Re\,\epsK$) or the
asymmetry between $\Kz_{t=0} \to e^+ X$ and $\Kzb_{t=0} \to e^- X$ (given by
$4\,{\rm Re}\, \epsilon_K$)~\cite{cplear}.  The latter asymmetry measuring \CP
violation in mixing is denoted by $A_{\rm SL}$ in the \B system, where it is
expected to be below the $10^{-3}$ level.

When both $\B^0$ and $\Bbar^0$ can decay to the same final state,
$f$, the time-dependent \CP asymmetry can be studied:
\beq\label{SCdef}
A_f(t) = {\Gamma[\Bbar^0(t)\to f] - \Gamma[B^0(t)\to f]\over
  \Gamma[\Bbar^0(t)\to f] + \Gamma[B^0(t)\to f] }
= S_f \sin(\Delta m\, t) - C_f \cos(\Delta m\, t)\,,
\eeq
where $\Delta m=m_{B_H}-m_{B_L}$ and $t$ is the proper decay time 
of the $\B$, and where we have assumed $\CPT$ invariance and neglected lifetime 
differences in the neutral $\B$-meson system (see
Refs.~\cite{Anikeev:2001rk} and \cite{pdg_cp} 
for a more general formulation of the time dependence).
The $S$ and $C$ coefficients can be calculated in terms of the $\B^0\Bbar^0$ 
mixing parameters and the decay amplitudes,
\beq\label{SClamdef}
S_f = {2\,\Im\,\lambda_f\over 1+|\lambda_f|^2}\,, \qquad
C_f = {1-|\lambda_f|^2 \over 1+|\lambda_f|^2}\,, \qquad
\lambda_f = \frac qp\, \frac{\Abar_f}{A_f}\,. 
\eeq
It is possible that $|q/p|=1$ and $|\lambda_f| = 1$, \ie, there  is no \CP
violation in either mixing or decay, but the \CP asymmetry in  Eq.~(\ref{SCdef})
is nonzero, because $\Im\lambda_f \neq 0$ ({\it \CP violation in the
interference between decay with and without mixing}, or
{\it mixing-induced \CP violation}).

For decays in which the final state is a \CP eigenstate and amplitudes with
one weak phase dominate, $A_{f_{CP}}$ measures a phase in the Lagrangian
theoretically cleanly, independent of hadronic matrix elements.  (However, $f$
does not have to be a \CP eigenstate for this type of \CP violation to occur,
nor for it to have a clean interpretation.)  For such decays, the \CP symmetry
of QCD implies $A_{f_{\CP}} = \eta_{f_{\CP}} A_{\ov{f}_{\CP}}$, where
$\eta_{f_{CP}}$ is the \CP eigenvalue of $f$, simplifying the evaluation of
$\lambda_f$. In such cases $C_{f_{\CP}} = 0$ and $S_{f_{\CP}} = \Im\,
\lambda_{f_{CP}} = \sin(\arg\lambda_{f_{\CP}})$, where $\arg \lambda_{f_{CP}}$
is the phase difference between the $\Bbar^0\to f$ and $\Bbar^0\to B^0\to f$ decay
paths. For the generic case, where several amplitudes with different weak
phases contribute, detailed knowledge of $\Abar_f/A_f$ is necessary to interpret 
the experimental measurements. In particular, the strong phases of the 
amplitudes, which usually result in hadronic uncertainties in the
interpretation of $S_f$, and may also give rise to $C_f \neq 0$, must
be determined. Note that if $C_f$ is small, it does not imply that $S_f$ 
provides clean information on short-distance physics. If there are 
amplitudes with different weak but small relative strong phases, then 
$C_f \approx 0$, but $S_f$ still depends on the hadronic physics.

\subsection{Physics Beyond the Standard Model}

There are many reasons to believe that there is physics beyond the SM.  The
evidence for dark matter implies not-yet-seen particles, the observed
matter-antimatter asymmetry of the Universe implies not-yet-observed \CP
violation, and neutrino masses make the existence of heavy right-handed
neutrinos an appealing scenario.  More aesthetic reasons include
the gauge hierarchy problem that requires new physics to 
stabilize the huge $M_{\rm Planck}/M_{\rm weak}$ ratio, grand unification
of the electroweak- and strong-interaction couplings (that appears 
naturally with Supersymmetry (SUSY) at the TeV scale), the
strong \CP problem that may imply the existence of an axion, and the evidence
for dark energy that may simply be a cosmological constant
in the Einstein equations.  Some of these may also relate to new flavor physics
that could have observable effects in low-energy experiments.

It is interesting to probe the flavor sector, because it is not well understood.
In the SM, the origin of quark-mass and mixing-angle hierarchies are not 
understood.  And if there is new physics at the TeV scale, as conjectured
from the gauge hierarchy problem and grand unification, we do not understand 
why it does not show up in flavor physics experiments.  A four-quark operator 
$(s\dbar )^2/\Lambda_{\rm NP}^2$ with ${\cal O}(1)$ coefficient would give a
contribution exceeding the measured value of $\epsK$ unless $\Lambda_{\rm
NP} \gsim 10^4\,\TeV$.  Similarly, $(d\bbar )^2/\Lambda_{\rm NP}^2$ yields
$\dmd$ above its measured value unless $\Lambda_{\rm NP} \gsim
10^3\,\TeV$.  Extensions of the SM typically have many new sources of \CP
violation and flavor-changing neutral-current interactions, and in most
scenarios,  to satisfy the experimental constraints, specific mechanisms 
must be introduced to suppress these.

Supersymmetric models are a good example.  Compared with the SM, their flavor
sector contains 59 new \CP-conserving parameters and 41 new \CP-violating
phases~\cite{Haber:1997if,Nir:2005js}, and many of them
have to be suppressed to not contradict the experimental data. The smallness of
$\epsK$ and $\Delta m_K$ have driven SUSY model building to a large extent,
and have given rise to several mechanisms that constrain the relevant soft 
SUSY-breaking terms.  Generic models with superpartner masses at the TeV 
scale also violate the current bounds on electric dipole moments, and fine 
tuning at the 1\% level is required to avoid conflict with the data.

We do not know whether there is new flavor physics at the TeV scale, so it is
hard, if not impossible, to make firm predictions. If the new flavor scale is at
the TeV scale (where the hierarchy problem suggests that some new physics should
exist), it could have observable effects in the (low-energy) flavor sector.  If
the flavor scale is much above the TeV scale, as is the case for example in
models with gauge-mediated SUSY breaking, then we expect no significant
deviations from the CKM picture. In any case, if we see new particles at the
Large Hadron Collider (LHC), it will be much clearer how to use flavor 
physics to learn about the new couplings.

\section{Theoretical Framework and Tools}

With the remarkable exception of the UT angles, the
experimental observables presently used to constrain $\rhobar$ and
$\etabar$ depend on hadronic matrix elements.  QCD is well established as the
theory of strong interaction, and it has been tested to high precision in the
perturbative regime where the coupling constant $\as$ is small.  However,
presently it is difficult to obtain quantitative predictions in the low-energy 
regime, except for a few special cases. Although it is beyond the scope 
of this review to discuss the wealth of approaches to nonperturbative QCD, it is
useful to recall a few general techniques to evaluate the matrix elements 
relevant to quark-flavor physics. 

The methods reviewed below all give controllable
systematic errors, by which we mean that the uncertainties can be incrementally
improved in a well-defined way, expanding in small parameters order by order.
Most of the model-independent theoretical tools utilize that some quark
masses are smaller while others are greater than $\lqcd$ (throughout this 
review $\lqcd$ denotes a typical hadronic scale, of order 500\,MeV). 
Expanding in the resulting small ratios can be used to simplify some of 
the hadronic physics.
However, depending on the process under consideration, the relevant hadronic
scale may or may not be much smaller than $m_b$.  For example, $f_\pi$,
$m_\rho$, and $m_K^2/m_s$ are all of order $\lqcd$, but their numerical values
span an order of magnitude.  In most cases experimental guidance is necessary to
determine how well the expansions work.

\subsection{Effective Hamiltonians for Weak Decays}

All flavor-changing interactions (except that of the top quark) are due to tree
and loop diagrams involving heavy virtual particles: $W$ bosons in the SM, or
not-yet-discovered particles in its extensions.  These particles propagate over
much shorter distances than $1/m_b$, so their interactions can be described by
local operators.  In principle, there is an infinite number of such operators.
The contributions of the higher dimensional ones are however suppressed by
increasing powers of $m_b/m_W$, so it is sufficient to consider the first few
operators.  The effective weak Hamiltonian can be written as
$H_W = \sum C_i(\mu)\, O_i(\mu)$, where $O_i$ are the lowest dimensional
operators contributing to a certain process and $C_i$ are their Wilson
coefficients, with perturbatively calculable scale dependences.  It is possible
to sum the large logarithms of $m_W^2/m_b^2$ by first calculating $C_i(\mu \sim
m_W)$ (matching) and then evolving the effective Hamiltonian down to a scale
$\mu \sim m_b$ using the renormalization group (running). This shifts
the large logarithms from the matrix elements of $O_i$ to $C_i$. In this last
step the operators can in general mix; for example, $C_2(m_W)$ affects all of
$C_{1-9}(m_b)$ below.

The simplest examples are semileptonic decays, where integrating out the
virtual $W$ in $b\to c\ell\bar\nu$, for instance, gives
\beq
H_W = \frac{G_F}{\sqrt 2}\, V_{cb}\, (\bar c b)_{V-A}\,
  (\bar\ell \nu)_{V-A}\,,
\eeq
where $V-A$ denotes the Dirac structure $\gamma^\mu(1-\gamma_5)$ between the
fermion fields.  In this case there is only one operator, and its coefficient is
scale independent because the (axial) vector current is (partially) conserved.
Semileptonic decays involving a $\bar\ell\ell$ pair and nonleptonic decays are
more complicated.  The $\Delta B=1$ effective Hamiltonian has both $\Delta S=0$
and 1 terms.  The $\Delta S=1$ part,
\beq\label{Hw}
H_W = \frac{G_F}{\sqrt 2} \sum_{p=u,c} V_{pb} V^*_{ps}\,
  \Big[ C_1 O_1^p + C_2 O_2^p + \sum_{i\geq3} C_i O_i \Big]\,,
\eeq
contains the operators
\beq\label{fullops}
\begin{array}{rclrcl}
O_1^u &=& (\bar u_\beta b_\alpha)_{V-A}\,
  (\bar s_\alpha u_\beta)_{V-A} \,, 
  & O_2^u &=& (\bar u b)_{V-A}\,
  (\bar s u)_{V-A}\,, \\
O_1^c &=& (\bar c_\beta b_\alpha)_{V-A}\,
  (\bar s_\alpha c_\beta)_{V-A} \,, 
  & O_2^c &=& (\bar c b)_{V-A}\,
  (\bar s c)_{V-A}\,, \\
O_3 &=& \sum_q (\bar s b)_{V-A}\,
  (\bar q q)_{V-A} \,,
  & O_{4} &=& \sum_q (\bar s_\beta b_\alpha)_{V-A}\,
  (\bar q_\alpha q_\beta)_{V-A} \,, \\
O_5 &=& \sum_q (\bar s b)_{V-A}\,
  (\bar q q)_{V+A} \,, 
  & O_6 &=& \sum_q (\bar s_\beta b_\alpha)_{V-A}\,
  (\bar q_\alpha q_\beta)_{V+A} \,, \\
O_7 &=& -\frac{e}{8\pi^2}\, m_b\, \bar s\, \sigma^{\mu\nu}
  F_{\mu\nu} (1 + \gamma_5) b \,, \qquad
  & O_8 &=& -\frac{g}{8\pi^2}\, m_b\, \bar s\, \sigma^{\mu\nu}
  G_{\mu\nu}^a T^a (1 + \gamma_5) b \,, \\
O_9 &=& \textstyle {e^2\over 8\pi^2}\, (\bar s b)_{V-A}\,
  (\bar\ell \ell)_V\,, \qquad
  & O_{10} &=& {e^2\over 8\pi^2}\, (\bar s b)_{V-A}\,
  (\bar\ell \ell)_A\,.
\end{array}
\eeq
The $\Delta S=0$ part is obtained by replacing $s\to d$ in Eqs.~(\ref{Hw}) and
(\ref{fullops}).  Here $O_{1,2}^u$ and $O_{1,2}^c$ are current-current
operators, $O_{3-6}$ are penguin operators with a sum over $q=u,d,s,c,b$ flavors
($\alpha,\beta$ are color indices).  The $C_i$ coefficients in Eq.~(\ref{Hw})
are known at next-to-leading logarithmic order~\cite{bbl}.  In nonleptonic \B decays,
four more operators corresponding electroweak penguin diagrams contribute to $H_W$:
\beq
\begin{array}{rclrcl}
O_7^{\rm ew} &=& \sum_q \frac32\, e_q\, (\bar s b)_{V-A}\,
  (\bar q q)_{V+A}\,, \qquad
  & O_8^{\rm ew} &=& \sum_q \frac32\, e_q\,
  (\bar s_\beta b_\alpha)_{V-A}\,
  (\bar q_\alpha q_\beta)_{V+A} \,, \\
O_9^{\rm ew} &=& \sum_q \frac32\, e_q\,
  (\bar s b)_{V-A}\, (\bar q q)_{V-A}\,, 
  & O_{10}^{\rm ew} &=& \sum_q \frac32\, e_q\,
  (\bar s_\beta b_\alpha)_{V-A}\,
  (\bar q_\alpha q_\beta)_{V-A} \,.
\end{array}
\eeq 

Non-SM physics can either modify the coefficients $C_i$ at the weak scale, or
give rise to additional operators (for example, those obtained by replacing
$(\bar s b)_{V-A}$ by $(\bar s b)_{V+A}$).  The goal is to find out if the data
show evidence for either type of modification.  The largest source of
theoretical uncertainty usually stems from our limited ability to compute the
hadronic matrix elements of such operators, and the rest of this section 
introduces methods that allow calculating them, or relating them between various
observables.

\subsection{Chiral Symmetry}

The $u$, $d$ and $s$-quark masses are small compared with $\lqcd$,
so it is useful to consider the $m_q\to 0$ limit ($q = u,d,s$) and 
treat corrections perturbatively. This is known as the chiral limit,
because the Lagrangian for the light quarks has a $\SU(3)_L \times
\SU(3)_R$ chiral symmetry, under which the left- and right-handed quarks
transform differently.  This symmetry is spontaneously broken to $\SU(3)_V$ by
the vacuum expectation value of the quark bilinears, $\langle \bar q_R^i
q_L^j\rangle = v \delta^{ij}$.  The eight broken generators are related to
Goldstone bosons, the three pions, four kaons, and the $\eta$ (for now, we
neglect $\eta$-$\eta^\prime$-$\pi^0$ mixing). Chiral symmetry relates 
different hadronic matrix elements to one another, and has very diverse 
applications in flavor physics.

Because the $u$ and $d$-quark masses are small, the $\SU(2)$ isospin symmetry
between the $u$ and $d$ is usually a very good approximation.  The corrections
to the chiral limit are suppressed by $(m_d-m_u)/\Lambda_{\chi\rm SB}$---where
$\Lambda_{\chi\rm SB} \approx 1\gev$ (of order $m_\rho$ or $4\pi f_\pi$) is
the chiral symmetry breaking scale---and are usually not larger than a few
percent at the amplitude level.  There are also explicit violations of chiral
symmetry, for example, due to weak or electromagnetic interactions. In certain
cases these effects can be enhanced as is the case, \eg, in the neutral 
versus charged $B$ (or $D$) meson lifetimes.  The full $\SU(3)$ symmetry is 
broken by $m_s/\Lambda_{\chi\rm SB}$, and is known to have typically $20-30\%$
corrections.  (The same is true for its $u$-spin and $d$-spin $\SU(2)$ subgroups,
which act on the $ds$ and $us$ pairs, respectively.)

Some of the most prominent cases of isospin symmetry in the context of 
the CKM matrix include relations between
amplitudes involving charged and neutral pions, the determination of $|V_{ud}|$
(Sec.~\ref{Vud}), and the extraction of the UT angle $\alpha$ from $B\to\pi\pi$
decays (Sec.~\ref{ut_alpha}).  Similarly, $\SU(3)$ symmetry and chiral
perturbation theory are key ingredients in determining $|V_{us}|$
(Sec.~\ref{Vus}).  It also relates form factors and certain matrix elements
involving pions and kaons to one another, a relation that has many applications. 
Recently, the $\SU(3)$ relations between nonleptonic decays have been 
extensively studied, because the $\pi\pi$, $K\pi$, and $KK$ amplitude 
relations give sensitivity to the UT angle $\gamma$ and possibly to 
new physics.  $\SU(3)$  has also been used as a bound on the SM-induced 
deviations of the time-dependent \CP asymmetries from $\sin2\beta$ in 
the penguin-dominated modes (see Sec.~\ref{ut_loop}).

\subsection{Heavy-Quark Symmetry and Heavy-Quark Effective Theory}
\label{sec:hqs}

In mesons composed of a heavy quark and a light antiquark (plus gluons and
$q\qbar $ pairs), the energy scale of strong interactions is small compared with
the heavy-quark mass.  The heavy quark acts as a static point-like color source
with fixed four-velocity, which cannot be altered by the soft gluons responsible
for confinement.  Hence the configuration of the light degrees of freedom (the
so-called ``brown muck") becomes independent of the spin and flavor (mass) of the
heavy quark, which, for $N_f$ heavy-quark flavors, results in a $\SU(2N_f)$ 
heavy-quark spin-flavor symmetry~\cite{Isgur:1989ed}.

Heavy-quark spin-flavor symmetry has many important implications for the 
spectroscopy and strong decays of
$B$ and $D$ mesons (for a review, see \eg~\cite{Manohar:2000dt}). It is especially
predictive for exclusive $B\to D^{(*)}\ell\nub$ semileptonic decays, which are
relevant for the determination of \Vcb (Sec.~\ref{Vcb}).  When the weak current 
suddenly changes the flavor (on a time scale $\ll \lqcd^{-1}$), momentum, 
and possibly the spin of the $b$-quark to a $c$-quark, the brown muck only 
notices that the four-velocity of the static color source has 
changed, $v_b \to v_c$.  Therefore, the form factors that
describe the wave-function overlap between the initial and final mesons become
independent of the Dirac structure of weak current, and depend only on a scalar
quantity, $w = v_b \cdot v_c$.  Thus all six $B\to D^{(*)}\ell\nub$ form
factors are related to a single Isgur-Wise function, $\xi(v_b\cdot v_c)$, which
contains all the low-energy nonperturbative hadronic physics relevant for these
decays.  Moreover, $\xi(1)=1$ because at zero recoil---where the $c$
quark is at rest in the $b$ quark's rest frame---the configuration of the 
brown muck does not change at all.

Deviations from the heavy-quark limit can be included using the heavy-quark
effective theory (HQET)~\cite{Georgi:1990um}, which provides a systematic
expansion in powers of $\as(m_Q)$ and $\lqcd/m_Q$ ($Q=b,c$).  The former
type of corrections are calculable perturbatively, whereas the latter ones 
can be parameterized by a minimal set of hadronic matrix elements that can 
be extracted from data and/or estimated using nonperturbative techniques.  
Heavy-quark spin symmetry also implies relations, in
combination with chiral symmetry, between, for example, $B\to\rho\ell\nub$
and $B\to K^*\ell^+\ell^-$ form  factors~\cite{Isgur:1990kf}.

\subsection{Factorization and Soft-Collinear Effective Theory}
\label{sec:scet}

Researchers have long known that in the decay $B\to M_1M_2$, if the meson $M_1$ 
that inherits the spectator quark from the $B$ is heavy and $M_2$ is light then
``color transparency" can justify
factorization~\cite{Bjorken:1988kk,Dugan:1990de}.  Traditionally, naive
factorization refers to the hypothesis that matrix elements of the four-quark 
operators can be estimated by grouping the quark fields into a pair that can
mediate $B\to M_1$ transition and into another pair that describes $\mbox{vacuum}
\to M_2$ transition.

Recently the development of the soft collinear effective theory
(SCET)~\cite{sceta,scetb} put these notions on a firmer footing. SCET
is designed to describe the interactions of energetic and low invariant-mass
partons in the $Q \gg \lqcd$ limit.  It introduces distinct fields for the
relevant degrees of freedom, and a power-counting parameter $\lambda$.  
There are two distinct theories, SCET$_{\rm I}$ in which
$\lambda=\sqrt{\lqcd/Q}$ and SCET$_{\rm II}$ in which $\lambda=\lqcd/Q$.  They
are appropriate for final states with invariant mass $Q\lambda$; \ie,
SCET$_{\rm I}$ for jets and inclusive $B\to X_s\gamma$, $X_u\ell\nub$,
$X_s\ell^+\ell^-$ decays ($m_X^2 \sim \lqcd Q$),
and SCET$_{\rm II}$ for exclusive hadronic final states ($m^2 \sim \lqcd^2$). 
It is convenient to use light-cone coordinates, decomposing momenta as
\beq
p^\mu = (\bar n\cdot p) {n^\mu\over2} + p_\perp^\mu + (n\cdot p) 
  {\bar n^\mu\over2} = p_-{n^\mu\over2} + p_\perp + p_+ {\bar n^\mu\over2}, 
\eeq
where $n^2 = \bar n^2 = 0$ and $n \cdot \bar n = 2$.  For a light quark moving
near the $n$ direction, $p_- \gg p_\perp \gg p_+$, and this hierarchy can 
be used to successively integrate out the less and less off-shell momenta.
Factorization has been proven in ($B\to D\pi^\pm$)-type
decays~\cite{BPSdpi} to all orders in $\as$ and at leading order in
$\lqcd/Q$ by decoupling the ultrasoft [$(p_-,
p_+, p_\perp) \sim Q(\lambda^2,\lambda^2,\lambda^2)$] gluons from the collinear
[$(p_-, p_+, p_\perp) \sim Q(1,\lambda^2,\lambda)$] Lagrangian
at leading order in $\lambda$.

The study of heavy to light transitions, which are particularly important for
\CP violation and CKM measurements, is complicated by the fact that one has to
work to subleading order in $\lambda$.  This is because collinear and ultrasoft
quarks cannot interact at leading order, only via the mixed ultrasoft-collinear
Lagrangian, ${\cal L}_{\xi q}^{(1)}$, which is suppressed by one power of
$\lambda$ and allowed to couple an ultrasoft and a collinear quark to a collinear
gluon~\cite{Beneke:2002ph}.  This is relevant for all processes in which the
spectator quark in the $B$ ends up in an energetic light meson. So far, maybe
the most surprising model-independent result at subleading order is proof of
factorization in the previously intractable $B\to D^0\pi^0$ type 
``color-suppressed" decays~\cite{Mantry:2003uz}.  In this case factorization 
means the systematic separation of the physics associated with different momentum 
scales, and  the factorization theorem can even accommodate a nonperturbative 
strong phase.

Other important applications include the exclusive semileptonic form factors,
and factorization in charmless two-body decays, both of which are subject to
intense discussions.  For semileptonic $B\to\pi\ell\nub$ and
$\rho\ell\nub$ decays, there are two contributions of
the same order in $\lqcd/Q$.  It is not yet clear whether or not the
factorizable parts that  violate the form factor relations~\cite{Charles:1998dr}
are smaller than the nonfactorizable contributions.  For charmless
decays~\cite{qcdf,bprs}, an additional complication is that the power counting
appropriate for the charm penguins has not been settled~\cite{pcdebate,cpengs}.

\subsection{The Operator Product Expansion for Inclusive Decays}
\label{sec:ope}

In the large $m_b$ limit, there is a simple argument based on a separation of
scales, that inclusive rates may be modeled by the decay of a free $b$-quark. 
The weak decay takes place on a time scale much shorter than the time it takes
the quarks in the final state to hadronize.  Once the $b$-quark has decayed, the
probability that the decay products hadronize is equal to unity, and 
the (unknown) probabilities of hadronization to specific final states
can be ignored.

\begin{figure}[t!]
\centerline{\includegraphics*[height=2.8cm]{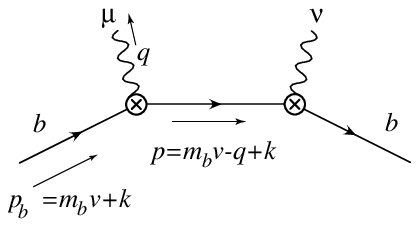} \hspace*{1cm}
  \includegraphics*[height=3cm]{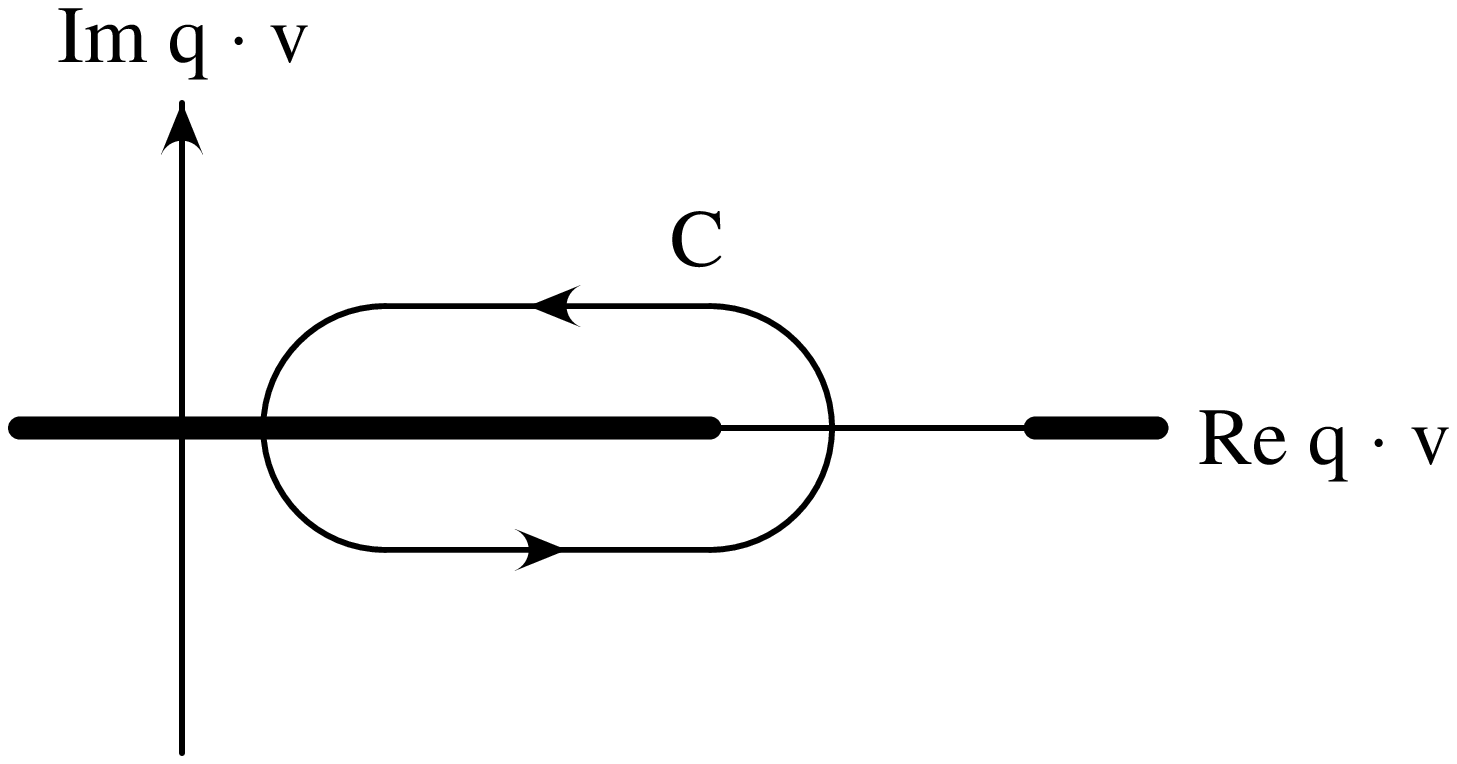}}
\caption{\it (Left) Forward-scattering amplitude whose imaginary part gives the
decay rate.  (Right) Analytic structure of the correlation function in the
complex $q\cdot v$ plane and the integration contour that gives the total
rate.}
\label{fig:opesketch}
\end{figure}

This argument can be formalized using an operator product expansion
(OPE)~\cite{OPE}.  To calculate the semileptonic rates, the
leptonic and the hadronic parts of the $\langle X\ell \nub\, | H\, |B\rangle$
matrix element can be separated.  Nonperturbative strong-interaction effects 
enter the rate via the hadronic tensor,
\beq
W^{\mu\nu} = \sum_X\, (2\pi)^3\, \delta^4(p_B-q-p_X)\,
  { \langle B | J^{\mu\dagger} | X \rangle\, 
  \langle X| J^\nu |B\rangle \over 2m_B } \,,
\eeq
where $J^\mu = \qbar \, \gamma^\mu P_L\, b$ for semileptonic $B$ decay.  The
optical theorem implies that $W^{\mu\nu}$ can be related to the discontinuity
across the cut of  the forward-scattering matrix element of the time-ordered
product,
\beq\label{1:top}
T^{\mu\nu} = - i \int \d^4 x\, e^{-iq\cdot x}\, { \langle B |\, 
  T [ J^{\mu\dagger}(x)\, J^\nu(0) ]\, |B\rangle \over 2m_B }\,,
\eeq
\ie, $W^{\mu\nu} = - \frac1\pi\, \Im\, T^{\mu\nu}$. When the energy release to
the hadronic final state is large, the time-ordered product can be expanded in
local operators, whose $B$-meson matrix elements can be parameterized using
HQET.  $T^{\mu\nu}$ is an analytic function with singularities that correspond
to on-shell hadronic intermediate states.  For semileptonic decays one can
deform the integration contour away from the physical region in the complex
$q\cdot v$ plane (see Fig.~\ref{fig:opesketch}), and asymptotic freedom ensures
that $T^{\mu\nu}$ can be reliably calculated far from its singularities.

In the $m_b\gg\lqcd$ limit, the OPE reproduces the $b$-quark decay result, and
shows that the leading power-suppressed corrections occur at order
$\lqcd^2/m_b^2$ and can be parameterized by two matrix elements, usually denoted
by $\lambda_1$ and $\lambda_2$.  The result is of the form
\beq\label{incl}
\d\Gamma = \bigg(\begin{array}{c}\mbox{$b$-quark} \\[-0.1cm] 
                                 \mbox{decay}
                 \end{array}
           \bigg) \times 
\bigg[ 1 + \frac0{m_b} + \frac{{\cal O}(\lambda_1,\lambda_2)}{m_b^2} + \ldots
  \bigg] \,.
\eeq
Most decay rates of interest have been computed including the $\lqcd^3/m_b^3$
corrections (which are parameterized by six additional hadronic matrix
elements), whereas the $b$-quark decay rates have been computed including 
perturbative corrections at order $\as$ and 
$\as^2\beta_0$ ($\beta_0 = 11 - 2n_f/3$ is the first coefficient of the 
QCD $\beta$-function, and the terms proportional to it often dominate at 
order $\as^2$).

The OPE can also be used to calculate inclusive nonleptonic decays, such
as heavy hadron lifetimes, the $B_s$ width difference, $\Delta\Gamma_s$, and the
semileptonic \CP asymmetry, $A_{\rm SL}$.  Whereas the structure of the results
is the same as for semileptonic decays, there is no ``external" current to the
strong interaction in these cases, so there is no variable that can be
used to move the integration contour away from the cuts.  Therefore, the OPE is
performed directly in the physical region, and we expect the accuracy of
these calculations to be worse than those for semileptonic decays.  Experimental
data are crucial to validate each of the applications; in particular, the unexpected
smallness of the $\Lambda_b$ to $B$ lifetime ratio may indicate that the nonleptonic 
calculations are more difficult to control. 

\subsection{Lattice QCD}
\label{lattice}

In lattice QCD (LQCD), the functional integral for correlation functions is 
computed using numerical Monte Carlo integration.  To keep the degrees of 
freedom and the required computational power at an acceptable level, a 
finite lattice spacing, $a$, is introduced to discretize space-time, and 
the calculation is performed within a finite volume.  The finite lattice spacing 
(finite volume) is an ultraviolet (infrared) cutoff, which can be dealt with
using field theory techniques to extract cutoff-independent quantities from the
results of the computations.  In practical calculations the ultraviolet cutoff is
typically approximately $a^{-1} \sim (2$--$3)\,\GeV$.  Consequently, $b$-quarks 
can be simulated only using actions based on HQET
and nonrelativistic QCD, where $m_b$ is removed from the dynamics using the
effective theory.  The matching between the lattice and the continuum is
in principle straightforward, but in practice it often introduces a 
significant uncertainty, because $\as(a^{-1})$ is not small, and few matching
calculations have been done beyond one loop.

The most difficult issues are related to how the fermionic part of the action is
computed.  Until a few years ago, to make the computations manageable, most
calculations used the quenched approximation, which amounts to neglecting
virtual quark loops.  It is not a systematic approximation to QCD and the error
associated with it is difficult to estimate. Recent advances in algorithms and
computers have allowed an increasing number of quantities to be computed with 
dynamical $u$, $d$, and $s$-quarks.  A remaining source of systematic uncertainty 
is often the chiral extrapolation.  The simulations with light $u$ and $d$-quarks
are computationally expensive, so the use of $m_{u,d}/m_s \sim 1/2$ is  not
uncommon. To date only the MILC collaboration has used  $m_{u,d}/m_s \sim 1/8$, 
which still has to be extrapolated to the physical values, $m_{u,d}/m_s \sim 1/25$. 
The results based on MILC's gauge configurations use improved ``staggered''
quarks.  There are several ways to discretize the quark action, and the cost of
simulations with light quarks is sensitive to this choice.  Unquenched
calculations with staggered quarks are much faster than with the Wilson, domain
wall, or overlap formulations.  In the staggered formulation each field produces
four quark ``tastes", so there are 16 light mesons.  The unphysical ones are
removed by taking the fourth root of the fermion determinant.  The validity of
this fourth-root trick has so far been proven only in perturbation theory.

Recently many new results have emerged in which all these systematic effects
are studied in detail.  The gold-plated quantities are those that contain at
most one hadron in both the initial and final state, not moving with
large velocities.  These include decay constants, bag parameters, and
semileptonic form factors, which are important for flavor physics.  For most of
these quantities, the currently available calculations using staggered fermions
obtain significantly smaller errors than those with Wilson and other
formulations.  The fastest LQCD computers now run at a computation rate of a few
tens of TeraFLOPs, and the  PetaFLOP barrier is expected to be passed by the end
of this decade.  This may allow LQCD to make predictions approaching the percent
level, especially if all ingredients of the calculation can be performed
numerically (including matching from the lattice to the continuum). With such
precision, several measurements in exclusive decays, which are sensitive to new
physics but are presently dominated by theoretical uncertainties in the
corresponding matrix elements, could become precise tests of the SM (such as
$B\to \rho\gamma$,  $B\to \pi\ell\bar\nu$, $\epsK$, etc.).

\section{Determining Magnitudes of CKM Elements}

Determining of the magnitudes of CKM elements uses a number of
sophisticated theoretical and experimental techniques, the complete discussion 
of which is beyond the scope of this review.  We concentrate on 
topics where recent developments have occurred.

\subsection{$|V_{ud}|$ from $\beta$-Decays}
\label{Vud}

The CKM matrix element $\Vud$ has been extracted by three different 
methods: superallowed nuclear $\beta$-decays with pure Fermi transitions 
($0^+\to 0^+$), neutron $\beta$-decay ($n\to p e\nueb$), and pion $\beta$-decay 
($\pi^+\to \piz e^+\nue(\gamma)$). All three methods involve the hadronic 
form factor of the vector current
$\langle f |\ubar\gamma_\mu d|i\rangle$, where $| i\rangle$ and $|f\rangle$
are the initial and final states of the transitions.
The neutron $\beta$-decay also involves an axial-vector form factor,
which requires external input for the ratio of the axial-vector to
the vector coupling constants. 
The form factors describe the long-distance hadronic effects that 
confine the quarks within hadrons. The normalization of the vector
current is fixed at the kinematic endpoint of zero-momentum transfer 
between initial and final states, and at the exact isospin limit 
($m_u = m_d$). Isospin-breaking corrections are suppressed by the 
quark mass difference divided by the hadronic scale. 

Superallowed nuclear $\beta$-decays give the best experimental precision.
The product of the integral over the emitted electron energy spectrum 
and the neutron lifetime is proportional to $|V_{ud}|^{-2}$.
Radiative and charge symmetry breaking corrections, which depend in 
part on the nuclear structure of the nucleus under consideration must
be applied. They lie between $3.1\%$ and $3.6\%$ for the nine superallowed
transitions that give the best sensitivity~\cite{savardetal}. Recent theoretical
improvements in the calculation of loop contributions, notably 
for the problematic $\gamma W$ box diagram, and a detailed error analysis
achieved a reduction in uncertainty by a factor of two to 
$1.9\times10^{-4}$~\cite{marcianosirlin}. With this, the world average 
is $|V_{ud}|_{0^+\to 0^+}=0.97377\pm0.00027$~\cite{vud_ckm2005}.

From a theoretical point of view, the neutron $\beta$-decay, or 
ultimately the pion $\beta$-decay, represent the cleanest determination 
of $|V_{ud}|$, because they are free from nuclear structure effects. 
Because no other decay channels exist, the decay rate can be obtained 
from lifetime measurements. The PERKEO-II experiment achieved significant 
improvement in this $|V_{ud}|$ determination owing to the precise 
measurement of $g_A/g_V=-1.2739\pm0.0019$~\cite{PERKEO}
(confirmed in Ref.~\cite{mund}), a value that is, 
however, significantly larger than earlier measurements. Several new 
experiments are planned or are in construction to remeasure this quantity.
Unfortunately, there is currently a problem with the second input to 
determine $|V_{ud}|$ from neutron $\beta$-decays, namely a large discrepancy 
between a new neutron lifetime measurement~\cite{serebrov} and the PDG 
average~\cite{PDG}. Such a discrepancy, if confirmed, would significantly 
impact the extracted value for $|V_{ud}|$, and hence CKM unitarity. 
Ignoring this measurement, the present world average from neutron $\beta$-decays
is $|V_{ud}|_{n_\beta}=0.9730\pm0.0004\pm0.0012\pm0.0002$~\cite{vud_ckm2005},
where the errors arise from the neutron lifetime, from $g_A/g_V$, 
and from radiative corrections, respectively. Note that the change
induced by the new lifetime measurement, if confirmed, would be as large as 
three times the total error quoted.

The pion $\beta$-decay $\pi^{+} \to \pi^{0} e^{+} \nu_{e}$ is an attractive 
candidate to extract $|V_{ud}|$ from the branching ratio of the decay and the 
pion lifetime. It is mediated by a pure vector transition. Unfortunately, 
because of the small branching 
ratio of $10^{-8}$, the statistical precision is not yet competitive with the 
other methods: $|V_{ud}|=0.9748 \pm 0.0025$~\cite{PIBETA,vud_ckm2005}.

\subsection{CKM Elements from (Semi)Leptonic Decays}

\subsubsection{Leptonic Decays}

The decay rate of the lightest pseudoscalar meson, $M$---which is 
composed of the quarks $q_d\bar q_u$---to a lepton antineutrino pair 
is given by
\beq
\Gamma\left(M^-\to \ell^-\nub_\ell\right) = {G_F^2\over 8\pi}\, 
\big|V_{q_u q_d}\big|^2 f_M^2 m_M
  m_\ell^2 \bigg(1 - {m_\ell^2\over m_M^2}\bigg)^{\!2} \,,
\eeq
where the decay constant $f_M$, defined by $i f_M p^\mu = \langle0| \bar q_u
\gamma^\mu\gamma_5 q_d |M(p)\rangle$, contains all the nonperturbative 
strong-interaction physics.  If $f_M$ is known, then leptonic $\pi$, $K$, $D$,
$D_s$, $B$, and $B_c$ decay rates determine $|V_{ud}|$, $|V_{us}|$, $|V_{cd}|$,
$|V_{cs}|$, $|V_{ub}|$, and $|V_{cb}|$, respectively.

Owing to the $m_\ell^2$ suppression, however, these rates are small for $\ell =
\mu$ or $e$, and the $\tau$ final state is harder to reconstruct
experimentally.  The current errors of the LQCD determinations of the decay
constants are around the 10\% level, so the observed leptonic $K$, $D$,
$D_s$ and $B$ decays do not give competitive direct determinations of the
corresponding CKM elements.  However, ratios of decay constants are easier to
determine in LQCD, and a recent calculation of $f_K/f_{\pi} = 1.198\pm 0.003
^{+0.016}_{-0.005}$~\cite{Bernard:2005ei} allows $|V_{us}/V_{ud}|$ to be 
extracted from the ratio of leptonic $K$ and $\pi$ decays, yielding
$|V_{us}| = 0.2245 ^{+0.0012}_{-0.0031}$~\cite{pdg_ckm}.  The ratios
$f_{D_s}/f_D$ and $f_{B_s}/f_B$ are also known much more precisely than the
individual decay constants.  However, the experimental errors are still sizable
for $D\to\ell\nub$, and because the $B_s$ meson is neutral, it cannot decay 
to a $\ell\nub$ pair.\footnote
{
	The hadronic physics relevant to the rare decay $B_s\to\ell^+\ell^-$  
	is also governed by $f_{B_s}$, and once the decay is measured at the 
	LHC, it will give a precise determination of $|V_{ub}|$ from 
	the double ratio $[{\cal B}(B\to\ell\nub) / {\cal B}(B_s\to\ell^+\ell^-)] 
        \times [{\cal B}(D_s\to\ell\nub) / {\cal
	B}(D\to\ell\nub)]$~\cite{Grinstein:1993ys}.
        This double ratio can also give a precise SM prediction for 
	$B_s\to\ell^+\ell^-$, and $B\to\ell\nub$ can even be replaced
	by $B\to\ell^+\ell^-$.
}

\begin{figure}[t!]
  \centerline{\includegraphics[width=8cm]{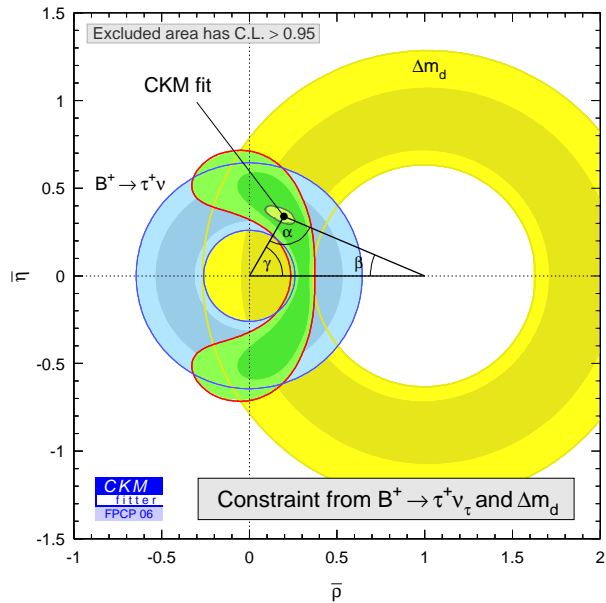}}
  \caption[.]{\label{fig:btaunu}\em
	Constraints of 95\% C.L. in the $\rhoeta$ plane from the 
	$\Bp\to\taup\nut$ branching fraction, $\dmd$, and the 
	combined use of these~\cite{ckmfitterFPCP06} (darker shades indicate 
	68\%~\CL regions)~\cite{btaunu}. The results are 
	compared with the global CKM fit (\cf\  Sec.~\ref{globalCKMfit}).
	}
\end{figure}

Metrologically useful experimental information exists for the decay 
$\Bp\to\taup\nut$, where averaging the experimental likelihoods  from Belle
(reporting a $4.2\sigma$ evidence for this decay) and  \babar~\cite{btaunu}
gives $(10.4\,^{+3.0}_{-2.7})\times10^{-5}$ for the branching fraction, which 
is consistent with the prediction, $(9.6\pm1.5)\times10^{-5}$, from the  global
CKM fit (not including the direct measurement)~\cite{ckmfitterFPCP06}. It 
is interesting to study the
constraint obtained from this measurement in conjunction with the  measurement
of the $\Bz\Bzb$ oscillation frequency, $\dmd$, which also  depends on $f_B$, so
that $f_B$ cancels in the combination. The  allowed regions obtained in the
$\rhoeta$ plane are shown in Fig.~\ref{fig:btaunu}. The remaining theoretical 
uncertainty stems from the bag parameter, $B_B$, that enters
the SM prediction of $\dmd$ [see Eq.~(\ref{eqof-dmd2})].

\subsubsection{$|V_{us}|$}
\label{Vus}

Traditionally, the magnitude of $V_{us}$ has been extracted from semileptonic $K$
decays, which can be analyzed using chiral perturbation theory.  The $K\to\pi\ell\nu$
amplitude can be expanded in powers of the $K$ and $\pi$ momenta and the quark
masses divided by a scale of order $\Lambda_{\chi\rm SB}$.  The
decay rates for $K=K^\pm,K^0$ and $\ell=e,\mu$ can be written
as~\cite{vud_ckm2005}
\beq
\Gamma(K\to \pi\ell\nub) = {G_F^2 m_K^5\over 128\pi^3}\, C\, I^{K\ell}\,
  \Vus^2 \left| f_+^{K^0\pi^-}\!(0)\right|^2 (1 + \ldots)\,,
\eeq
where $C = 1$ ($1/2$) for $K^0$ ($K^\pm$) decays, $I^{K\ell}$ is a phase-space
integral depending on the shape of the form factor, $f_+(q^2)$, and the dots
stand for isospin-breaking and electroweak corrections.  It is convenient to
factor out $f_+^{K^0\pi^-}\!(0) \equiv f_+(0)$, because the Ademolo-Gatto
theorem~\cite{Ademollo:1964sr} implies that there is no ${\cal O}(p^2)$ 
correction to $f_+(0)$, \ie, $f_+(0) = 1 + f_{p^4} +
f_{p^6}$.  Most of the ${\cal O}(p^4)$ corrections are known in terms of
physical parameters, and beyond this the estimates are model dependent.  The
original value, $f_+(0) = 0.961\pm0.008$~\cite{Leutwyler:1984je}, is still
broadly accepted, although some other calculations differ by as much as 2\%,
with uncertainties around 1\%~\cite{vud_ckm2005}.  It will be interesting to see
whether unquenched LQCD results confirm this estimate. Significant progress
on the experimental side has taken place recently. After a high-statistics
measurement of the ${\cal B}(K^+ \rightarrow \pi^0 e^+ \nu)$~\cite{Sher:2003fb},
a suite of measurements of neutral-kaon branching
ratios~\cite{Alexopoulos:2004sw}, of form factors~\cite{Yushchenko:2004zs}, 
and of lifetime~\cite{Ambrosino:2005vx} followed.  From semileptonic kaon 
decays one obtains $\Vus = 0.2257 \pm 0.0021$. 

The $\tau$ spectral function into $|\Delta S|=1$ final states can be used to 
determine the $s$-quark mass and the CKM element $|V_{us}|$~\cite{jaminetal}. 
Although the determination of $\overline m_s(m_\tau)$ is strongly 
model dependent, the determination of $|V_{us}|$ appears on safe 
grounds, because it relies on the nonweighted spectral function with maximum 
inclusiveness and a better control of the perturbative expansion. Present 
data give~\cite{davierRMP} $|V_{us}|=0.2204 \pm 0.0028 \pm 0.0003 \pm 0.0001$, 
where the first error is experimental, the second theoretical and the 
third due to $m_s$ (taken from LQCD).

\subsubsection{$|V_{cs}|$ and $|V_{cd}|$}

For the dominant semileptonic $D$ decays, $D\to K^{(*)}\ell\nu$, neither 
heavy-quark nor chiral symmetry is particularly useful.  Hence, until recently, 
the only precise direct constraint involving $|V_{cs}|$ was from the leptonic
branching ratio of the $W$, which measured the sum of CKM elements in the first
two rows (corrected for effects from perturbative QCD), 
$\sum_{i,j\in\{u,c,d,s,b\}} |V_{ij}|^2 = 1.999 \pm 0.026$~\cite{LEPEW2005}
(compared with the expected value of two), providing a precise test of 
unitarity.

Fortunately, the $D\to K\ell\nu$ and $\pi\ell\nu$ decays are well suited for
LQCD calculations, because the charm quark is neither too heavy nor too light,
and the maximal momenta of the final state $K$ or $\pi$ is not too large. 
Together with the CLEO-c program, it promises significant advances both in the
measurements of $|V_{cs}|$ and $|V_{cd}|$ as well as in our understanding of the
robustness of LQCD calculations from which \B physics will also benefit.

The direct determination of $|V_{cs}|$ and $|V_{cd}|$ is possible from
semileptonic $D$ or leptonic $D_s$ decays, relying on the form factors (which
depend on the invariant mass of the lepton pair, $q^2$).  Recently, the first 
three-flavor unquenched LQCD calculations for $D\to K \ell \nub$ and
$D\to \pi \ell \nub$ were published~\cite{Aubin:2004ej}.  Using these
calculations, and the isospin-averaged charm semileptonic widths measured by
CLEO-c, one obtains~\cite{Artuso:2005jw}
\beq\label{VcdVcs}
|V_{cd}| = 0.213 \pm 0.008 \pm 0.021\,, \qquad
|V_{cs}| = 0.957 \pm 0.017 \pm 0.093\,,
\eeq
where the first errors are experimental, and the second theoretical due
to  the form factors.\footnote
{
	The most precise measurement of $|V_{cd}|$ still comes from charm 
	production in neutrino scattering.  The PDG quotes 
	$|V_{cd}| = 0.230 \pm 0.011$~\cite{pdg_ckm}, an uncertainty that will be
	hard to reduce, whereas that from Eq.~(\ref{VcdVcs}) will improve with
	better LQCD calculations and data.
}
Note that both $|V_{cs}|$ and $|V_{cd}|$ are determined much more precisely if
CKM unitarity is assumed, with uncertainties around 0.0002 and 0.001,
respectively (see Table~\ref{tab:fitResultsII} on
p.~\pageref{tab:fitResultsII}).  Therefore, these measurements do not affect the
global CKM fit at present.

\subsubsection{$|V_{cb}|$}
\label{Vcb}

The CKM element $|V_{cb}|$ can be extracted from exclusive and inclusive semileptonic
$B$ decays.  The determination from exclusive $B\to D^{(*)} \ell \nub$ uses an
extrapolation of the measured rates to zero recoil, $w=1$
(\ie, maximal momentum transfer to the
leptons, $q^2_{\rm max}$). The rates can be written schematically as
\beq\label{rates}
{\d\Gamma(B\to D^{(*)} \ell\nub)\over \d w} = (\mbox{known terms})\times
  |V_{cb}|^2\times \cases{ (w^2-1)^{1/2}\, {\cal F}_*^2(w)\,,\ \
  & for $B\to D^*$, \cr
  (w^2-1)^{3/2}\, {\cal F}^2(w)\,, & for $B\to D$\,. \cr}
\eeq
Heavy-quark symmetry (see Sec.~\ref{sec:hqs}) determines the rate at $w=1$, that
is, in the $m_Q \gg \lqcd$ limit ($Q=b,c$), ${\cal F}(w) = {\cal F}_*(w) =
\xi(w)$, and in particular ${\cal F}_{(*)}(1) = 1$, allowing for a 
model-independent determination of $|V_{cb}|$.  The corrections to the 
heavy-quark limit can be organized in expansions in $\as(m_Q)$ and $\lqcd/m_Q$, 
which take the form
\beqa\label{F1}
{\cal F}_*(1) &=& 1 + c_A(\as) + {0\over m_Q} 
  + {(\mbox{LQCD or models})\over m_Q^2} + \ldots \,, \nn\\
{\cal F}(1) &=& 1 + c_V(\as) 
  + {(\mbox{LQCD or models})\over m_Q} + \ldots \,.
\eeqa
Heavy-quark symmetry implies that the leading-order results are unity, whereas
the absence of the $\lqcd/m_Q$ correction to ${\cal F}_*(1)$ is due to Luke's
theorem~\cite{Luke}.  The perturbative corrections $c_A = -0.04$ and $c_V =
0.02$ are known to order $\as^2$~\cite{Czar}, and higher-order terms are below
the $1\%$ level.  The terms indicated by ``LQCD or models'' are known 
only from quenched LQCD~\cite{latticeDDs} or phenomenological models.  
The experimental results for the rates, extrapolated to  $w=1$
are~\cite{hfag2005}
\beq\label{zerorec}
|V_{cb}|\, {\cal F}_*(1) = (37.6 \pm 0.9) \times 10^{-3} \,,\qquad
|V_{cb}|\, {\cal F}(1) = (42.2 \pm 3.7) \times 10^{-3} \,.
\eeq
Using ${\cal F}_*(1) = 0.91 \pm 0.04$, yields $|V_{cb}| = (41.3 \pm 1.0_{\rm
exp} \pm 1.8_{\rm th}) \times 10^{-3}$.  The $B\to D\ell\nub$ data are
consistent with this, but to make a real test, smaller experimental errors and 
unquenched LQCD calculations are needed.

The inclusive determination of \Vcb is based on the OPE (see
Sec.~\ref{sec:ope}).  The state of the art is that the semileptonic rate, as
well as moments of the lepton energy and the hadronic invariant-mass spectra
with varying cuts have been computed to orders $\lqcd^3/m_b^3$ and
$\as^2\beta_0$.  Their dependence on $m_{b,c}$ and the parameters that occur at
subleading orders in $\lqcd/m_b$ are different. The expressions for a large
number of observables are of the form
\beqa
\Gamma(B\to X_c\ell\nub) &=& {G_F^2 |V_{cb}|^2\over 192\pi^3}\,
  \bigg({m_\Upsilon\over 2}\bigg)^{\!5}\, (0.534)
  \bigg[1 - 0.22 \bigg({\Lambda_{1S}\over 500\,\MeV}\bigg) \nn\\
&-& 0.011 \bigg({\Lambda_{1S}\over 500\,\MeV}\bigg)^{\!2}
  - 0.052 \bigg({\lambda_1\over (500\,\MeV)^2}\bigg)
  - 0.071 \bigg({\lambda_2\over (500\,\MeV)^2}\bigg) \nn\\
&+& 0.096\varepsilon - 0.030 \varepsilon^2_{\rm BLM} 
  + 0.015 \varepsilon \bigg({\Lambda_{1S}\over 500\,\MeV}\bigg) + \ldots \bigg] \,,
\eeqa
where $m_\Upsilon$ is the $\Upsilon(1S)$ mass.
The precise determination of $|V_{cb}|$ requires the use of an appropriate
short-distance quark mass, the so-called ``threshold"
mass~\cite{Hoang:1998hm,Bigi:1997fj}, and $\Lambda_{1S} \equiv m_\Upsilon/2 -
m_b^{1S}$ is related to one of these,
$m_b^{1S}$~\cite{Hoang:1998hm,Hoang:1999zc}. The $\lqcd^2/m_b^2$ corrections are
parameterized by $\lambda_{1,2}$.  The $\lqcd^3/m_b^3$ terms are known and are
parameterized by six more nonperturbative matrix elements.  For the perturbative
corrections $\varepsilon \equiv 1$ shows the order in the expansion, and the BLM
subscript shows that only the terms with the highest power of $\beta_0$ are
known~\cite{Brodsky:1982gc}.  
Such formulas are fit to approximately $90$ (correlated) observables obtained 
from the lepton energy and hadronic invariant-mass spectra. The fits determine
$|V_{cb}|$ and the hadronic parameters, and their consistency provides a
powerful test of the theory.  The fits have been performed in several schemes
and give~\cite{Bauer:2004ve,Buchmuller:2005zv},
\beq
|V_{cb}| = (41.7 \pm 0.7) \times 10^{-3} ,
\eeq
where the central values were averaged and the errors quoted in each paper
were retained.
The same fits simultaneously determine the quark masses, $m_b^{1S} =
(4.68\pm 0.03)\, \GeV$ and $\ov m_c(\ov m_c) = (1.22\pm0.06)\,
\GeV$~\cite{Bauer:2004ve,Hoang:2005zw}, which correspond to $\ov m_b(\ov m_b) =
(4.18 \pm 0.04)\, \GeV$ and $m_c^{1S} = (1.41\pm 0.05)\, \GeV$.  The $m_b-m_c$
mass difference is even better
constrained~\cite{Bauer:2004ve,Buchmuller:2005zv}.

\subsubsection{$|V_{ub}|$}
\label{Vub}

Of the measurements of $|V_{ub}|$ from exclusive decays, $B\to\pi\ell\nub$ is
the most advanced, as both experiment and LQCD calculations are under
the best control.  The determination relies on measuring the rate and
calculating the form factor $f_+(q^2)$,
\beq
{\d\Gamma(\Bbar^0\to\pi^+\ell\nub)\over \d q^2} 
  = {G_F^2 |\vec p_\pi|^3\over 24\pi^3}\, |V_{ub}|^2\, |f_+(q^2)|^2 \,.
\eeq
Unquenched calculations of $f_+$ exist only for $q^2>16\,\GeV^2$ (small
$|\vec p_\pi|$)~\cite{Okamoto:2004xg,Shigemitsu:2004ft}, where the available
statistics is reduced because the phase space is proportional to $|\vec p_\pi|^3$. 
Averaging the LQCD calculations, and using data only in the $q^2>16\,\GeV^2$
region, gives $|V_{ub}| = (4.13 \pm 0.62) \times
10^{-3}$~\cite{hfag2005,Okamoto:2005zg}.

Some of the current $|V_{ub}|$ determinations use model-dependent
parameterizations of $f_+(q^2)$ to extend the LQCD results to a larger part
of the phase space, or to combine them with QCD sum-rule calculations at small
$q^2$ (which tend to give smaller values for $|V_{ub}|$~\cite{Ball:2005tb}). 
These model-dependent ingredients can be avoided using constraints on the shape
of  $f_+(q^2)$ that follow from dispersion relations and the knowledge of
$f_+(q^2)$ at a few values of $q^2$~\cite{Boyd:1994tt}.  The recent LQCD
results revitalized this area~\cite{Arnesen:2005ez,Fukunaga:2004zz}.  Using the
LQCD calculations of $f_+$ at large $q^2$, the experimental measurements, and
the dispersion relation to constrain the shape of $f_+(q^2)$, gives $|V_{ub}| =
(3.92 \pm 0.52) \times 10^{-3}$~\cite{Arnesen:2005ez}, using the full $q^2$
range.

The determination of $|V_{ub}|$ from inclusive semileptonic $B$ decay is more
complicated than that of $|V_{cb}|$, because of the large $B \to X_c \ell \nub$
background.  Much of the recent experimental progress is related to the fact
that the \B-factories produce pure $B\Bbar$ pairs, where the full
reconstruction of one $B$ allows the measurement of the energy and momenta of
both the leptonic and the hadronic systems in the $X_u \ell \nubar$ decay of the
other $B$ (the so-called ``\B-beam'' technique). This provides several handles
to suppress the $B \to X_c \ell \nub$ background. The total $B \to X_u \ell
\nub$ rate is known theoretically at the 5\%
level~\cite{Hoang:1998hm,Bigi:1997fj}, but the cuts used in most experimental
analyses to remove the $B\to X_c\ell\nub$ background complicate the theory.  The
local OPE used to extract \Vcb is applicable for  $B\to X_u\ell\nub$ if the
typically allowed $m_X$ and $E_X$ of the final state satisfy
\beq\label{converge}
m_X^2 \gg E_X \lqcd \gg \lqcd^2 \,.
\eeq
However, most of the current \Vub determinations restrict the phase space to 
(a subset of) the $m_X< m_D$ region.  As a result, there are three qualitatively 
different regions. 

If $m_X \sim \lqcd$, the final state is dominated by resonances, and  it is not
possible to compute inclusive quantities reliably.   If $m_X^2 \sim E_X \lqcd
\gg \lqcd^2$, the OPE becomes an expansion in terms of $b$-quark light-cone
distribution functions (sometimes termed shape functions).  This is the case for
many measurements to date, such as the rate for $E_\ell > (m_B^2-m_D^2)/(2m_B)$,
$m_X<m_D$, or $P_X^+(\equiv E_X-|\vec p_X|) < m_D^2/m_B$.  At leading order one
such nonperturbative function occurs~\cite{Neubert:1993ch,Bigi:1993ex}, which
can be determined from the  $B\to X_s\gamma$ photon spectrum.  At order
$\lqcd/m_b$ there are several new functions~\cite{Bauer:2001mh}, which cannot be
extracted from the data, and are modelled.  The hadronic physics 
parameterized by functions is a significant complication compared with the
determination of $|V_{cb}|$, where it is encoded in a few hadronic matrix
elements. Moreover, at order $\alpha_s$, moments of the shape function are no
longer given simply by local hadronic matrix
elements~\cite{Bauer:2003pi,Bosch:2004th}.  One can either fit the shape
function from $B\to X_s\gamma$, or directly relate the $B\to X_u \ell \nub$
partial rates to weighted integrals of the $B\to X_s\gamma$
spectrum~\cite{Leibovich:1999xf}.  Finally, if $m_X^2 \gg E_X \lqcd \gg
\lqcd^2$, the OPE converges, and the first few terms give reliable results. 
This is the case for the rate for $q^2 > (m_B-m_D)^2$~\cite{Bauer:2000xf}, but
the expansion parameter is then $\lqcd/m_c$.  The dependence on the shape
function can be kept under control by combining $q^2$ and $m_X$ cuts.  The 
shape-function dependence can also be reduced by extending the measurements into the
$B \to X_c \ell \nub$ region. Recent analyses could measure the $B\to X_u \ell
\nub$ rates for $|\vec p_e| \ge 1.9$\,GeV~\cite{Bornheim:2002du} and for $m_X <
2.5\,\GeV$~\cite{Aubert:2006qi}.
Averaging these inclusive measurements, the Heavy Flavor Averaging Group (HFAG)
obtains $|V_{ub}| = (4.38 \pm 0.19 \pm 0.27) \times 10^{-3}$~\cite{hfag2005}
using~\cite{Bosch:2004th}, which indicates a slight tension with the CKM fit 
for $|V_{ub}|$ dominated by  the $\sin2\beta$ measurement (\cf\ 
Table~\ref{tab:fitResultsII} on  p.~\pageref{tab:fitResultsII}).

\subsection{CKM Elements from Rare Loop-Mediated $B$ Decays}
\label{sec:loops}

To measure $|V_{td}|$ and $|V_{ts}|$, one has to rely on loop-mediated rare $B$
decays, or rare $K$ decays discussed in Sec.~\ref{kaons_rare}, or $B\Bbar$
oscillations discussed in Sec.~\ref{b_oscillation}. All of them provide 
important tests of the SM, as different transitions can receive different 
contributions from new physics.

The inclusive branching fraction ${\cal B}(B\to X_s\gamma) =
(3.39\,^{+0.30}_{-0.27})\times10^{-4}$~\cite{hfag2005} is sensitive to $|
V_{tb}V_{ts}^*|$ via penguin diagrams with a top quark. A large
contribution to the rate stems from the $b\to c \cbar s$ four-quark
operator, $O_2^c$ in Eq.~(\ref{fullops}), mixing into the electromagnetic
penguin operator, $O_7$. Using unitarity, $V_{cb}V_{cs}^*= -
V_{tb}V_{ts}^* - V_{ub}V_{us}^*$, the measured rate implies $|
V_{tb}V_{ts}^*| = (39.0\pm 3.1)\times10^{-3}$~\cite{pdg_ckm}.

The theoretical uncertainties can be reduced by taking ratios of processes that
are equal in the flavor $\SU(3)$ limit to determine $|V_{td}/V_{ts}|$.  The
exclusive rare decays suffer from larger theoretical uncertainties owing to
unknown hadronic form factors, but exclusive branching fractions are easier to 
measure. Recently, Belle observed the first $b \to d \gamma$ signals in 
$B \to (\rho/\omega) \gamma$ decays. We define the ratios of branching 
fractions
\beq
{{\cal B}(B\to \rho\gamma) \over 
  {\cal B}(B\to K^*\gamma)}
  = \bigg|{V_{td}\over V_{ts}}\bigg|^2 
  \bigg({m_B - m_\rho \over m_B - m_{K^*}}\bigg)^{\!3} \times
  \cases{ \frac12 (\xi_{V^0\gamma})^{-2}\,,  &  for $V^0\gamma$\,, \cr
  (\xi_{V^\pm\gamma})^{-2}\,,  &  for $V^\pm\gamma$\,, \cr}
\eeq
where $\xi_{V^0\gamma}$ and $\xi_{V^\pm\gamma}$ denote 
$\SU(3)$-breaking corrections.
The ratio of the neutral rates gives the theoretically cleaner determination of
$|V_{td}/V_{ts}|$, because weak annihilation enters in the $\rho^\pm\gamma$
mode, but only $W$ exchange contributes to $\rho^0\gamma$. (Although
both annihilation and exchange are suppressed by $\lqcd/m_b$, they are hard to
estimate, and exchange is color-suppressed compared with annihilation.) Averaging
the results from \babar and Belle~\cite{btoVgamma,hfag2005} gives
${\cal B}(\Bz\to\rho^0\gamma)/{\cal B}(\Bz \to K^{*0}\gamma)=0.0095\pm0.0045$~\cite{hfag2005}.  
Using the estimate $\xi_{V^0\gamma} = 1.2 \pm 0.1$ for this
average~\cite{Grinstein:2000pc}  implies $|V_{td}/V_{ts}| = 0.16 \pm 0.05$. 
Although the theoretical uncertainty for this determination of $|V_{td}/V_{ts}|$
will not compete with that for $\dmd/\dms$, it provides an important
test of the SM, as new physics  could contribute differently to these decays and
to $B\Bbar$ mixing (\cf\ right plot of Fig.~\ref{fig:wadms}).

\subsection{Neutral $B$-Meson Oscillation}
\label{b_oscillation}

The $\BzBzb$ oscillation frequency is governed by the mass difference
$\dmd$ between the two $\Bz$ mass eigenstates, $B_{H}$ and $B_{L}$. It 
is defined as a positive number and has been measured by many experiments 
leading to an overall $1\%$ precision, $\dmd=(0.507\pm0.005)\ps^{-1}$
(dominated by the measurements at the \B-factories)~\cite{hfag2005}. 
In analogy to $|\epsk|$ (see below), $\BzBzb$ oscillation in the SM is 
described by FCNC box diagrams. However, 
in contrast to $|\epsk|$, where the large quark-mass-dependent hierarchy 
in the FCNC amplitude is offset by the tiny CKM matrix element $|V_{td}V_{ts}^*|$, 
the  $\Delta B=2$ box diagrams are dominated by intermediate 
top quarks. This implies that $\dmd$ is determined by short-distance physics
up to the matrix element of the $(\dbar  \gamma^\mu(1-\gamma_5)b)^2$ operator, 
parameterized by $f_{B_d}^2B_d$,
\beq
\label{eqof-dmd2}
   \dmd = \frac{\GF^2}{6 \pi^2}\etaB m_{B_d}f_{B_d}^2B_d
          m_W^2 S(x_t) \left|V_{td} V_{tb}^*\right|^2\,.
\eeq
Here $\etaB = 0.551 \pm 0.007$ is a perturbative QCD correction to the 
Inami-Lim function $S(x_t)$ (with $x_t=
\mtRun^2/m_{W}^2$)~\cite{InamiLim}, and $f_{B_d} \sqrt{B_d}$ is 
taken from LQCD. Much progress has been achieved in this domain 
with unquenched calculations becoming available (see 
Sec.~\ref{lattice}). A fixed value of the CKM factor $|V_{td} V_{tb}^*|^2$
occurring in  Eq.~(\ref{eqof-dmd2}) describes approximately a circle around
$(1,0)$ in  the $\rhoeta$ plane, to which an elliptic distortion appears only at
order $\lambda^6$.

In the SM, the mass difference $\dms$ between the heavy and the light $\Bs$ 
mass eigenstates has only ${\cal O}(\lambda^2)$ dependence on the Wolfenstein
parameters $\rhobar$, $\etabar$. A measurement of $\dms$ is nevertheless
useful for CKM metrology within the SM, because it leads to an improvement 
in the constraint from the $\dmd$ measurement on $|V_{td} V_{tb}^*|^2$. Its 
SM prediction is given by
\beq
\label{eqof-dms2}
   \dms = \frac{\GF^2}{6 \pi^2}\etaB m_{B_s} \xi^2 \fbd^2 B_d
          m_W^2 S(x_t) \left|V_{ts} V_{tb}^*\right|^2\,,
\eeq
where the parameter $\xi = \fbs \sqrt{B_s}/\fbd \sqrt{B_d}$ quantifies
$\SU(3)$-breaking corrections to the matrix elements, which can be calculated 
more accurately in LQCD than the matrix elements themselves. Hence 
a measurement of $\dms$ reduces the uncertainty of $\fbd \sqrt{B_d}$. 
LQCD calculations using Wilson fermions have to work with light-quark 
masses of order $100\mev$, so calculations for $\Bd$ 
mesons need to be extrapolated to the chiral (massless) limit. This is 
not necessary for the $\Bs$ due to the relatively heavy strange quark. 
The use of staggered fermions allows LQCD calculations to be performed
with significantly smaller light-quark masses, which implicates smaller 
systematic uncertainties.

Because of the large ratio $|V_{ts}/V_{td}|^2$ and $\SU(3)$-breaking corrections, 
$\BszBszb$ oscillation occurs approximately 35 times faster than $\BzBzb$ 
oscillation. Also, measurements of the time-integrated mixing probability $\chi_s$ 
indicate a value close to its maximum, $1/2$, requiring very fast oscillation.
Excellent proper-time resolution and large data samples are needed to resolve 
$\BszBszb$ oscillation in the detector. Because only pairs of the two 
lightest \B mesons can be produced in \FourS decays, $\BszBszb$ oscillation 
can be probed only at high-energy $\epem$ or hadron colliders.

Lower limits on $\dms$ have been obtained by the LEP, SLD and Tevatron
experiments. A convenient approach to averaging the various results 
is the amplitude method (\cite{Roussarie}; see also the studies
in Refs.~\cite{ckmfitter2004,toy}). It consists of introducing an 
ad hoc amplitude coefficient, $\Ampli$, placed in front of the cosine 
modulation term that describes the time-dependent mixing asymmetry
$(N_{{\rm unmixed}}(t)-N_{{\rm mixed}}(t))/(N_{{\rm unmixed}}(t)+N_{{\rm mixed}}(t))=\Ampli \cdot\cos(\dms t)$.
The advantage of this indirect probe for oscillation 
is that the dependence on $\Ampli$ is linear so that $\Ampli$ is 
Gaussian distributed. The measured amplitude must reach $\Ampli=1$ at 
the true values of $\dms$ (if within reach of the experimental sensitivity), 
and zero elsewhere. The preliminary combined $95\%~\CL$ lower limit 
was $\dms>16.6\ps^{-1}$~\cite{hfag2005}. 

\begin{figure}[!t]
  \centerline{	\includegraphics[width=0.49\textwidth]{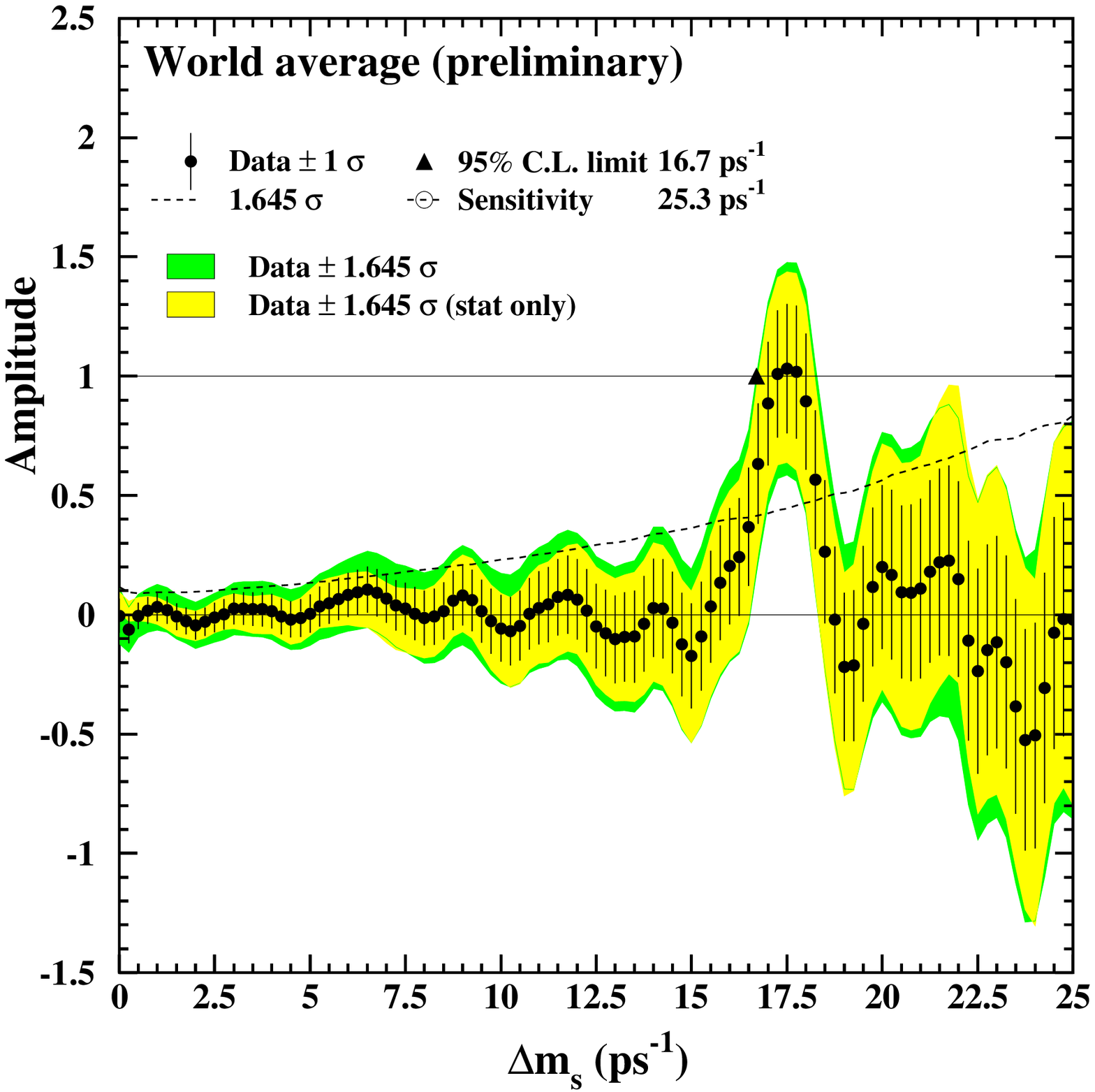} \hfill
  \raisebox{-17pt}{\includegraphics[width=0.51\textwidth]{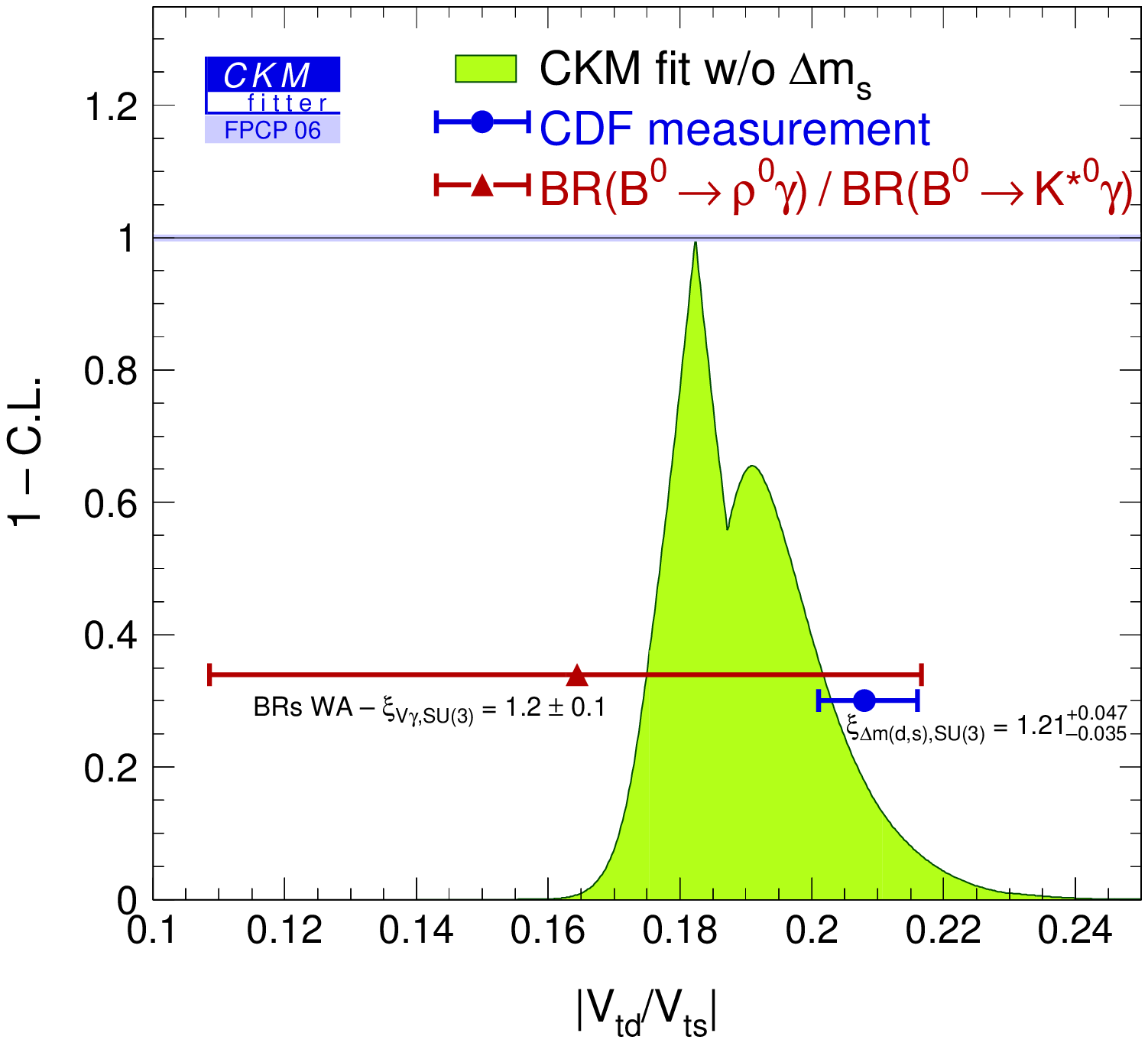}}}
  \caption[.]{\label{fig:wadms}\em
	Left: preliminary world-average amplitude spectrum as a function 
	of $\dms$~\cite{hfag2005}, dominated by the 
	recent CDF measurement~\cite{cdfbsbsbar}.
	Right: $|V_{td}/V_{ts}|$ determined by CDF from the 
	ratio $\dmd/\dms$ (full circle)~\cite{cdfbsbsbar}, 
	from the ratio
	$\BR(\Bz\to\rho^0\gamma)/\BR(\Bz\to K^{*0}\gamma)$ 
	(triangle) (see Sec.~\ref{sec:loops}), 
	and from the global CKM fit (shaded region; not 
	including $\dms$ in the fit (see Sec.~\ref{globalCKMfit}).
	}
\end{figure}

Shortly before finalizing this review, the \dzero and CDF Collaborations
reported new preliminary results on $\BszBszb$ oscillation using data 
samples corresponding to an integrated luminosity of $1\invfb$.
Using the charge of the muon in a partially reconstructed $\Bs\to
D_s^-\mup\num$  decay to tag the flavor of the $\Bs$ meson, \dzero sets
the $90\%$~\CL interval $17\invps<\dms<21\invps$~\cite{d0bsbsbar}.
Because the upper bound is close to the sensitivity limit of the measurement,
larger \dms values are excluded only at the $2.1\sigma$ level. 
Using hadronic and semileptonic $\Bs$ decays, CDF finds~\cite{cdfbsbsbar} 
\beq
	\dms = \left(17.31\,^{+0.33}_{-0.18}\pm0.07\right)\invps\,,
\eeq
where the first error is statistical and the second is systematic. The 
probability that the observed signal is a background fluctuation is 
quoted as 0.2\% (5\%) for CDF (\dzero). The value of \dms\ is in agreement 
with the prediction from the global CKM fit, $(21.7\,^{+5.9}_{-4.2})\invps$, 
obtained without using the measurement in the fit (\cf\  
Table~\ref{tab:fitResultsI} on p.~\pageref{tab:fitResultsI}). 
The preliminary world-average amplitude scan is 
shown in the left plot of Fig.~\ref{fig:wadms}. The combined signal 
significance amounts to $3.8\sigma$ at $17.5\invps$~\cite{hfag2005}.
CDF infers from its measurement of $\dms$ that the ratio of CKM elements 
$|V_{td}/V_{ts}| = 0.208\,^{+0.008}_{-0.006}$ (dominated by the theoretical 
uncertainty on $\xi=1.21\,^{+0.047}_{-0.035}$, used by CDF). A comparison
with the CKM fit is given in the right plot of Fig.~\ref{fig:wadms}.
This measurement provides the first strong indication that
$\BszBszb$ mixing is probably SM-like. 

\section{CKM Matrix Constraints from Kaon Physics}

The neutral-kaon system constrains the UT through $\KzKzb$ 
mixing, through indirect and direct \CP violation,\footnote
{
       ``Indirect" \CP violation refers to situations in which the \CP-violating
       phase can be chosen to appear only in the mixing amplitudes, whereas
       ``direct" \CP violation occurs when a \CP violating phase must appear in
       decay amplitudes. The observables ${\rm Im}\,\epsK$ and ${\rm
       Im}\,\epsPr$  measure \CP violation in the interference between decays
       with and without mixing, whereas  ${\rm Re}\,\epsK$ and ${\rm
       Re}\,\epsPr$, measure \CP violation in mixing and in decay,
       respectively.
} 
and through the rare decays $\Kp\to\pip\nu\nub$ and the yet unobserved 
$\KL\to\piz\nu\nub$. In this case, contrary to Eq.~(\ref{Bpq}), the 
convention is to label the states by their lifetimes, $|\KSL\rangle
= p_K |K^0\rangle \pm q_K |\Kbar^0\rangle$ (experimentally, $m_{\KL} >
m_{\KS}$). The SM prediction for $\Delta m_K$ suffers from badly controlled
long-distance contributions to the mixing amplitudes.  Nonperturbative physics
with large hadronic uncertainties also impede a reliable SM prediction of direct
\CP violation. The above-mentioned rare decays are  cleaner and will give
precise constraints as soon as they are measured with  sufficient accuracy. They
are discussed in Sec~\ref{kaons_rare}.

\subsection{Indirect \CP Violation}

The most precise measurement of the \CP-violation parameter $\epsK$ uses
$\eta_{+-}$ and $\eta_{00}$---which are the ratios of $\KL$ and $\KS$ 
amplitudes to a pair of pions, 
$\eta_{ij} = A(\KL\to\pi^i\pi^j) / A(\KS\to\pi^i\pi^j)$---so that
\beq
   \epsK = \frac{2}{3}\, \eta_{+-} + \frac{1}{3}\, \eta_{00}\,.
\eeq
This is motivated by the so-called $\Delta I=\frac12$ rule, \ie, that the
amplitude to two pions in an $I=0$ state is approximately 20 times larger than the
amplitude to two pions in $I=2$.  The definition of $\epsK$ ensures that only
the dominant hadronic amplitude contributes, so \CP violation in decay gives
negligible contribution to $\epsK$, and so $\epsK \approx
[1-\lambda_{\pi\pi(I=0)}]/2$ (and $2\Re\, \epsK \approx 1-|q_K/p_K|$).  The
phase of $\epsK$ is independent of CKM elements (and of new physics), and is
approximately $\pi/4$ owing to to $2\Delta m_K \approx \Delta\Gamma_K$. The current
world average is $|\epsK| = (2.221 \pm 0.008)\times 10^{-3}$~\cite{Alexopoulos:2004sw}.  Other
observables measure $\left|\epsK\right|$ with less precision. Among these are
the charge asymmetry in semileptonic $\KL$ decays, $\delta_L$, or decay-plane
asymmetry in  $\KL\to \pi^+\pi^-e^+e^-$ decays.

Within the SM, $\epsK$ is induced by $\Delta S=2$ transitions mediated
by box diagrams. They are related to the CKM elements via a local hadronic
matrix element, parameterized by the bag parameter $B_K$,
\beq
\langle
       \Kzb|(\sbar\gamma^{\mu}(1-\gamma_5)d)^2|\Kz
\rangle
 = \frac{8}{3}\, m_{K}^2 f_{K}^2 B_{K}(\mu)\,,
\eeq
where $\mu$ is the renormalization scale, usually chosen to be $2\gev$.
Note that $B_K=1$ corresponds to the vacuum insertion approximation.
The renormalization scale 
and scheme-independent bag parameter (sometimes denoted by $\widehat B_K$)
is hereafter denoted by $B_K$, which differs from $B_K(\mu)$ by a series 
in $\as$ that cancels the scale and scheme dependences (similar to $B_d$ 
introduced in Sec.~\ref{b_oscillation}). The kaon decay constant,
$f_{K}=(159.8\pm1.5)\mev$~\cite{PDG}, is extracted from the leptonic decay rate
$\Gamma(\Kp\to\mup\nu_{\mu})$.  The most reliable prediction of $B_K$ is
obtained from LQCD.  At present, calculations are performed assuming
$\SU(3)$ symmetry and using mostly the quenched approximation. The world
average is $B_{K} = 0.79 \pm 0.04 \pm 0.09$~\cite{Dawson:2005za,ckm2005}, where
the first error combines statistical and accountable systematic
uncertainties\footnote
{
	The term ``accountable systematic uncertainties'' denotes uncertainties 
	that have been evaluated in a systematic manner. Most of the experimental	
	systematic uncertainties fall into this category. An example of an
	unaccountable (\ie, to a large extent arbitrary) theoretical 
	uncertainty is the quenched approximation used in some LQCD
	calculations.
} 
and the second is theoretical. Analytical approaches 
based on the large-$N_c$ expansion  of QCD find a smaller value for $B_{K}$ in 
the chiral limit~\cite{BK_Nc}.  Large chiral corrections could be the origin 
of this difference. At  present, $B_K$ is the primary source of theoretical 
uncertainty in the SM prediction of $\epsK$.

Neglecting the real part of the off-diagonal element, $M_{12}$, of the 
neutral-kaon mixing matrix, one obtains~\cite{bbl}.
\beqn
\label{eq_epsk}
|\epsK|
        &=& \frac{G_{F}^2 m_{W}^2 m_{K} f_{K}^2}
               {12 \sqrt{2}\, \pi^2 \Delta m_{K}}\, B_{K}\, \Big\{
              \eta_{cc} S(x_{c},x_{c}) \Im[(V_{cs} V_{cd}^*)^2]
              + \eta_{tt} S(x_{t},x_{t}) \Im[(V_{ts} V_{td}^*)]^2 \nn\\
        &&\hspace{3.25cm}{}
              + 2\eta_{ct} S(x_{c},x_{t})\,
                 \Im[V_{cs} V_{cd}^* V_{ts} V_{td}^*] \Big\}\,,
\eeqn        
where $\Delta m_{K}=(3.490 \pm 0.006)\times 10^{-12}\mev$~\cite{PDG},
and $S(x_i,x_j)$, with $x_i=m_i^2/m_W^2$ ($i=c,t$), are 
Inami-Lim functions~\cite{InamiLim}. The running quark masses are 
evaluated in the $\MSbar$ scheme. The parameters $\eta_{ij}$ in 
Eq.~(\ref{eq_epsk}) are next-to-leading-order QCD corrections to 
the Inami-Lim functions~\cite{Nierste}, where $\eta_{cc}$ is the 
parameter with the largest uncertainty.

\subsection{Direct \CP Violation}

A nonzero value for the \CP-violation parameter $\epsPr$, defined by
\beqn
        \epsPr = \frac13\, (\eta_{+-} - \eta_{00})
\eeqn
establishes direct \CP violation, because $\epsPr \approx (\lambda_{\pi^0\pi^0} -
\lambda_{\pi^+\pi^-}) / 6$. The  experimentally convenient observable ${\rm
Re}(\epe)$ is determined  by the measurement of the double ratio
\beq
       \frac{\Gamma(\KL\to\piz\piz)}{\Gamma(\KL\to\pip\pim)}
        \frac{\Gamma(\KS\to\pip\pim)}{\Gamma(\KS\to\piz\piz)}
       = \bigg| {\eta_{00} \over \eta_{+-} }\bigg|^2
        \simeq 1 - 6\, \Re\!\left(\frac{\epsPr}{\epsK}\right)\,,
\eeq
which cancels many experimental uncertainties.
Evidence for direct \CP violation was first seen by the NA31
Collaboration~\cite{na31}, and was firmly established by
NA48~\cite{na48} and KTeV~\cite{ktev},
with the world average ${\rm Re}(\epe)=(16.7\pm 1.6)\times 10^{-4}$.

Because ${\rm Re}(\epe)$ is proportional to  ${\rm
Im}(V_{td}V_{ts}^*)=A^2\lambda^5\etabar + \mathcal{O}(\lambda^7)$, it measures
$\etabar$. However, the SM prediction of ${\rm Re}(\epe)$ suffers from large
uncertainties, because the dominant gluonic and electromagnetic penguin
contributions tend to cancel each other~\cite{Flynn:1989iu}. Detailed
calculations at next-to-leading order~\cite{Buras93,Roma93} have been 
carried out for the
coefficients of the two dominant hadronic parameters, $B_6^{(1/2)}$ (gluonic
penguins) and $B_8^{(3/2)}$ (electroweak penguins), where the superscripts
denote the isospin change in the $K\to \pi\pi$ transition. Because no reliable
LQCD calculation for $B_6^{(1/2)}$ is available, the measurement of ${\rm
Re}(\epe)$ is usually not used to constrain CKM elements.

\subsection{$K^+\to\pi^+\nu\nub$ and $\KL\to\pi^0\nu\nub$}
\label{kaons_rare}

The E787 experiment at the Brookhaven National Laboratory (BNL) observed 
two events of the very rare decay $K^+\to \pi^+\nu\nub$, resulting in
$\BR(K^+\to\pi^+\nu\nub)=(1.57^{+1.75}_{-0.82}) \times 10^{-10}$~\cite{E787},
which, owing to the small expected background rate ($0.15\pm 0.03$ events),
effectively excludes the null hypothesis. One additional event was 
observed near the upper kinematic limit by the successor 
experiment, BNL-E949~\cite{E949}. They quote the combined branching 
fraction $\BR(K^+\to\pi^+\nu\nub)=(1.47^{\,+1.30}_{\,-0.89})\times 10^{-10}$.

In the SM, the branching fraction at next-to-leading order is given
by~\cite{BuBu}
\beq
\label{eq_kpppnunu}
    \BR(K^+\to\pi^+\nu\nub)
        = r_{K^+}
          \frac{3\alpha^2\, \BR(K^+\to\pi^0 e^+ \nu)}
               {2\pi^2\, |V_{us}|^2 \sin^4\theta_{\rm W}}\,
        \Big| X(x_t)V_{td}V^*_{ts} + X_c V_{cd}V^*_{cs} \Big|^2 ,
\eeq
where $r_{K^+} = 0.901$ corrects for isospin
breaking~\cite{kaonmarciano}. $X(x_t) = \eta_X X_0(x_t)$ is an Inami-Lim
function for the dominant top-quark  contribution, $X_0$, corrected by a
perturbative QCD factor, $\eta_X$.  The second term contains the charm-quark
contribution.

Expressing Eq.~(\ref{eq_kpppnunu}) in terms of the Wolfenstein parameters gives
\beq
\label{BRkpinn}
        \BR(K^+ \to \pi^+ \nu \nub) = \kappa_{+} A^{4} X(x_t)^2
                                           \frac{1}{\sigma}
        \left[(\sigma \etabar)^2 + (\rho_0 - \rhobar)^2\right]\,,
\eeq
with $\sigma = 1 + \lambda^2 + \mathcal{O}(\lambda^4)$ and
\beq
        \rho_0 = 1 + \frac{P_{0}}{A^{2} X(x_{t})}\,.
\label{BRkpinn1}
\eeq
It provides an almost elliptic constraint in the $\rhoeta$ plane, centered near
$(\rhobar=1,\etabar=0)$. The parameter $\kappa_+ = \lambda^8\,
[3\alpha^2 \BR(K^+\to\pi^0 e^+ \nu)] / [2\pi^2 \sin^4\theta_{\rm W})$ is such
that $\BR(K^+ \to \pi^+ \nu\nub)$ depends on $(A\lambda^2)^4$, 
which is constrained by $\Vcb$. The parameter $P_0=P_c+\delta P_c$ quantifies 
the charm-quark contribution, which until recently provided the largest theoretical 
error owing to the $m_c$ and scale dependences. The short-distance part, 
$P_c=0.38\pm0.04$, is now calculated to next-to-next-to-leading order~\cite{Buras:2005gr}. 
The long distance contributions of $c$ and $u$ loops are estimated to be
$\delta P_c=0.04\pm0.02$ and are expected to be under good control~\cite{Isidori:2005xm}.

In the SM, the decay $\KL \to \pi^0 \nu \nub$ is theoretically clean,
because the final state is almost completely
\CP-even~\cite{Littenberg:1989ix}, so the decay proceeds dominantly
through \CP violation in the interference of decay with and without
mixing~\cite{Grossman:1997sk,Buchalla:1998ux}. The amplitude is dominated by the
top-quark contribution. The theoretical prediction of the branching fraction is
given by~\cite{BuBu}
\beq
        \BR(\KL \to \pi^{0} \nu \nub) 
        = \kappa_{L}  \left[\frac{{\rm Im}(V_{td}V_{ts}^*)}{\lambda^5}\,
                 X(x_{t}) \right]^{\!2}
        = \kappa_{L} A^4\etabar^2 X^2(x_{t}) + \mathcal{O}(\lambda^4)\,,
\eeq
where $\kappa_{L}=\kappa_{+}(r_{K_L}\tau_{K_L})/(r_{K^+}\tau_{K^+})=(2.12\pm
0.03)\times10^{-10}$~\cite{PajuoMayor}, and $r_{K_L}=0.944$ accounts for
isospin breaking~\cite{kaonmarciano}. As for the charged mode, the branching
fraction is proportional to $(A\lambda^2)^4$, which would, however, cancel in 
the ratio of rates. The constraint in the $\rhoeta$ plane from a future
measurement of $\BR(\KL \to \pi^{0} \nu \nub)$ would correspond to two
horizontal bands at a certain value of $\pm|\etabar|$.

The interpretation of rare $K$ decays involving lepton pairs is more 
complicated than for the similar $B$ decays, because in most  channels
long-distance contributions are important. For example, $K^\pm\to  \pi^\pm
\ell^+\ell^-$ is dominated by a virtual photon converting into $\ell^+\ell^-$,
whereas $\KL\to \ell^+\ell^-$ is dominated by virtual two-photon contribution.
More interesting are the $K^0\to \pi^0 \ell^+\ell^-$ modes. Similar to $\KL\to
\pi^0 \nu\nubar$, the \CP-violating  amplitude dominates in $\KL\to \pi^0
e^+e^-$, whereas there are comparable \CP-violating and \CP-conserving
contributions in $\KL\to \pi^0\mu^+\mu^-$~\cite{Buchalla:2003sj}.  A more
precise measurement of ${\cal B}(\KS\to  \pi^0 e^+e^-)$ would allow the clean
extraction of the relevant short-distance physics from $\KL\to \pi^0 e^+e^-$.

\section{Unitarity Triangle Angle Measurements}
\label{ut_intro}

The observation of a nonzero Jarlskog parameter~\cite{jarlskog} in the SM
leads to the requirement that the angles of the UT
(and equivalently of all six unitarity triangles of the CKM matrix) be
nonzero (modulo $\pi$). The UT angles $\alpha$, $\beta$ and 
$\gamma$~(\ref{eq:utdefinitions}) are all accessible from the \B sector,
albeit with different sensitivity and purity. Whereas the measurements of 
$\beta$ (the leading experimental observable here is $\stwob$) and $\gamma$, 
through \B decays in charmonium and open charm, respectively, are 
theoretically clean, the measurement of $\alpha$ in charmless \B decays 
relies on theoretical assumptions. Because the measurements of $\alpha$
and $\gamma$ involve interference with transitions governed by the small
CKM matrix element $V_{ub}$, they require larger data samples than when
measuring $\stwob$.

The experimental techniques to measure the UT angles also change radically from
one to another. The measurements of $\alpha$ and $\beta$ require $\Bz\Bzb$
mixing and therefore use neutral \B mesons, whereas the measurements of
$\gamma$ use interference between $b\to u$ and $b\to c$ decay amplitudes, and
can be done with both neutral and charged \B decays.

In the following we neglect \CP violation in \Bd mixing, which has been searched
for with both flavor-specific and inclusive \Bd decays in samples where the
initial  flavor state is tagged. The current world average is 
$|q/p| = 1.0018 \pm 0.0017$~\cite{ASL,ASLnot4S}, whereas the deviation from unity 
is  expected to be $|q/p|- 1 \approx 0.0003$~\cite{Beneke:2003az}, and around
$\lambda^2$ times smaller in \Bs mixing. See Sec.~\ref{new_physics} for an
application of this search in the context of new physics studies.

\subsection{$\beta$ from \B Decays to Charmonium Final States}
\label{ut_beta_cc}

In $b \to c \cbar s$ quark-level decays, the time-dependent \CP violation 
parameters measured from the interference between decays with  and without
mixing are $S_{c\cbar s}=- \etacp \stb$ and $C_{c\cbar s} = 0$, to a very good
approximation.  The theoretically cleanest case if $B\to J/\psi \KSL$, where
\beq\label{BpsiKlam}
\lambda_{\psi \KSL} 
= \mp \bigg( { V_{tb}^* V_{td} \over V_{tb} V_{td}^*} \bigg)
  \bigg( {V_{cb} V_{cs}^* \over V_{cb}^* V_{cs}} \bigg)
  \bigg( {V_{cs} V_{cd}^* \over V_{cs}^* V_{cd}} \bigg) 
= \mp e^{-2i\beta} \,,
\eeq
and so $\Im \lambda_{\psi \KSL} = S_{\psi \KSL} = \pm \sin2\beta$ (see
Sec.~\ref{sec:CPviolation}).  The sign is from $\eta_{\psi \KSL} = \mp 1$,
the first factor is the SM value of $q/p$ in $\Bz\Bzb$ mixing, the second is $\ov
A/A$, and the last one is $p_K/q_K$.  In the absence of $\Kz\Kzb$ mixing there
could be no interference between $\Bzb\to \psi \Kzb$ and $B^0\to \psi K^0$.

\babar~\cite{babar_stb} and \belle~\cite{belle_stb} have both used the $\etacp=-1$ 
modes $\jpsi \KS$, $\psi(2S) \KS$, $\chi_{c1} \KS$ and $\eta_c \KS$, as well as 
$\jpsi \KL$, which has $\etacp = +1$, and $\jpsi K^{*0}(892)(\to\KS\piz)$, which is 
found from an angular analysis to have $\etacp$ close to $+1$ (the \CP-odd
fraction amounts to $0.217\pm0.010$~\cite{hfag2005}). In the latest result 
from \belle, only $\jpsi \KS$ and $\jpsi \KL$ are used. The world average 
reads~\cite{hfag2005}
\beq
\label{eq:stwob}
	\stwob = 0.687 \pm 0.032\,,
\eeq
giving for the angle $\beta$ within $[0,\pi]$ the solutions
$(21.7\,^{+1.3}_{-1.2})^\circ$ and $(68.3 \, ^{+1.2}_{-1.3})^\circ$,
where the first number is compatible with the result from
the global CKM fit without the measurement of $\beta$,
$(24.4^{+2.6}_{-1.5})^\circ$ and $\stwob_{\rm CKM}=0.752\,^{+0.057}_{-0.035}$
(\cf\   Sec.~\ref{globalCKMfit}).  As expected in the SM, no direct \CP
violation has been observed in these modes.

In $b \to c \cbar d$ quark-level decays, such as $\Bz \to \jpsi \pi^{0}$ or 
$\Bz \to D^{(*)}D^{(*)}$, unknown contributions from (not CKM suppressed) 
penguin-type diagrams, carrying a different weak phase than the tree-level 
diagram, compromises the clean extraction of $\stb$. Consequently, they 
are not included in the $\stb$ average.

\subsubsection{Resolving the four-fold ambiguity}

Despite the agreement of Eq.~(\ref{eq:stwob}) with the SM, it is still possible 
that, because of contributions from new physics, the correct value of $\beta$ 
is one of the three solutions not compatible with the SM. The measurement of 
the sign of $\ctb$ eliminates two of the solutions.\footnote
{
        The invariance $\beta\to\pi+\beta$ remains. It cannot be lifted without 
	theoretical input on a strong phase~\cite{quinngrossman}.
}
$\B$-meson decays to the vector-vector final state $J/\psi K^{*0}$, where
three helicity states of the vector mesons mix \CP-even and \CP-odd
amplitudes, are also 
mediated by the $b \to c \cbar s$ transition. When the final state
$K^{*0}\to\KS\pi^0$ is used, a time-dependent transversity analysis can be
performed allowing sensitivity to both $\stwob$ and
$\ctwob$~\cite{dunietz_etal}. Here, $\ctb$ enters as a factor in the
interference between the \CP-even and \CP-odd amplitudes. In principle, 
the sign of $\ctb$ in this analysis is ambiguous owing to an 
incomplete determination of the strong phases occurring in the three
transversity amplitudes. \babar resolves this ambiguity by inserting the known
variation~\cite{lassKpi} of the rapidly changing $P$-wave phase relative to the
slowly changing $S$-wave phase with the invariant mass  of the $K\pi$ system in
the vicinity of the $\Kstarz(892)$ resonance.\footnote
{
	The result for the strong phase is in agreement with the prediction
	obtained from $s$-quark helicity conservation~\cite{suzuki}.
}
The result from \babar, $\ctwob=3.32 \,^{+0.76}_{-0.96} \pm 0.27$, although seemingly
significant, exhibits a strongly non-Gaussian behavior, and the confidence level
for a positive value (as expected in the SM) is only $86\%$~\cite{babar_jpsikst}.
The result from \belle is compatible with zero~\cite{belle_jpsikst}.

Bondar \ea~\cite{b_cud_bondar} suggested another idea to resolve the four-fold 
ambiguity of $\beta$. For quark-level transitions $b \to c\ubar d$, where
the final state is a $\CP$ eigenstate (such as $D_{\CP}\pi^0$), the time dependence 
is given by the usual formulas, with the sine coefficient sensitive to
$\stwob$. When a multibody $D$ decay, such as  $D \to \KS\pi^+\pi^-$ is
used instead, a time-dependent analysis of the Dalitz plot  of the neutral $D$
decay allows the extraction of both $\stwob$ and  $\ctwob$. 
The Belle Collaboration~\cite{b_cud_belle} has perfomed such an analysis, 
giving $\beta=(16 \pm 21 \pm 12)^\circ$, which rules out the non-SM solution at  
$97\%$~\CL. Although HFAG has not attempted a combination of the two results 
on $\ctwob$, it is not premature to conclude that the SM solution is 
largely favored by the data.

\subsection{$\alpha$ from Charmless \B Decays}
\label{ut_alpha}

Unlike $\Bz\to \jpsi \Kz$, for which amplitudes with weak phases
different from the dominant tree phase are doubly CKM suppressed, multiple weak
phases must be considered in most of the analyses of $B$ decays to final states
without charm. This complication makes the extraction of the CKM couplings from
the experimental observables considerably more difficult, although
richer. The decays most sensitive to $\alpha$ are $\Bz \to \pip\pim$,
$\rho^\pm\pi^\mp$, and $\rho^+\rho^-$ ($\alpha$ measures $\pi-\beta-\gamma$ in
the SM and in any model with unitary $3\times3$ CKM matrix). The extraction of
$\alpha$ in the presence of unknown penguin amplitudes requires  an isospin
analysis~\cite{grolon} for $\pi\pi$, $\rho\rho$, and  a Dalitz-plot
analysis~\cite{SnyderQuinn}  for $\rho^\pm\pi^\mp$, which are briefly introduced
below. Relying on flavor  symmetries, in particular $\SU(2)$, does not represent
a severe theoretical limitation.  However, it certainly creates model-dependent
uncertainties from flavor-symmetry  breaking so that --- neglecting statistical
considerations --- the measurement of $\alpha$ is not of the same quality as the
measurements of $\stwob$ and $\gamma$.

In naive factorization, there is a hierarchy between penguins in modes
where the $\Bz$ decays to two pseudoscalar particles (such as $\pip\pim$), 
to a pseudoscalar and a vector particle ($\rho^\pm\pi^\mp$), and to two vector 
particles ($\rho^+\rho^-$)~\cite{aleksanRR}. This is due to the Dirac structure 
of the $(V-A)\times(V+A)$ penguin operators, which do not contribute when the meson
that does not receive the spectator quark  (the ``upper'' meson) is a
vector, as in $B^0\to\rho^+\pi^-$ and $B^0\to\rho^+\rho^-$. Similarly, these
operators interfere constructively (destructively) with $(V-A)\times(V-A)$ 
penguin operators when the upper meson is a pseudoscalar and the lower meson
is a pseudoscalar (vector), as in $B^0\to\pi^+\pi^-$ ($B^0\to\rho^-\pi^+$). 
This qualitative picture\footnote
{
	Because $(V-A)\times(V+A)$ operators 
	are power suppressed in the QCD factorization approach, 
	the naive hierarchy may receive large corrections~\cite{BN}.
} 
reproduces quite well the experimental results. Using the value of $\alpha$ from
the global CKM fit, the penguin-to-tree ratios are $0.23^{+0.41}_{-0.10}$,
$0.05^{+0.07}_{-0.05}$, $0.03^{+0.09}_{-0.03}$ and  $0.10^{+0.02}_{-0.03}$, for
$\pip\pim$, $\rho^+\rho^-$, $\rho^+\pim$ and  $\rho^-\pip$,
respectively~\cite{ckmfitter2004}, where penguin and tree correspond to the
$\Ppipi$ and $\Tpipi$ amplitudes defined in Eq.~(\ref{eq:apipi}) below.

\subsubsection{Relations Between Transition Amplitudes}
\label{amplitude}

\begin{figure}[t!]
  \centerline{
	\includegraphics[width=5cm]{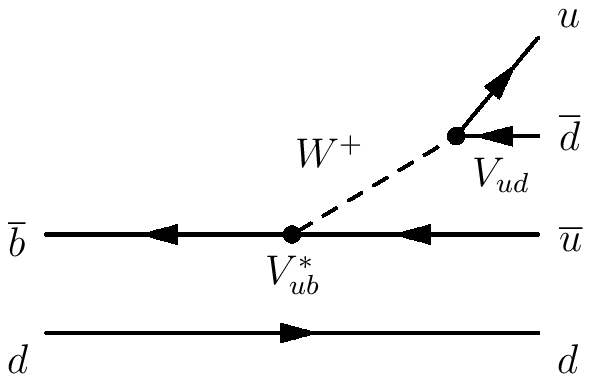}
        \hspace{2cm}
	\includegraphics[width=5cm]{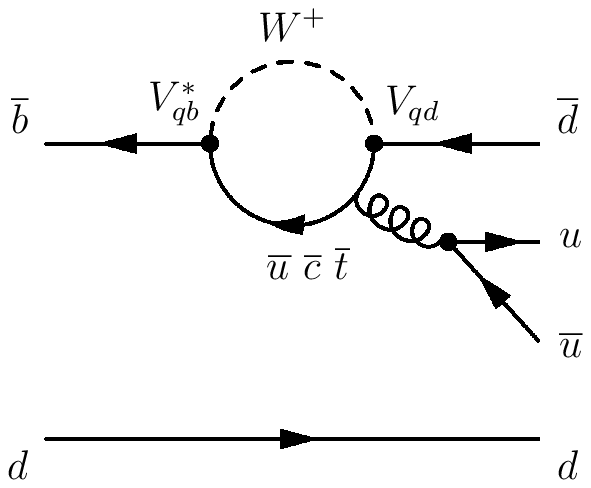}
  }
  \caption[.]{\label{fig:B0pippim}\em
        Tree (left) and penguin (right) diagrams for the decay
        $\Bz\to\pi^+\pi^-$.}
\end{figure}

The general form of the $B^0\to\pi^+\pi^-$ decay amplitude, accounting for the
tree and penguin diagrams that correspond to the three up-type
quark flavors ($u,c,t$) occurring in the $W$ loop (see Fig.~\ref{fig:B0pippim}),
reads
\beq
\label{eq:utgeneral}
   \Apipi \equiv  A(B^0\to\pi^+\pi^-)
                          =   V_{ud}V_{ub}^* M_u
                            + V_{cd}V_{cb}^* M_c
                            + V_{td}V_{tb}^* M_t\,,
\eeq
and the \CP-conjugated amplitude can be similarly written. Using unitarity, 
this can be written as\footnote
{
	In Eq.~(\ref{eq:apipi}), the $c$-quark loop is eliminated
        using the unitarity relation. In Charles \ea's~\cite{ckmfitter2004} 
	definition, this corresponds to the c-convention.
	Note that the particular choice of which amplitude to remove
	in the definition of a total transition amplitude is arbitrary
	and does not have observable physical implications.
}
\beq
\label{eq:apipi}
        \Apipi = V_{ud}V_{ub}^*\Tpipi + V_{td}V_{tb}^*\Ppipi\,,
\eeq
with $\Tpipi= M_u - M_c$ and $\Ppipi = M_t - M_c$. We refer to 
the $T^{+-}$ and $P^{+-}$ 
amplitudes as ``tree'' and ``penguin'', respectively, although it is 
implicitly understood that both of them receive various contributions 
of distinct topologies, which are mixed under hadronic rescattering.
The time-dependent \CP-violating asymmetry is given by
Eqs.~(\ref{SCdef}) and (\ref{SClamdef}), with the \CP-violation parameter 
$\lambda_{\pip\pim}=(q/p)(\Abarpipi/\Apipi)$, since $\pi^+\pi^-$ is a \CP
eigenstate with eigenvalue $+1$. In the absence of penguin contributions
$\lambda_{\pip\pim}=e^{2i\alpha}$, and hence $\Spipi=\sta$ and 
$\Cpipi=0$. In general, however, the 
phase of $\lambda_{\pip\pim}$ is modified by the interference between the 
penguin and the tree amplitudes, which can also lead to direct \CP violation.
It is customary to define an effective angle $\alphaeff$ that incorporates 
the penguin-induced phase shift 
$\lambda_{\pip\pim}=|\lambda_{\pip\pim}|\,e^{2i\alphaeff}$,
so that the sine coefficient becomes $\Spipi = \sqrt{1-\Cpipi^2} \staeff$. 
Twice the effective angle $\alphaeff$ corresponds to the relative phase 
between $e^{-2i\beta}\Abarpipi$ and $\Apipi$. 

The isospin invariance of the strong interaction relates the amplitudes of  the
various $B\to\pi\pi$ decays to each other.  A pair of pions in a $B\to \pi\pi$
decay must be in a zero angular momentum state, and then, because of Bose
statistics, they must have even isospin.  Consequently, $\pi^\pm\pi^0$ is in a
pure isospin-2 state, free of QCD penguin amplitudes, which contribute only to
the isospin-0 final states. Gronau \&  London showed that the measurements of
rates and  \CP-violating asymmetries of the charged and two neutral $\pi\pi$
final states provide  sufficient information to extract the angle $\alpha$ as
well as the various  tree and penguin amplitudes~\cite{grolon}. Unfortunately,
this  isospin analysis  is plagued by a 16-fold ambiguity for
$\alpha\in[0,2\pi]$, \ie, 16 (not necessarily) different values of $\alpha$ 
reproduce the same set of observables. 

Using the same convention as in Eq.~(\ref{eq:apipi}), one can write
for the remaining $\B\to\pi\pi$ decays
\beqn
  \sqrt{2}\Apippiz \:\equiv\:  \sqrt{2}A(B^+\to\pi^+\pi^0)
                        &=&   V_{ud}V_{ub}^*\Tpipiz
                            + V_{td}V_{tb}^*\PpippizEW\,, \nonumber\\
\label{eq:b0pi0pi0}
  \sqrt{2}\Apizpiz \:\equiv\:  \sqrt{2}A(B^0\to\pi^0\pi^0)
                        &=&   V_{ud}V_{ub}^*\Tcpizpiz
                            + V_{td}V_{tb}^*\Ppizpiz\,,
\eeqn
and the \CP-conjugated modes can be similarly written. The subscript C
stands for a color-suppressed amplitude, and the subscript EW 
stands for the electroweak penguin amplitude
contributing to $\pi^+\pi^0$. The $\PpippizEW$ contributions are
usually neglected in the amplitude fits performed by the experiments, 
so that the $\Delta I=3/2$ decay $B^+\to\pi^+\pi^0$ is mediated by a pure 
tree amplitude.\footnote
{
	The bulk part of the shift of $\alpha$ due to $\PpippizEW$ 
	can be estimated model independently~\cite{NRPew,GPYew,BFPEW},
	and amounts to $-2.1^\circ$~\cite{ckmfitter2004}.
} 
Applying the isospin relations~\cite{grolon}
\beq
\label{eq:isospin}
  \Apippiz    = \frac{1}{\sqrt{2}}\Apipi + \Apizpiz\,, \quad{\rm and}\quad
  \Abarpippiz = \frac{1}{\sqrt{2}}\Abarpipi + \Abarpizpiz\,,
\eeq
the amplitudes~(\ref{eq:b0pi0pi0}) can be rearranged and, \eg, $\Tpipiz$ 
and $\Ppizpiz$ eliminated. Isospin-breaking effects have been examined 
in Refs.~\cite{su2break} and are expected to be smaller than the 
current experimental precision on $\alpha$. 

Applying the isospin relations reduces the number of unknowns in the 
$\B\to\pi\pi$ isospin analysis to six, which aligns with the number of 
experimental observables: three branching ratios, one mixing-induced \CP
asymmetry ($S_{\pi^+\pi^-}$), and two direct \CP asymmetries ($C_{\pi^+\pi^-}$
and $C_{\pi^0\pi^0}$). Owing to the missing tracks in most of the $\Bz\to\piz\piz$
events, $S_{\pi^0\pi^0}$ cannot be measured in the foreseeable future. In
absence of isospin breaking, direct \CP violation cannot occur in
$\Bp\to\pip\piz$. 
A small value for the branching fraction to $\pi^0\pi^0$ would indicate that 
the penguin contribution cannot be large~\cite{GrQu}. This observation led  
to the development of isospin bounds on  $\deltaAlpha =
\alpha-\alphaeff$~\cite{GrLoSiSi,theseJerome}. Applying  the optimal bound is
equivalent to constraining $\alpha$ using the isospin relations without using
the measurement of $C_{\pi^0\pi^0}$.

\subsubsection{Isospin Analysis of $\B\to\pi\pi$ Decays}

All six observables used in the isospin analysis have been measured in 
$\B\to\pi\pi$ decays. The past disagreement between \babar and \belle 
for the \CP asymmetries in $\Bz\to\pip\pim$ (\belle saw evidence for 
both mixing-induced and direct \CP asymmetries, whereas \babar did not) 
has been alleviated recently. The world averages for the \CP asymmetries 
and branching fractions measured by \babar, Belle and CLEO (for the 
branching fractions) have been compiled by the HFAG~\cite{allpipi,hfag2005}.
A large $\BRpizpiz/\BRpipi$ ratio is found, indicating significant 
penguin or color-suppressed tree contributions, and representing a 
challenge for the theoretical 
understanding of the decay dynamics. This ratio implies only a loose
upper bound $|\deltaAlpha|<40^\circ$ at $95\%$~\CL, if 
$\Cpizpiz$ is not used~\cite{ckmfitterFPCP06}.

\begin{figure}[t!]  
  \centerline{	\includegraphics[width=8cm]{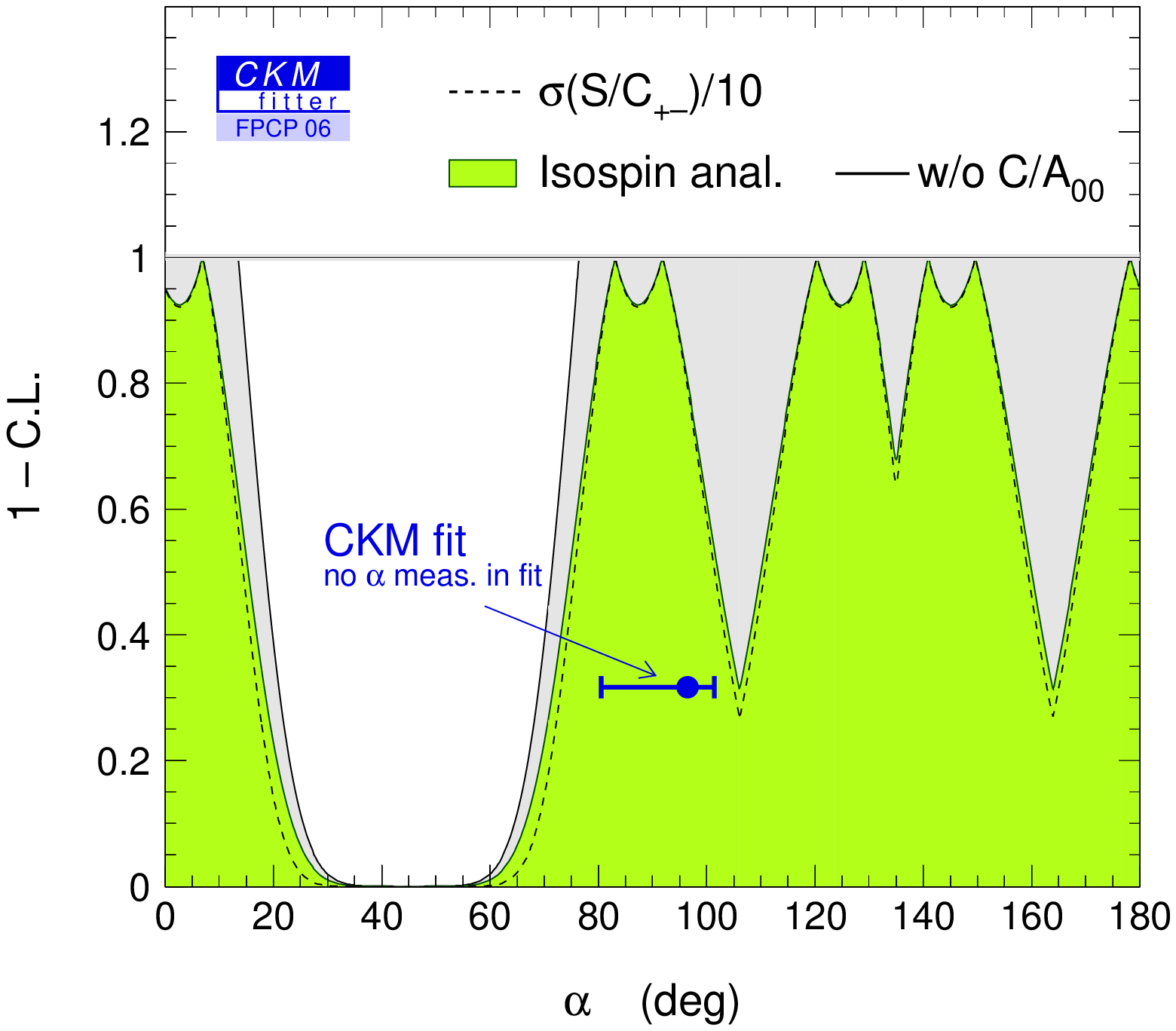} \hfil
		\includegraphics[width=8cm]{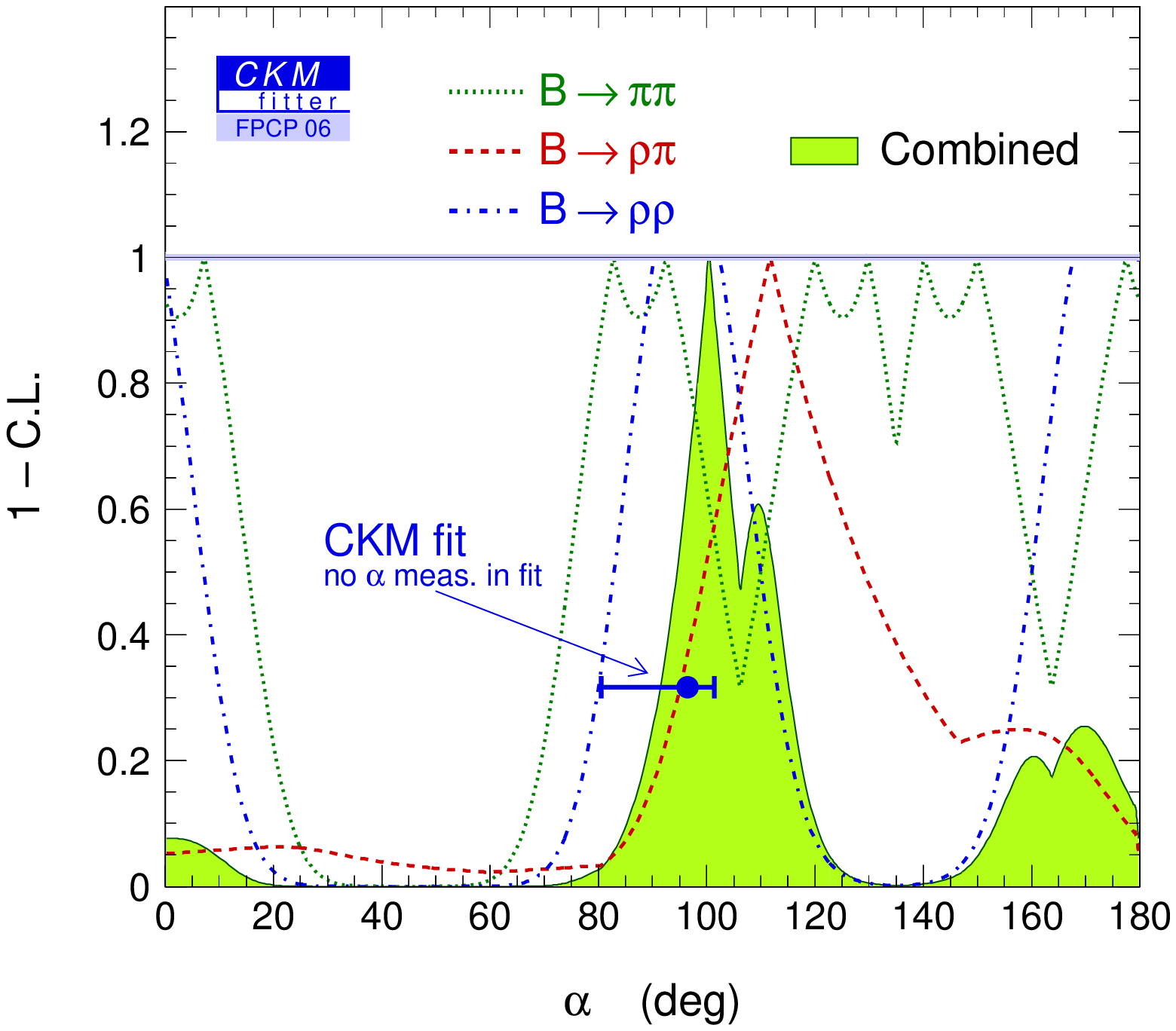}}
  \vspace{-0.2cm}
  \caption[.]{\label{fig:alpha_all}\em
	Left: confidence level for the UT angle $\alpha$
 	obtained from the $\B\to\pi\pi$ isospin analysis under various 
	conditions: full isospin analysis (dark shaded), isospin analysis without
	using $\Cpizpiz$ (light shaded), and full isospin analysis 
	assuming perfect knowledge of the parameters $\Spipi$ and
	$\Cpipi$ (dashed line). Right: 
	confidence level for the UT angle $\alpha$ from 
	the isospin analyses of $\B\to\pi\pi$ (dotted) and $\B\to\rho\rho$
	(dashed-dotted), and the Dalitz plot analysis of $\B\to\rho\pi$ (dashed).
	The shaded region shows the combined constraint from all three 
	measurements~\cite{ckmfitterFPCP06}. }
\end{figure}

The confidence level for $\alpha$ from $\B\to\pi\pi$ is shown in the 
left plot of Fig.~\ref{fig:alpha_all}. Three cases are distinguished: 
$a)$ the full isospin analysis using 
the present world averages (shaded), $b)$ the incomplete isospin 
analysis without $\Cpizpiz$ (solid line), and $c)$ the full isospin 
analysis assuming perfect knowledge of $\Spipi$ and $\Cpipi$ (dashed line). 
The present constraint on $\alpha$ from $\pipi$ is weak, and future 
improvement will be driven almost entirely by the measurement of $\Cpizpiz$.

\subsubsection{Isospin Analysis of $\B\to\rho\rho$ Decays}

Similar to $\B\to\pi\pi$, the angle $\alpha$ can be extracted from 
the measurement of time-dependent \CP asymmetries in $\Bz\to\rhop\rhom$,
together with an isospin analysis including the full $\B\to\rho\rho$ system,
to determine the penguin pollution. The specific interest of this mode
is the potentially small penguin contribution, which is suggested by 
the smallness of the experimental upper bound on the $\Bz \to\rho^0\rho^0$ 
branching fraction when compared with the branching fractions of 
$\Bz \to \rho^+\rho^-$ and  $\Bp \to\rho^+\rho^0$~\cite{allrhorho,hfag2005}.
In addition, both direct and mixing-induced \CP-violating asymmetries 
of the $\Bz \to\rho^0\rho^0$ decays are experimentally 
accessible once the decay has been seen.

The analysis is complicated by the presence of three helicity  states
for the two vector mesons. One corresponds to longitudinal  polarization ($A_0$)
and is \CP-even. Two helicity states ($A_\pm$) are  transversely polarized and
are admixtures of \CP-even and  \CP-odd amplitudes. Because helicity flips
are suppressed in these decays, the relative proportions  of the helicity 
amplitudes in naive factorization are expected to be~\cite{koernergold}
\beq
	A_0:A_-:A_+ \approx 1	
			:\frac{\Lambda_{\rm QCD}}{m_b}
			:\left(\frac{\Lambda_{\rm QCD}}{m_b}\right)^{\!2}\,.
\eeq
The dominance of the longitudinal polarization has been confirmed by experiment
in the measured $\B\to\rho\rho$ decays. The isospin analysis can therefore, 
without significant loss of sensitivity, be applied to the longitudinally 
polarized states only.\footnote
{
	Owing to the finite width of the $\rho$ meson, $I=1$ contributions can
	occur in $\B\to \rho\rho$ decays~\cite{ligetirhorho}. Although no
	explicit calculation is available, one expects these effects to be of
	order $(\Gamma_\rho/m_\rho)^2\sim4\%$. At present the experiments also
	neglect effects from the interference with the radial excitations of
	the   $\rho$, with other $\pi\pi$ resonances, and with a nonresonant
	component. The latter two effects are included in the systematic errors 
	quoted by the experiments.
}

The confidence level obtained from the $B \to \rho\rho$ system for $\alpha$
is given in the right plot of Fig.~\ref{fig:alpha_all} (dashed-dotted
line). The reduction of $\BR(\Bp\to\rho^+\rho^0)$ by the latest \babar\
measurement improved the consistency of the isospin analysis, which now 
exhibits a closed isospin triangle, and deteriorated somewhat the resulting
constraint on $\alpha$. Choosing the solution consistent with the SM 
gives $\alpha_{\rho\rho}=(100\,^{+13}_{-20})^\circ$~\cite{ckmfitterFPCP06}. 
Also shown in the plot is the prediction from the CKM fit, where the direct 
measurement of $\alpha$ has been excluded from the fit. It exhibits excellent 
agreement with the measurement. The $95\%$~\CL bound on $\deltaAlpha$ is 
$|\deltaAlpha| < 17^{\circ}$, reflecting a small penguin pollution. 

\subsubsection{Dalitz Plot Analysis of $\B\to\rho\pi$ Decays}

Unlike  $\pip\pim$ and $\rho^+\rho^-$, the final state $\rho^{\pm}\pi^{\mp}$ 
is not a \CP eigenstate, and four flavor-charge configurations
$(\Bz(\Bzb) \to \rho^{\pm}\pi^{\mp})$ must be considered.  
The corresponding isospin analysis~\cite{Lipkinetal} is unfruitful
with the present statistics, because two pentagonal amplitude relations with 
a total of 12 unknowns have to be solved. However, as Snyder and 
Quinn~\cite{SnyderQuinn} pointed out, the necessary degrees of freedom 
to constrain $\alpha$ without ambiguity (except for the irreducible 
$\alpha \to \alpha+\pi$), can be obtained by including in the analysis 
the (known) strong-phase modulations of the interfering $\rho$
resonances in the Dalitz plot. 

The \babar~\cite{babarRhopiDalitz} Collaboration has performed such 
an analysis, where the interfering \Bz decay amplitudes into 
$\rho^+\pim$, $\rho^-\pip$ and $\rho^0\piz$ have been included.
To avoid mirror solutions for the fit parameters and non-Gaussian observables,
\babar\  fits 16 interdependent coefficients that are bilinears
of the charged and neutral $\rho$ form factors.\footnote
{
  	The full set of bilinears comprises 27 observables out of which 
	\babar\  neglects nine related to \CP asymmetries in $\Bz\to\rho^0\piz$,
	which has only a small event yield.
} 
Each coefficient is related in a 
unique way to the quantities of interest, the tree-level and penguin-type 
amplitudes, and $\alpha$. These are obtained in a least-squares
fit to the measured coefficients. Systematic uncertainties are
dominated by the lack of knowledge of the tails of the $\rho$ form factor, 
where higher $\rho$ excitations occur and precise \B decay data are scarce.  
Ignoring the mirror solution at $\alpha + 180^\circ$, \babar\  finds
$\alpha_{\rho\pi}= (113^{\,+27}_{\,-17}\pm6)^\circ$, whereas only a marginal 
constraint on $\alpha$ is obtained beyond two standard deviations.

The confidence level for $\alpha$ obtained from the $B \to \rho\pi$ Dalitz
analysis is given in the right plot of Fig.~\ref{fig:alpha_all} (dashed
line). Also shown is the combined constraint from all three measurements (shaded), 
giving 
\beq
	\alpha = \left(100\,^{+15}_{-9}\right)^{\!\circ}\,
\eeq
for the SM solution.
This measurement is in agreement with the expectation
$\alpha_{\rm CKM}=(97\,^{+5}_{-16})^\circ$ from the global 
CKM fit (see Table~\ref{tab:fitResultsI} on 
p.~\pageref{tab:fitResultsI})~\cite{ckmfitterFPCP06}.

\subsection{$\gamma$ from \B Decays to Open Charm}
\label{sec:gamma}

The golden methods to determine $\gamma$ at the \B-factories utilize the
measurement of direct \CP violation in $\Bp\to D\Kp$ decays, where the neutral
$D$ meson can be both $\Dz$ and $\Dzb$ (and where $\Dz$ also stands for $\Dstarz$). 
The $\Dz$ corresponds to the leading $\bbar \to \cbar$ transition, whereas the $\Dzb$ 
is produced by a CKM- and color-suppressed $\bbar \to \ubar$ transition. 
If the final state is chosen so that both $\Dz$ and $\Dzb$ can contribute, the 
two amplitudes interfere, and the resulting observables are sensitive to the 
UT angle $\gamma$, the relative weak phase between the two $\B$ decay amplitudes. 

Among the many methods that exploit this interference, the experiments concentrate 
on the reconstruction of the neutral $D$ in a \CP eigenstate (GLW)~\cite{glw}, 
in other final states common to $\Dz$ and $\Dzb$ such
as $K^\mp\pi^\pm$ (ADS)~\cite{ads}, or in the self-conjugate
three-body final  state $\KS \pi^+\pi^-$ (GGSZ)~\cite{gamma_dalitz}. All
variations are sensitive to the same \B decay parameters and can therefore be
treated in a combined fit to extract $\gamma$. For the GLW analysis, the relations 
between the experimental observables, the rates $R_{\CP^\pm}$ and charge asymmetries
$A_{\CP^\pm}$ of the \CP-even and \CP-odd modes, and the model parameters are
\beq
	R_{\CP^\pm} = 1+r_B^2\pm2 r_B \cos\!\delta_B\cos\!\gamma\,, \qquad
	A_{\CP^\pm} = \pm\frac{2 r_B \sin\!\delta_B\sin\!\gamma}{R_{\CP^\pm}}\,,
\eeq
where $\delta_B$ is the (\CP-conserving) relative strong phase between the 
interfering amplitudes, and $\rB^{(*)} = |A(\Bp\to\Dz{}^{(*)}\Kp) /
A(\Bp\to\Dzb{}^{(*)}\Kp)|$ is the  ratio of the \B decay amplitudes. The same
relations apply for the GGSZ analysis,  where the variations of additional
strong phases and amplitude ratios  throughout the Dalitz plot of the three-body
decay are extracted from high statistics $D$ decay data and modeled by the
experiments.  For the ADS analysis two additional unknowns (phase and amplitude
ratio) appear because the neutral $D$ mesons do not decay to \CP eigenstates.
High-statistics charm  decays can be used as external constraints for the
amplitude ratio. Moreover, only  two observables (rate and asymmetry) per \B
decay mode exist for ADS. The above equations show that the feasibility of the
$\gamma$ measurement depends crucially on the size of $r_B$ (expected to be
roughly $\rB \sim 0.1$, if  naive color suppression holds). Instead of fitting
the model parameters ($\gamma$, $r_B$, $\delta_B$) directly, the experiments
proceed by determining the Cartesian coordinates $(x_\pm,y_\pm)=(r_B
\cos(\delta_B\pm\gamma),r_B \sin(\delta_B\pm\gamma))$, which have the advantage
of being Gaussian distributed, uncorrelated and unbiased  (the parameter $r_B$ is
positive definite and hence exhibits a fit bias toward larger values, if its
central value is in the vicinity of zero within errors), and simplify the
averaging of the various measurements.

Results are available from both \babar\ and \belle\ on the GLW analysis using 
the decay  modes $\Bp \to D\Kp$, $\Bp \to \Dstar\Kp$ and $\Bp \to
D\Kstarp$~\cite{GLWanalysis}. For ADS~\cite{ADSanalysis} only \babar\  uses all
three decays. Here, the decays $\Dstar \to D\pi^0$ and  $\Dstar \to D\gamma$
have a relative strong-phase shift and must be studied 
separately~\cite{gershonbondar}. The world averages for the GLW and ADS 
measurements are provided by the HFAG~\cite{hfag2005}. So far none of these
measurements have obtained a significant signal for the suppressed amplitude.

Both experiments have studied the GGSZ analysis using the same \B decay 
modes as for the GLW and ADS analyses, and using in all cases the subsequent 
decay $\Dz(\Dzb) \to \KS\pi^+\pi^-$. Both experiments perform a full 
frequentist statistical analysis to 
determine the physical parameters. Exploiting the three-body Dalitz plot 
enhances the sensitivity to the suppressed amplitude compared to the 
(quasi-)two-body decays used for GLW and ADS. This leads to a significant 
determination of $r_B$ and hence $\gamma$ with this method. Using the same 
three modes $D^{(*)}K^{(*)+}$, the experiments report the GGSZ measurements 
$\gamma=(67 \pm 28 \pm 13 \pm 11)^\circ$ (\babar~\cite{babar_ggsz}) and
$\gamma=(53\,^{+15}_{-18} \pm 3 \pm 9)^\circ$ (Belle~\cite{belle_ggsz}), 
where the first errors are statistical, the second are systematic, and the 
third are due to the Dalitz plot model.

\begin{figure}[t!]
  \centerline{\includegraphics[width=8cm]{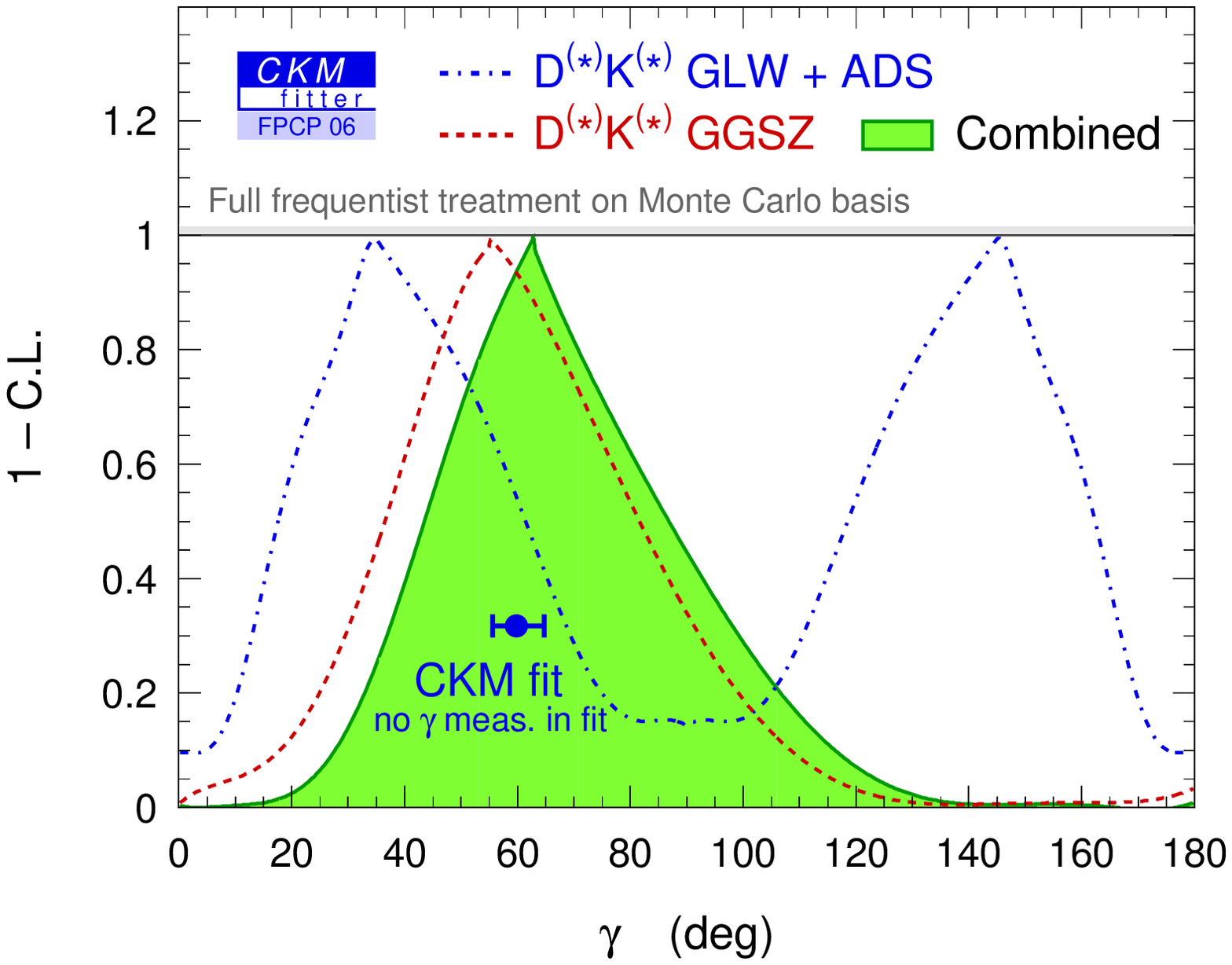} \hfil
	      \includegraphics[width=8cm]{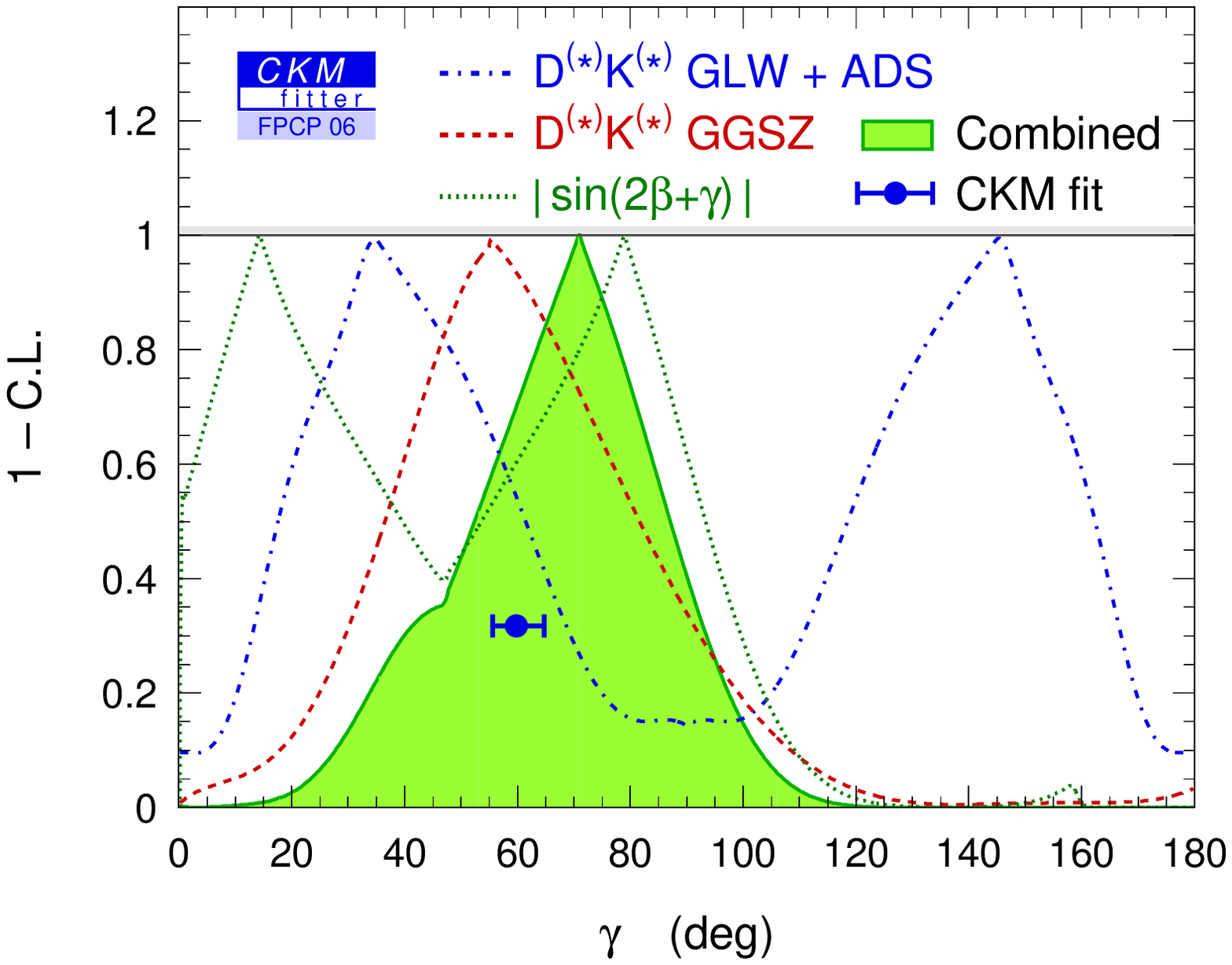}}

  \caption[.]{\label{fig:gamma_all}\em
	Left: confidence level for the UT angle $\gamma$ from 
	the measurement of direct \CP asymmetries in $\B\to D^{(*)}K^{(*)}$
	decays (left). Shown are the individual results for the combined GLW and ADS
	analysis (dashed-dotted), for GGSZ (dashed), and the combination of
	both analyses (shaded). Right:
	adding to these the measurement of $2\beta+\gamma$ 
	from time-dependent analyses of neutral \B decays to open charm,
	together with the precise $\stwob$ measurement.
	A theoretical error of $15\%$ has been used to account for 
	SU(3) breaking. The plots are taken from 
	Ref.~\cite{ckmfitterFPCP06}.}
\end{figure}

The CKMfitter Group has performed a combined analysis of all GLW, ADS and GGSZ 
results using a frequentist statistical framework. Including the former 
two methods leads to a reduction in the average $r_B$ values of all modes, 
which are $r_B(DK)=0.10\pm0.03$, $r_B(D^*K)=0.10\,^{+0.03}_{-0.05}$ and 
$r_B(DK^*)=0.11\,^{+0.08}_{-0.11}$~\cite{ckmfitterFPCP06}. All values 
are in agreement with the 
naive expectation from CKM and color suppression. The smaller $r_B$
values deteriorate the constraint on $\gamma$ obtained from the GGSZ 
analyses. The preliminary overall $\gamma$ average is~\cite{ckmfitterFPCP06}
\beq
	\gamma=\left(62\,^{+35}_{-25}\right)^{\!\circ}\,,
\eeq
which is in agreement with the prediction 
$\gamma_{\rm CKM}=(60\,^{+5}_{-4})^\circ$ from the global CKM fit (where the 
direct $\gamma$ measurement has been excluded from the fit)~\cite{ckmfitterFPCP06}.
The GLW, ADS and GGSZ confidence levels for $\gamma$ as well as the combined 
constraint are shown in Fig.~\ref{fig:gamma_all}. Also shown is the expectation
from the global CKM fit.

\subsection{$2\beta+\gamma$ from \B Decays to Open Charm}

Similar to the decay $\Bz\to\rho^\pm\pi^\mp$, which is not a \CP eigenstate
but sensitive to $\alpha$ because both final states can be reached by
both neutral \B flavors, interference between decays with and without
mixing can occur in $\Bz\to D^{(*)\pm}\pi^\mp(\rho^\pm)$. A time-dependent 
analysis of these decays is sensitive to $\sin(2\beta+\gamma)$, because
the CKM-favored $\bbar\to\cbar$ decay amplitude interferes with the CKM-suppressed 
$\b\to\u$ decay amplitude with a relative weak-phase shift $\gamma$.
In these $\bbar\to\cbar( u\dbar),\ \ubar(c\dbar)$ quark-level transitions 
no penguin contributions are possible, because
all quarks in the final state are different. Hence there is no direct $\CP$ 
violation. Furthermore, owing to the smallness of the ratio of the magnitudes 
of the CKM-suppressed and CKM-favored amplitudes (denoted $r_f$),
to a very good approximation, $C_f = - C_{\ov{f}} = 1$
(using $f = D^{(*)-}h^+$, $\ov{f} = D^{(*)+}h^-$ $h = \pi,\rho$),
and the coefficients of the sine terms are given by
\beq
  S_f = - 2 r_f \sin( 2\beta + \gamma - \delta_f )\,, \qquad
  S_{\ov{f}} = - 2 r_f \sin( 2\beta + \gamma + \delta_f ),
\eeq
where $\delta_f$ is the strong-phase shift due to final-state 
interaction between the decaying mesons. Hence, the weak phase $2\beta+\gamma$
can be cleanly obtained from measurements of $S_f$ and $S_{\ov{f}}$, 
although external information on at least $r_f$ or $\delta_f$ is 
necessary. It can be obtained from the corresponding flavor-tagged branching 
fractions, or from similar modes that are easier to measure. These can be 
ratios of branching fractions of the charged $B^+\to D^{(*)+}\pi^0$ to the 
neutral CKM-favored decay, or ratios involving self-tagging decays with 
strangeness like $B^0\to D_s^{(*)+}\pi^-$. Corrections for $\SU(3)$ breaking 
in the latter case generate a significant theoretical uncertainty,
which is in general hard to quantify. Naively, one can estimate
$r^{(*)}\sim|V_{cd}^*V_{ub}/V_{ud}V_{cb}^*|\simeq0.02$. At present,
the most precise estimate of $r_f$ obtained from $\SU(3)$-corrected ratios
are~\cite{babardstarpiF} $r_{D\pi}= 0.013\pm0.002$ and 
$r_{D^*\pi} = 0.019\pm0.0029$, where  a $15$--$30\%$ theoretical
uncertainty is usually added to account for $\SU(3)$ breaking.

The constraint on $2\beta+\gamma$ obtained using these amplitude ratios
(assuming a $15\%$ uncertainty on $\SU(3)$ breaking), and the world averages 
of the experimental observables~\cite{sin2bpganalysis,hfag2005}, is depicted
by the dotted line in the right plot of Fig.~\ref{fig:gamma_all}. The 
measured values for $\stwob$ (and $\ctwob$) have been used to infer $\gamma$
from this. The shaded curve gives the overall constraint on $\gamma$ obtained 
when averaging these results with those given in Sec.~\ref{sec:gamma}.
The $2\beta+\gamma$ measurements help to exclude large values of $\gamma$.

\subsection{Weak-Phase Information from Direct \CP Violation in \B Decays}

The KM mechanism causes ``direct'' \CP violation in the decay, 
as soon as at 
least two amplitudes with different strong and weak phases contribute. 
Because virtual loops are present in all meson 
decays, ``some'' (possibly unobservable) amount of direct \CP violation 
{\em always} occurs. Owing to the large weak phases arising in \B decays,
direct \CP violation should be more prominent here than, \eg, in the kaon 
system. This has been confirmed by the measurement of the
direct \CP asymmetry $A_{K^+ \pi^-} = -0.115 \pm 0.018$ in 
$\Bz\to\Kp\pim$ decays~\cite{directcpkpi}. Evidence for direct \CP
violation in neutral \B decays also exists for $\Bz\to\pip\pim$ 
($3.7\sigma$ significance) and $\Bz\to\rho^\pm\pi^\mp$ ($3.4\sigma$ 
significance)~\cite{hfag2005}. Recently, the first evidence for direct \CP
violation in charged \B decays emerged from the mode $\Bp\to\Kp\rho^0$
with a charge asymmetry of $0.31\,^{+0.12}_{-0.11}$~\cite{directcprhok}.
With the data samples at the \B-factories increasing, we expect the
discovery of more and more rare-decay modes with significant \CP 
violation in the decay.

From the point of view of the weak-phase extraction, the required 
conspiracy between competing amplitudes of similar size and the 
occurrence of strong phases, represent serious obstacles. A reliable and 
model-independent calculation of direct \CP violation is not possible at present, 
and estimates based on factorization are plagued by large uncertainties.
However, flavor symmetries in particular
isospin can be exploited to (essentially) assess model independently direct 
\CP violation. In $\B$ decays 
to $\pi\pi$, $\rho\pi$ and $\rho\rho$, the measurements of direct 
\CP-violating asymmetries (independent of whether they are
compatible with zero or not) are essential inputs to the isospin
analyses. In the $K\pi$ system the corresponding isospin analysis
used to extract $\gamma$~\cite{Lipkinetal} is fruitless at present, 
and affected by possibly large isospin-breaking corrections from 
electroweak penguins, which cannot be taken into account model independently
as is the case in the $\pi\pi$ and $\rho\rho$ isospin analyses. 

Although a quantitative prediction is difficult, direct \CP violation 
can be a powerful probe for new physics in decays where negligible
\CP asymmetries are expected. This is the case for all $\B$ decays 
dominated by a single decay amplitude. Prominent examples are 
penguin-dominated decays, such as $b\to s\gamma$ or
$\B\to K^{(*)}\ell^+\ell^-$, where a significant nonzero direct
\CP violation would unambiguously indicate new physics.

\section{The Global CKM Fit}
\label{globalCKMfit}

   The global CKM fit consists of maximizing a likelihood built upon 
   relevant experimental measurements and their SM predictions,
   which depend on the parameters of the theory. Some of these 
   parameters, such as quark masses or matrix elements, are 
   experimentally or theoretically constrained, whereas others are 
   unknown. These unknowns contain the four Wolfenstein 
   parameters, but also, for instance, hadronic quantities that occur in 
   the determination of the UT angles $\alpha$ and $\gamma$. To avoid
   uncontrolled biases, the unknown parameters should be treated as 
   free parameters of the theory in a frequentist statistical 
   sense. Such an approach has been adopted by the CKMfitter 
   Group~\cite{ckmfitterOrig,ckmfitter2004}, whose results we use here.
   
   Many of the observables 
   presently used to constrain the CKM matrix depend on hadronic matrix
   elements and hence are subject to significant theoretical uncertainties.
   The treatment of these theoretical uncertainties has a large amount of 
   arbitrariness, and different choices lead to different
   results. The CKMfitter Group advocates the 
   Rfit approach~\cite{ckmfitterOrig,ckmfitter2004}.
   In this approach the ranges spanned by the model-dependent theoretical errors 
   are scanned to maximize the agreement between theory and experiment, which 
   corresponds to a minimization of the exclusion confidence level at a given point 
   in parameter space that is to be tested. Other approaches exist in the 
   literature~\cite{ckmreport,bayesian}.

\subsection{Fit Inputs and Initial Tests of Unitarity}

What is often termed the ``standard global CKM fit'' includes 
only those observables for which the SM predictions can be considered 
quantitatively under control, and that lead to a significant constraint 
on the CKM parameters. These comprise the measurements of $\Vud$, $\Vus$, $\Vub$, 
$\Vcb$, $|\epsK|$, $\dmd$, $\dms$,\footnote
{
	The results shown in this section include the new \dzero and CDF analyses 
	of $\Bs\Bsb$ mixing~\cite{d0bsbsbar,cdfbsbsbar}.
}
$\stwob$, $\ctwob$, $\alpha$, $\gamma$ and $\BR(\Bp\to\tau^+\nut)$. The inputs 
used by the fit are given in Table~\ref{tab:fitInputs}.

Using the independently measured moduli of the CKM elements mentioned in the
previous sections, the unitarity of the CKM matrix can be checked. We obtain
$|V_{ud}|^2 + |V_{us}|^2 + |V_{ub}|^2 -1 = -0.0008 \pm 0.0011$ (first row),
$|V_{cd}|^2 + |V_{cs}|^2 + |V_{cb}|^2 -1 = -0.03 \pm 0.18$ (second row), and
$|V_{ud}|^2 + |V_{cd}|^2 + |V_{td}|^2 - 1 = 0.001 \pm 0.005$ (first column).
The sum in the second column, $|V_{us}|^2 + |V_{cs}|^2 +
|V_{ts}|^2$ is practically identical to that in the second row, as the errors
in both cases are dominated by $|V_{cs}|$.  For the second row, a more stringent
test is obtained from the measurement of $\sum_{u,c,d,s,b} |V_{ij}|^2$ minus the
sum in the first row, giving $|V_{cd}|^2 + |V_{cs}|^2 + |V_{cb}|^2 = 1.000 \pm
0.026$.

It is an unexpected success of the \B-factory experiments that, beyond
the precision measurement of $\stwob$, measurements of $\alpha$ and $\gamma$
became available. Remarkably, the most powerful determinations of the latter two
angles involve measurements that were not considered before the data
became available. Choosing the SM solution for each of the angles, their sum 
reads~\cite{ckmfitterFPCP06}
\beq
	\alpha+\beta+\gamma=\left(186\,^{+38}_{-27}\right)^{\!\circ}\,, 
\eeq
which corresponds to a closed triangle verifying the three-generation 
KM mechanism.

\begin{table}[!t]
\begin{center}
\setlength{\tabcolsep}{0.0pc}
{\small
\begin{tabular*}{\textwidth}{@{\extracolsep{\fill}}lcccc}\hline
        &       &       &       \mc{2}{c}{Errors}       \\
  \rs{Parameter}        & \rs{Value $\pm$ Error(s)}
                & \rs{Reference}
                & GS
                & TH \\
\hline
  $\Vud$ ($\beta$-decays)  
                & $0.97377\pm0.00027$
                & \cite{ckm2005}
                & $\star$ & -- \\
  $\Vus$ ($K_{\ell3}$ and $K_{\mu2}$)
                & $0.2257\pm0.0021$
                & \cite{PDG}
                & $\star$ & -- \\
  $\Vub$ (inclusive)         
                & $(4.45 \pm 0.23 \pm 0.39) \times 10^{-3}$
                & \cite{hfag2005,ckmfitterFPCP06}
                & $\star$ & $\star$ \\
  $\Vub$ (exclusive)         
                & $(3.94 \pm 0.28 \pm 0.51) \times 10^{-3}$
                & \cite{PDG}
                & $\star$ & $\star$ \\
  $\Vcb$ (inclusive)   
                & $(41.70\pm 0.70) \times 10^{-3}$
                & \cite{PDG}
                & $\star$ & -- \\
  $\Vcb$ (exclusive)   
                & $(41.2\pm 1.7)\times10^{-3}$
                & \cite{hfag2005}
                & $\star$ & -- \\
\hline
  $|\epsk|$        
                & $(2.221\pm0.008)\times10^{-3}$
                & \cite{Alexopoulos:2004sw,ckmfitterFPCP06}
                & $\star$ & -- \\
  $\dmd$           
                & $(0.507 \pm 0.004)~{\rm ps}^{-1}$
                & \cite{hfag2005}
                & $\star$ & -- \\
  $\dms$           
                & $\begin{array}{c}
                   \mbox{Amplitude spectrum (incl.\ \dzero), and} \\[-0.1cm] 
                   \mbox{CDF $\log{\cal L}$ parameterization}\end{array}$
                & \cite{ckm2005}
                & $\star$ & -- \\
\hline
  $\stbwa$ & $0.687 \pm 0.032$
                & \cite{hfag2005}
                & $\star$ & -- \\
\hline
  $S_{\pi\pi}^{+-}$, $C_{\pi\pi}^{+-}$, $C_{\pi\pi}^{00}$, $\BR_{\pi^{+/0}\pi^{-/0}}$ 
		& Inputs to isospin analysis
                & \cite{hfag2005}
                & $\star$ & -- \\
  $S_{\rho\rho,L}^{+-}$, $C_{\rho\rho,L}^{+-}$, $\BR_{\rho^{+/0}\rho^{-/0},L}$ 
		& Inputs to isospin analysis
                & \cite{hfag2005}
                & $\star$ & -- \\
  $B^0 \to (\rho \pi)^0 \to 3\pi$ & Time-dependent Dalitz plot analysis
                & \cite{BABAR-dalitz3pi}
                & $\star$ & -- \\
\hline
  $B^{-}\rightarrow D^{(*)} K^{(*)-}$ & Inputs to GLW analysis
                & \cite{hfag2005}
                & $\star$ & -- \\
  $B^{-}\rightarrow D^{(*)} K^{(*)-}$ & Inputs to ADS analysis
                & \cite{hfag2005}
                & $\star$ & -- \\
  $B^{-}\rightarrow D^{(*)} K^{(*)-}$ & GGSZ Dalitz plot analysis
                & \cite{hfag2005}
                & $\star$ & -- \\
\hline
  $\BR(B^-\to\tau^-\nutb)$ & Experimental likelihoods
		& \cite{btaunu}
		& $\star$ & -- \\
\hline
  $\mcRun(m_c)$    & $(1.24\pm0.037\pm0.095)\gev$
                & \cite{Buchmuller:2005zv}
                & $\star$ 
                & $\star$ \\
  $\mtRun(m_t)$    
                & $(162.3\pm2.2)\gev$
                & \cite{TopCDFD0Tev}
                & $\star$ 
                & -- \\
  $f_K$            
                & $(159.8\pm1.5)\mev$
                & \cite{PDG}
                & --
                & -- \\
\hline
  $B_K$            
                & $0.79\pm0.04\pm0.09$
                & \cite{ckm2005}
                & $\star$ 
                & $\star$ \\
  $\as(m_Z^2)$
                & $0.1176\pm0.0020$
                & \cite{PDG}
                & -- 
                & $\star$ \\
  $\eta_{ct}$      
                & $0.47\pm0.04$
                & \cite{Nierste}
                & --
                & $\star$ \\
  $\eta_{tt}$
                & $0.5765\pm0.0065$
                & \cite{Nierste}
                & --
                & -- \\
  $\etaB(\MSbar)$
                & $0.551\pm0.007$ 
                & \cite{bbl}
                & --
                & $\star$ \\

  $\fbd$
                & $(191\pm27)\mev$
                & \cite{ckm2005}
                & $\star$ 
                & -- \\
  $B_d$
                & $1.37\pm0.14$
                & \cite{ckm2005}
                & $\star$ 
                & -- \\
  $\xi^{(a)}$
                & $1.24 \pm 0.04 \pm 0.06$
                & \cite{ckm2005}
                & $\star$ 
                & $\star$ \\
\hline
\end{tabular*}
}
\end{center}
\vspace{-0.3cm}
{\footnotesize $^{(a)}$anticorrelated theory error with $\fbd\sqrt{B_d}$.}
\vspace{-0.1cm}
\caption{\label{tab:fitInputs} \em
        Inputs to the standard CKM fit. 
        If not stated otherwise, when two errors given, the 
        first is statistical and accountable systematic and the 
        second stands for  theoretical uncertainties.
        The last two columns indicate whether errors are treated
	Gaussian (GS) or theoretical (TH) within Rfit~\cite{ckmfitter2004}.
	Table taken from Ref.~\cite{ckmfitterFPCP06}.
	}
\end{table}

Various theory parameters enter the SM predictions that relate the measurements
to the fundamental CKM parameters determined by the fit. Those with the largest
uncertainties are hadronic matrix elements obtained with LQCD, quark masses
determined using LQCD and QCD sum-rule analyses, and coefficients calculated 
in perturbative QCD. The values used for these parameters in the CKM fit are 
also given in Table~\ref{tab:fitInputs}.

\subsection{Fit Results}

\begin{table}[t!]
 \begin{center}
 \setlength{\tabcolsep}{0.8pc}
 {\normalsize
 \begin{tabular*}{\textwidth}{@{\extracolsep{\fill}}lccc}\hline&&& \\[-0.3cm]
 Observable & central $\pm$ $\CL\equiv1\sigma$ & 
 $\pm$ $\CL\equiv2\sigma$ & $\pm$ $\CL\equiv3\sigma$  \\[0.15cm]
\hline
 $\lambda$                                                              & $   0.2272 ^{\,+   0.0010}_{\,-   0.0010}$ & $^{\,+   0.0020}_{\,-   0.0020}$ & $^{\,+   0.0030}_{\,-   0.0030}$
 \\[0.00cm]
 $A$                                                                    & $    0.809 ^{\,+    0.014}_{\,-    0.014}$ & $^{\,+    0.029}_{\,-    0.028}$ & $^{\,+    0.044}_{\,-    0.042}$
 \\[0.00cm]
 $\rhobar$                                                             & $    0.197 ^{\,+    0.026}_{\,-    0.030}$ & $^{\,+    0.050}_{\,-    0.087}$ & $^{\,+    0.074}_{\,-    0.133}$
 \\[0.00cm]
 $\etabar$                                                             & $    0.339 ^{\,+    0.019}_{\,-    0.018}$ & $^{\,+    0.047}_{\,-    0.037}$ & $^{\,+    0.075}_{\,-    0.057}$
 \\[0.00cm]
 \hline
 $J$~~$[10^{-5}]$                                                       & $     3.05 ^{\,+     0.18}_{\,-     0.18}$ & $^{\,+     0.45}_{\,-     0.36}$ & $^{\,+     0.69}_{\,-     0.54}$
 \\[0.00cm]
 \hline
 $\stwoa$                                                        & $    -0.25 ^{\,+     0.17}_{\,-     0.15}$ & $^{\,+     0.49}_{\,-     0.28}$ & $^{\,+     0.71}_{\,-     0.42}$
 \\[0.00cm]
 $\stwoa$ {\small (meas.\ not in fit)}                            & $    -0.23 ^{\,+     0.55}_{\,-     0.16}$ & $^{\,+     0.72}_{\,-     0.32}$ & $^{\,+     0.83}_{\,-     0.45}$
 \\[0.00cm]
 $\stwob$                                                         & $    0.716 ^{\,+    0.024}_{\,-    0.024}$ & $^{\,+    0.048}_{\,-    0.049}$ & $^{\,+    0.074}_{\,-    0.075}$
 \\[0.00cm]
 $\stwob$ {\small (meas.\ not in fit)}                             & $    0.752 ^{\,+    0.057}_{\,-    0.035}$ & $^{\,+    0.105}_{\,-    0.073}$ & $^{\,+    0.135}_{\,-    0.112}$
 \\[0.00cm]
 $\alpha$~~(deg)                                                        & $     97.3 ^{\,+      4.5}_{\,-      5.0}$ & $^{\,+      8.7}_{\,-     14.0}$ & $^{\,+     13.7}_{\,-     20.7}$
 \\[0.00cm]
 $\alpha$~~(deg) {\small (meas.\ not in fit)}                            & $     96.5 ^{\,+      4.9}_{\,-     16.0}$ & $^{\,+      9.9}_{\,-     21.2}$ & $^{\,+     14.6}_{\,-     25.3}$
 \\[0.00cm]
 $\alpha$~~(deg) {\small (direct meas.\ only)}                           & $    100.2 ^{\,+     15.0}_{\,-      8.8}$ & $^{\,+     22.7}_{\,-     18.2}$  & $^{\,+     32.0}_{\,-     28.1}$
 \\[0.00cm]
 $\beta$~~(deg)                                                         & $    22.86 ^{\,+     1.00}_{\,-     1.00}$ & $^{\,+     2.03}_{\,-     1.97}$ & $^{\,+     3.22}_{\,-     2.93}$
 \\[0.00cm]
 $\beta$~~(deg) {\small (meas.\ not in fit)}                             & $     24.4 ^{\,+      2.6}_{\,-      1.5}$ & $^{\,+      5.1}_{\,-      3.0}$ & $^{\,+      6.9}_{\,-      4.5}$
 \\[0.00cm]
 $\beta$~~(deg) {\small (direct meas.\ only)}                            & $     21.7 ^{\,+      1.3}_{\,-      1.2}$ & $^{\,+      2.6}_{\,-      2.4}$& $^{\,+      4.1}_{\,-      3.6}$
 \\[0.00cm]
 $\gamma\simeq\delta$~~(deg)                                            & $     59.8 ^{\,+      4.9}_{\,-      4.1}$ & $^{\,+     13.9}_{\,-      7.9}$ & $^{\,+     20.8}_{\,-     12.1}$
 \\[0.00cm]
 $\gamma\simeq\delta$~~(deg) {\small (meas.\ not in fit)}                & $     59.8 ^{\,+      4.9}_{\,-      4.2}$ & $^{\,+     14.1}_{\,-      8.0}$ & $^{\,+     21.0}_{\,-     12.3}$
 \\[0.00cm]
 $\gamma\simeq\delta$~~(deg) {\small (direct meas.\ only)}               & $       62 ^{\,+       35}_{\,-       25}$ & $^{\,+       61}_{\,-       39}$& $^{\,+       100}_{\,-       54}$
 \\[0.00cm]
 $\beta_s$~~(deg)                                                       & $    1.045 ^{\,+    0.061}_{\,-    0.057}$ & $^{\,+    0.151}_{\,-    0.114}$ & $^{\,+    0.238}_{\,-    0.177}$
 \\[0.00cm]
 $\sin(2\beta_s)$                                                       & $   0.0365 ^{\,+   0.0021}_{\,-   0.0020}$ & $^{\,+   0.0053}_{\,-   0.0040}$ & $^{\,+   0.0083}_{\,-   0.0062}$
 \\[0.00cm]
 \hline
 $R_u$                                                                  & $    0.391 ^{\,+    0.015}_{\,-    0.015}$ & $^{\,+    0.031}_{\,-    0.029}$ & $^{\,+    0.049}_{\,-    0.044}$
 \\[0.00cm]
 $R_t$                                                                  & $    0.872 ^{\,+    0.033}_{\,-    0.028}$ & $^{\,+    0.095}_{\,-    0.054}$ & $^{\,+    0.143}_{\,-    0.082}$
 \\[0.00cm]
 \hline
 $\Delta m_d$~~$({\rm ps}^{-1})$ {\small (meas.\ not in fit)}            & $    0.394 ^{\,+    0.097}_{\,-    0.097}$ & $^{\,+    0.219}_{\,-    0.132}$ & $^{\,+    0.361}_{\,-    0.162}$
 \\[0.00cm]
 $\Delta m_s$~~$({\rm ps}^{-1})$                                        & $    17.34 ^{\,+     0.49}_{\,-     0.20}$ & $^{\,+     0.65}_{\,-     0.35}$ & $^{\,+     0.78}_{\,-     0.49}$
 \\[0.00cm]
 $\Delta m_s$~~$({\rm ps}^{-1})$ {\small (meas.\ not in fit)}            & $     21.7 ^{\,+      5.9}_{\,-      4.2}$ & $^{\,+      9.7}_{\,-      6.8}$ & $^{\,+     13.1}_{\,-      9.1}$
 \\[0.00cm]
 $\epsilon_K$~~$[10^{-3}]$ {\small (meas.\ not in fit)}                  & $     2.46 ^{\,+     0.63}_{\,-     0.88}$ & $^{\,+     1.05}_{\,-     1.05}$ & $^{\,+     1.50}_{\,-     1.20}$
 \\[0.00cm]
 \hline
 $f_{B_d}$~~(MeV) {\small (lattice value not in fit)}                   & $      183 ^{\,+       10}_{\,-       10}$ & $^{\,+       21}_{\,-       20}$ & $^{\,+       34}_{\,-       28}$
 \\[0.00cm]
 $\xi$ {\small (lattice value not in fit)}                              & $    1.061 ^{\,+    0.122}_{\,-    0.047}$ & $^{\,+    0.213}_{\,-    0.083}$ & $^{\,+    0.324}_{\,-    0.119}$
 \\[0.00cm]
 $B_K$ {\small (lattice value not in fit)}                              & $    0.722 ^{\,+    0.251}_{\,-    0.084}$ & $^{\,+    0.348}_{\,-    0.157}$ & $^{\,+    0.461}_{\,-    0.216}$
 \\[0.00cm]
 \hline
 $m_c$~~$({\rm GeV}/c^2)$ {\small (meas.\ not in fit)}                   & $     0.81 ^{\,+     0.93}_{\,-     0.36}$ & $^{\,+     1.08}_{\,-     0.36}$ & $^{\,+     1.23}_{\,-     0.81}$
 \\[0.00cm]
 $m_t$~~$({\rm GeV}/c^2)$ {\small (meas.\ not in fit)}                   & $      150 ^{\,+       27}_{\,-       21}$ & $^{\,+       57}_{\,-       35}$ & $^{\,+       79}_{\,-       46}$
 \\[0.00cm]
 \hline
 $\Kzpiznn$~~$[10^{-11}]$                                               & $     2.58 ^{\,+     0.48}_{\,-     0.40}$ & $^{\,+     1.01}_{\,-     0.68}$ & $^{\,+     1.53}_{\,-     0.93}$
 \\[0.00cm]
 $\Kppipnn$~~$[10^{-11}]$                                               & $      7.5 ^{\,+      1.8}_{\,-      2.0}$ & $^{\,+      2.5}_{\,-      2.4}$ & $^{\,+      3.2}_{\,-      2.7}$
 \\[0.00cm]
 \hline
 $\Bptaun$~~$[10^{-5}]$                                                 & $      9.7 ^{\,+      1.3}_{\,-      1.3}$ & $^{\,+      2.9}_{\,-      2.5}$ & $^{\,+      4.6}_{\,-      3.6}$
 \\[0.00cm]
 $\Bptaun$~~$[10^{-5}]$ {\small (direct meas.\ only)}                    & $     10.4 ^{\,+      3.0}_{\,-      2.7}$ & $^{\,+      3.0}_{\,-      5.1}$ & $^{\,+      3.0}_{\,-      7.1}$
 \\[0.00cm]
 $\Bpmun$~~$[10^{-7}]$                                                  & $     4.32 ^{\,+     0.58}_{\,-     0.57}$ & $^{\,+     1.27}_{\,-     1.12}$ & $^{\,+     2.05}_{\,-     1.62}$
 \\[0.00cm]
 \hline
 \end{tabular*}
 }
 \end{center}
 \vspace{-0.5cm}
 \caption[.]{\label{tab:fitResultsI}\em
       	Numerical results of the global CKM fit (I)~\cite{ckmfitterFPCP06}. 
	The errors correspond to one, two and three standard deviations, 
	respectively. 
	}
\end{table}
 \begin{table}[!t]
 \begin{center}
 \setlength{\tabcolsep}{0.0pc}
 {\small
 \begin{tabular*}{\textwidth}{@{\extracolsep{\fill}}lccc}\hline&&& \\[-0.3cm]
 Observable & central $\pm$ $\CL\equiv1\sigma$ & 
 $\pm$ $\CL\equiv2\sigma$ & $\pm$ $\CL\equiv3\sigma$  \\[0.15cm]
 \hline
 $|V_{ud}|$                                                             & $  0.97383 ^{\,+  0.00024}_{\,-  0.00023}$ & $^{\,+  0.00047}_{\,-  0.00047}$ & $^{\,+  0.00071}_{\,-  0.00071}$
 \\[0.00cm]
 $|V_{us}|$                                                             & $   0.2272 ^{\,+   0.0010}_{\,-   0.0010}$ & $^{\,+   0.0020}_{\,-   0.0020}$ & $^{\,+   0.0030}_{\,-   0.0030}$
 \\[0.00cm]
 $|V_{ub}|$~~$[10^{-3}]$                                                & $     3.82 ^{\,+     0.15}_{\,-     0.15}$ & $^{\,+     0.31}_{\,-     0.29}$ & $^{\,+     0.49}_{\,-     0.44}$
 \\[0.00cm]
 $|V_{ub}|$~~$[10^{-3}]$ {\small (meas.\ not in fit)}                    & $     3.64 ^{\,+     0.19}_{\,-     0.18}$ & $^{\,+     0.39}_{\,-     0.36}$ & $^{\,+     0.60}_{\,-     0.55}$
 \\[0.00cm]
 $|V_{cd}|$                                                             & $  0.22712 ^{\,+  0.00099}_{\,-  0.00103}$ & $^{\,+  0.00199}_{\,-  0.00205}$ & $^{\,+  0.00300}_{\,-  0.00307}$
 \\[0.00cm]
 $|V_{cs}|$                                                             & $  0.97297 ^{\,+  0.00024}_{\,-  0.00023}$ & $^{\,+  0.00048}_{\,-  0.00047}$ & $^{\,+  0.00071}_{\,-  0.00071}$
 \\[0.00cm]
 $|V_{cb}|$~~$[10^{-3}]$                                                & $    41.79 ^{\,+     0.63}_{\,-     0.63}$ & $^{\,+     1.26}_{\,-     1.27}$ & $^{\,+     1.89}_{\,-     1.90}$
 \\[0.00cm]
 $|V_{cb}|$~~$[10^{-3}]$ {\small (meas.\ not in fit)}                    & $     44.9 ^{\,+      1.2}_{\,-      2.8}$ & $^{\,+      2.4}_{\,-      5.7}$ & $^{\,+      3.8}_{\,-      7.7}$
 \\[0.00cm]
 $|V_{td}|$~~$[10^{-3}]$                                                & $     8.28 ^{\,+     0.33}_{\,-     0.29}$ & $^{\,+     0.92}_{\,-     0.57}$ & $^{\,+     1.38}_{\,-     0.86}$
 \\[0.00cm]
 $|V_{ts}|$~~$[10^{-3}]$                                                & $    41.13 ^{\,+     0.63}_{\,-     0.62}$ & $^{\,+     1.25}_{\,-     1.24}$ & $^{\,+     1.87}_{\,-     1.86}$
 \\[0.00cm]
 $|V_{tb}|$                                                             & $ 0.999119 ^{\,+ 0.000026}_{\,- 0.000027}$ & $^{\,+ 0.000052}_{\,- 0.000054}$ & $^{\,+ 0.000078}_{\,- 0.000082}$
 \\[0.00cm]
 \hline
 $|V_{td}/V_{ts}|$                                                      & $   0.2011 ^{\,+   0.0081}_{\,-   0.0065}$ & $^{\,+   0.0230}_{\,-   0.0127}$ & $^{\,+   0.0345}_{\,-   0.0195}$
 \\[0.00cm]
 $|V_{ud}V_{ub}^*|$~~$[10^{-3}]$                                        & $     3.72 ^{\,+     0.15}_{\,-     0.14}$ & $^{\,+     0.30}_{\,-     0.29}$ & $^{\,+     0.48}_{\,-     0.43}$
 \\[0.00cm]
 $\arg\left[V_{ud}V_{ub}^*\right]$~~(deg)                               & $     59.8 ^{\,+      4.9}_{\,-      4.0}$ & $^{\,+     13.9}_{\,-      7.8}$ & $^{\,+     20.9}_{\,-     12.1}$
 \\[0.00cm]
 $\arg\left[-V_{ts}V_{tb}^*\right]$~~(deg)                              & $    1.043 ^{\,+    0.061}_{\,-    0.057}$ & $^{\,+    0.151}_{\,-    0.114}$ & $^{\,+    0.238}_{\,-    0.176}$
 \\[0.00cm]
 $|V_{cd}V_{cb}^*|$~~$[10^{-3}]$                                        & $     9.49 ^{\,+     0.15}_{\,-     0.15}$ & $^{\,+     0.30}_{\,-     0.30}$ & $^{\,+     0.45}_{\,-     0.45}$
 \\[0.00cm]
 $\arg\left[-V_{cd}V_{cb}^*\right]$~~(deg)                              & $   0.0339 ^{\,+   0.0021}_{\,-   0.0020}$ & $^{\,+   0.0050}_{\,-   0.0040}$ & $^{\,+   0.0077}_{\,-   0.0060}$
 \\[0.00cm]
 $|V_{td}V_{tb}^*|$~~$[10^{-3}]$                                        & $     8.27 ^{\,+     0.33}_{\,-     0.29}$ & $^{\,+     0.93}_{\,-     0.57}$ & $^{\,+     1.38}_{\,-     0.85}$
 \\[0.00cm]
 $\arg\left[V_{td}V_{tb}^*\right]$~~(deg)                               & $   -22.84 ^{\,+     1.00}_{\,-     0.99}$ & $^{\,+     1.98}_{\,-     2.02}$ & $^{\,+     2.93}_{\,-     3.21}$
 \\[0.00cm]
 \hline
 $\sin\theta_{12}$                                                      & $   0.2272 ^{\,+   0.0010}_{\,-   0.0010}$ & $^{\,+   0.0020}_{\,-   0.0020}$ & $^{\,+   0.0030}_{\,-   0.0030}$
 \\[0.00cm]
 $\sin\theta_{13}$~~$[10^{-3}]$                                         & $     3.82 ^{\,+     0.15}_{\,-     0.15}$ & $^{\,+     0.31}_{\,-     0.30}$ & $^{\,+     0.49}_{\,-     0.44}$
 \\[0.00cm]
 $\sin\theta_{23}$~~$[10^{-3}]$                                         & $    41.78 ^{\,+     0.63}_{\,-     0.63}$ & $^{\,+     1.26}_{\,-     1.26}$ & $^{\,+     1.90}_{\,-     1.89}$
 \\[0.00cm]
 \hline
 \end{tabular*}
 }
 \end{center}
 \vspace{-0.5cm}
 \caption[.]{\label{tab:fitResultsII}\em
	Numerical results of the global CKM fit (II)~\cite{ckmfitterFPCP06}. 
	The errors correspond to one, two and three standard deviations,
	respectively.
	}
\end{table}

The global CKM fit results in the Wolfenstein parameters are given in 
Table~\ref{tab:fitResultsI} (top four lines).
The goodness-of-fit, evaluated with the use of Monte Carlo methods~\cite{ckmfitter2004}, 
is satisfying, so one can assume the 
correctness of the SM and proceed with the metrology of the CKM 
parameters and the related quantities. 
It is interesting to compare the direct measurements of the UT angles with
the predictions from the CKM fit that do not include these measurements. The 
measurement of $\stwob$ outperforms by far the indirect prediction. Also,
the precision of the $\alpha$ measurement exceeds that of the CKM prediction 
so that its inclusion in the fit improves the knowledge of the apex of the UT.
For $\gamma$, the experimental precision is not yet sufficient to 
improve the CKM metrology. It is remarkable that the 
prediction of $\stwobs$ has an absolute precision of 0.002 (owing to 
its CKM suppression), so a measurement of it at the LHC will 
represent a very sensitive probe for physics beyond the SM. Also given are 
the sides of the UT, and the angles of the standard parameterization of the 
CKM matrix.

Once the Wolfenstein parameters are fit, determining fully consistent 
confidence levels for all related observables is straightforward.
It is also possible 
to derive constraints on the matrix elements of neutral $K$ and $B$ 
mixing, which in turn are in agreement with the 
LQCD results. The determination of $\xi$ through the $\dms$ measurement 
in the CKM fit is not yet competitive with the LQCD determination,
owing to the currently insufficient precision of the $\gamma$ 
and $\alpha$ measurements.
It represents a remarkable success of flavor physics to 
correctly reproduce the scales of the charm and top masses through
FCNC loop processes only. Also listed in Table~\ref{tab:fitResultsI} 
are the predictions for the rare kaon decays to pions and 
neutrino-antineutrino pairs (including the next-to-next-to-leading-order
calculation of the 
charm contribution to the charged decay~\cite{Buras:2005gr}) and for 
rare leptonic $\B$ decays. (The comment ``meas.\ (or lattice value) 
not in fit'' in some rows indicates that for the corresponding 
determination the direct measurement of the quantity (or the LQCD prediction) 
has been excluded from the fit.)

\begin{figure}[!t]
  \centerline{\includegraphics[width=9cm]{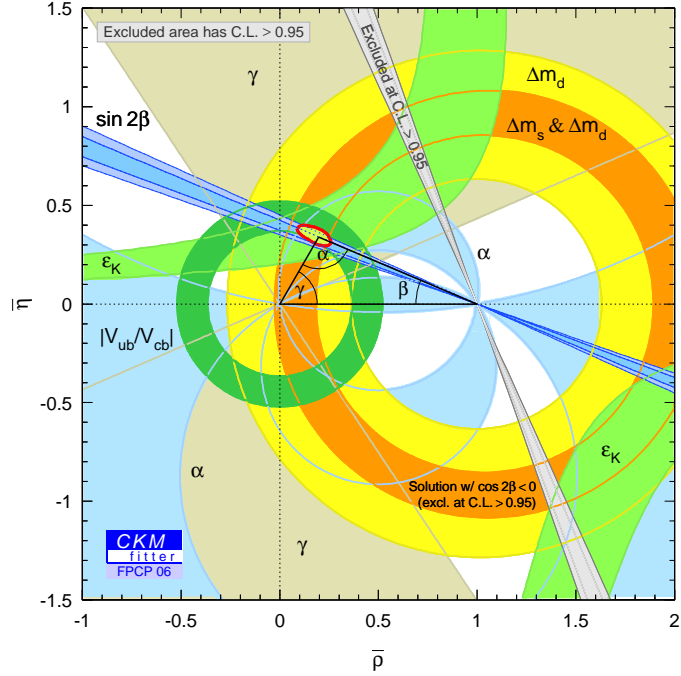}}
  \caption[.]{\label{fig:global_ckmfit}\em
        Confidence levels in the $\rhoeta$ plane for the 
        global CKM fit. The shaded areas indicate $95\%$~\CL allowed 
	regions~\cite{ckmfitterFPCP06}. 
	}
\end{figure}

Finally, Table~\ref{tab:fitResultsII} gives the standard CKM fit results for the
absolute values of the CKM elements and the magnitudes and phases of the sides
of the non-rescaled UT.  The results of the fit are shown in the $\rhoeta$ plane
in Fig.~\ref{fig:global_ckmfit}. The outer contour of the combined fit 
corresponds to $95\%$~\CL exclusion. Also shown are the $95\%$~\CL regions  for
the individual constraints entering the fit (the constraint from
$\Bp\to\taup\nut$ is not shown, although it is included in the fit). 

\begin{figure}[!t]
  \centerline{\includegraphics[width=.5\textwidth]{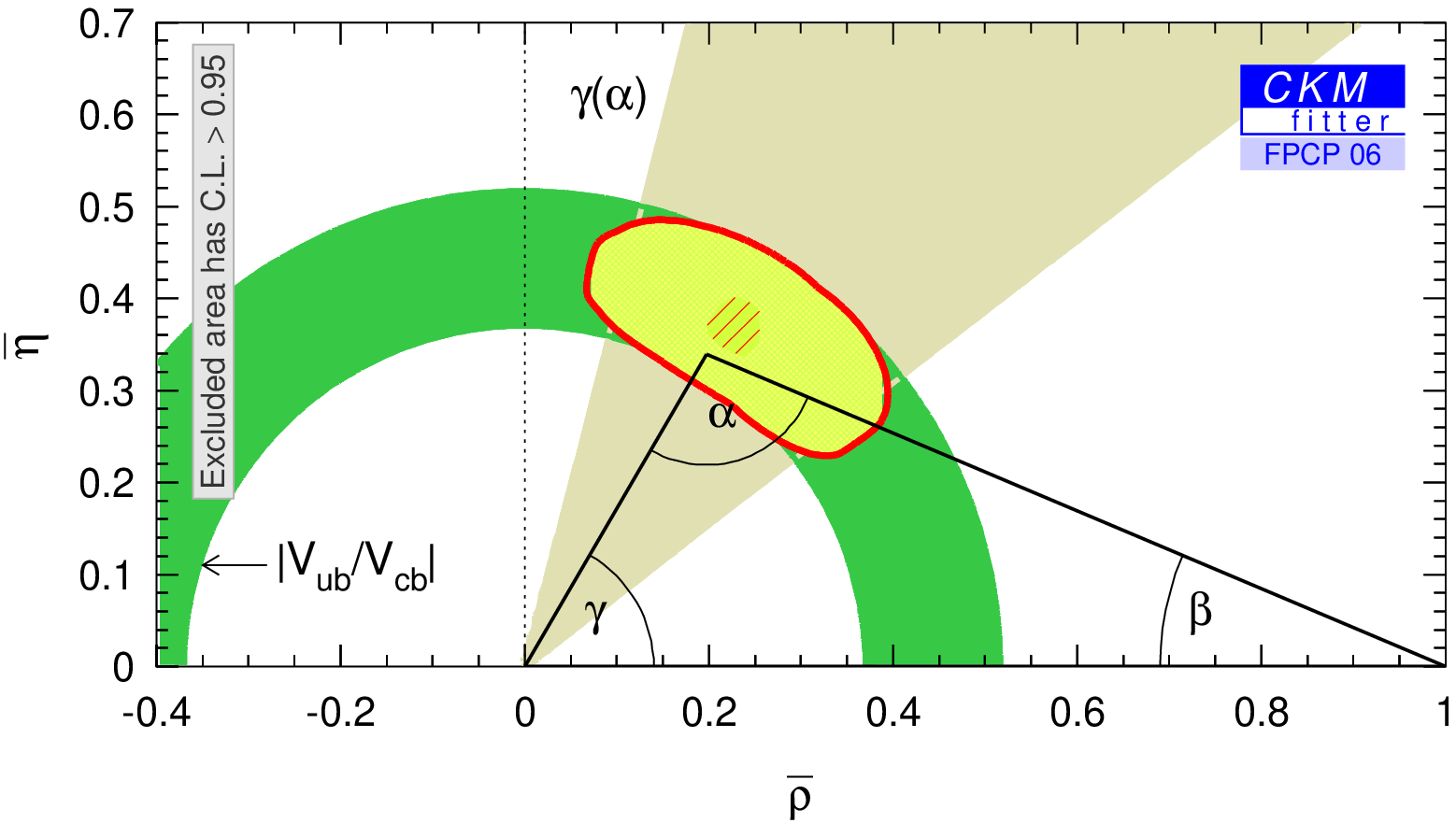}
	\includegraphics[width=.5\textwidth]{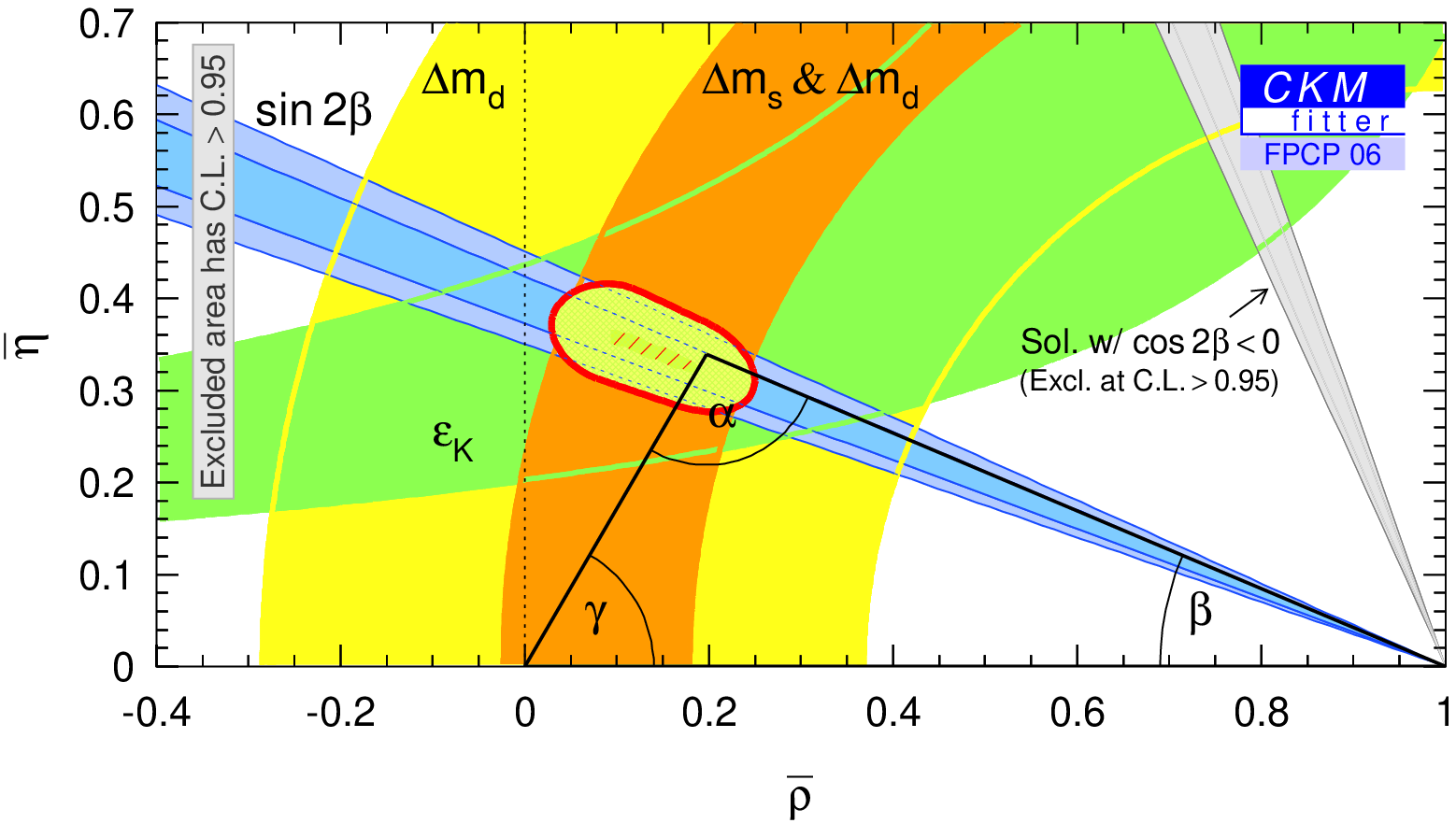}}  
  \caption[.]{\label{fig:global_ckmfit_treeloop}\em
        Confidence levels in the $\rhoeta$ plane for global
	CKM fits using only tree-level (left) and loop-induced (right)
	inputs. The shaded areas indicate $95\%$~\CL allowed regions. 
	}
\end{figure}

The global CKM fit performed so far contains all relevant information  collected
by the experiments. From the new physics perspective, it is  interesting to
confront the measurements according to their sensitivity to new physics
contributions. Figure~\ref{fig:global_ckmfit_treeloop}  shows on the left
plot the constraints that originate from mainly tree-level processes, together
with their combined fit. The right plot shows  the constraints from
loop-induced processes. To fix the length scale of the UT and the constraints
on $\lambda$ and $A$ from the tree-level determinations of the CKM elements
$\Vud$, $\Vus$ and $\Vcb$ are used.  If $\gamma$ is extracted from the
measurement of $\alpha$ using $\beta$  from mixing-induced \CP violation as
input, it is effectively a tree-level  quantity, because the isospin analysis
isolates the $\Delta I=3/2$ component in the decay amplitude, which is assumed
to be standard (this corresponds to the SM4FC case defined
in Ref.~\cite{ckmfitter2004}).  Consequently, the constraint for $\gamma$ that enters
the tree-level plot  is the average of the direct measurement of $\gamma$ via
open-charm processes,  and the value obtained from $\pi-\alpha-\beta$, from
which new physics in mixing cancels. This provides the first determination of
\rhoeta\ from (effectively) tree-level processes.  Good agreement is observed
between the tree-level and loop-induced constrained fits.

\section{The CKM Matrix and the Search for New Physics}
\label{new_physics}

\subsection{New Physics Extensions of the Standard Model and their Relation to CKM}

If the constraints of the SM are lifted, $K$, $B$, and $D$ decays are described
by many more parameters than just the four CKM parameters and the $W$, $Z$ and
quark masses.  The most general effective Lagrangian at lowest order contains
approximately one hundred flavor-changing operators, and the observable effects of any
short-distance, high-energy physics are encoded in their Wilson coefficients. 
Therefore, all measurements sensitive to different short-distance physics are
important,  and overconstraining the magnitudes and phases of CKM elements could
detect deviations from the SM.  For example, $\Delta m_d$, $\Gamma(B\to
\rho\gamma)$, and $\Gamma(B\to X_d\ell^+\ell^-)$ are all proportional to
$|V_{td}V_{tb}^*|^2$ in the SM, however, they may receive unrelated
contributions from new physics.  Similar to the measurements of $\sin2\beta$
from tree- and loop-dominated modes, testing such correlations, and not simply
the unitarity of the CKM matrix, provides the best sensitivity to new physics.

In the kaon sector both \CP-violating observables, $\epsK$ and 
$\epsPr$, are tiny, so models in which all sources of \CP violation 
are small are viable.  Owing to the measurement of $\stwob$ we know that \CP
violation is an ${\cal O}(1)$ effect and only the flavor mixing is
suppressed between the three quark generations.  Hence, many models with
spontaneous \CP violation that predicted small \CP-violating 
effects are excluded.  However, model-independent statements
for the constraints imposed by the CKM measurements on new physics are hard to make
because most models that allow new flavor physics contain a large number of
new parameters.

There are large classes of models in which the biggest effects of new physics
occur in transitions involving second- and third-generation fermions, thus
escaping the bounds from kaon physics (see, \eg, Refs.~\cite{Cohen:1996vb} 
and \cite{Agashe:2005hk}).  For example, in SUSY GUTs, the
observed near-maximal $\nu_\mu-\nu_\tau$ mixing may imply large mixing between
$s_R$ and $b_R$, and between their supersymmetric partners 
$\tilde s_R$ and $\tilde b_R$~\cite{Chang:2002mq}. 
Although the mixing of the right-handed quarks does not enter the CKM matrix, the
mixing among the right-handed squarks is physical.  It may entail sizable
deviations of the time-dependent \CP asymmetries in penguin-dominated modes from
$\stwob$ discussed in the next section, and an enhancement, compared with the 
SM value, of $\dms$.  Another recently popular scenario with large effects in flavor
physics involves Higgs-mediated FCNCs in the large $\tan\beta$ region of SUSY
($\tan\beta$ is the ratio of the vacuum expectation values of the two Higgs
doublets, and $\tan\beta \sim m_t/m_b$ is motivated by the unification of top
and bottom Yukawa couplings predicted by some GUT models).  Neutral Higgs-mediated 
FCNC couplings that occur at one loop increase rapidly with $\tan\beta$
and can exceed other SUSY contributions by more than an order of magnitude,
giving an enhancement of ${\cal B}(B_{d,s}\to \mu^+\mu^-)$ proportional to
$\tan^6\beta$~\cite{Babu:1999hn}.  This scenario also predicts a suppression of
$\Delta m_s$ proportional to $\tan^4\beta$~\cite{Buras:2001mb}. As discussed
below (see Fig.~\ref{fig:hdsd}), $\Bz\Bzb$ mixing may still receive a
significant positive or negative new physics contribution even after
the measurement of $\Delta m_s$.

The existing data on rare \B decays set severe constraints on (SUSY) model
building.  In particular ${\cal B}(B\to X_s\gamma)$ provides  strong parameter
bounds.  Both the SUSY and the SM contributions enter at the one-loop level, and the
charged Higgs diagrams always enhance the rate, whereas the chargino contributions
can reduce it if $\mu>0$ ($\mu$ is the coefficient of the $H_uH_d$ term in the
superpotential, and $H_{u,d}$ are the Higgs fields that couple to the up- and 
down-type fermions).  One reason that $B\to X_s\gamma$ is the only $B$ decay that
enters the bounds in many analyses (\eg, the constraints on dark matter
candidates~\cite{Ellis:2005mb}, usually the lightest neutralino) is because such
studies often assume the constrained MSSM (MSUGRA), which implies minimal flavor
violation~\cite{Ciuchini:1998xy}, and relate the charged Higgs and the
chargino contributions to each other.  
The $B\to X_s\gamma$ rate can be enhanced significantly
even without new sources of flavor violation, \ie, with all flavor changing
interactions proportional to CKM matrix elements.  Nontrivial flavor physics due
to \tev-scale new physics would indicate deviations from this parameterization,
which could weaken the bounds from $B\to X_s\gamma$.  Then the constraints can
become more complex, and other $B$ decays would also be important in
constraining the new physics parameters.

As mentioned above, not all \CP-violating measurements can be interpreted as
constraints on the $\rhoeta$ plane.  For example, probing with $\Bs\to J/\psi \phi$
whether $\beta_s$, the order $\lambda^2$ angle of the squashed UT defined in
Eq.~(\ref{betasdef}), agrees with its SM prediction is an important 
independent test of the theory.  Other comparisons
between the SM correlations in $B$ and $K$ physics can come from future
measurements of $\KL \to \pi^0 \nu\nub$ and $K^+ \to \pi^+ \nu\nub$. 
These loop-induced decays are sensitive to new physics and will allow a
determination of $\beta$ independent of its value measured in $B$
decays~\cite{Buchalla:1994tr}.

\subsection{\CP Asymmetries in Loop-Dominated Modes}
\label{ut_loop}

The FCNC $b \to s$ transition is mediated
by penguin diagrams.  It can have any up-type quark in the loop, so 
its amplitude can be written as
\beqa
  \label{eq:ut_loop}
    A_{b \to s} & = & 
    m_t V_{tb}V^*_{ts} + m_c V_{cb}V^*_{cs} + m_u V_{ub}V^*_{us} \nn\\
    & = & (m_c - m_t) V_{cb}V^*_{cs} + (m_u - m_t) V_{ub}V^*_{us}
    = {\cal O}(\lambda^2) + {\cal O}(\lambda^4) \,,
\eeqa
where the unitarity of the CKM matrix has been used in the second step. In the
SM, the amplitude is dominated by the first, $V_{cb}V^*_{cs}$, term, which has
the same weak phase as the amplitude in $\Bz\to J/\psi \Kz$ decay. We expect
$||\Abar/A| - 1| = {\cal O}(\lambda^2)$, and the time-dependent \CP\ asymmetry
parameters are given to a similar accuracy by $S_{b \to s q\qbar } \approx -
\etacpf \stwob$ and $C_{b \to s q\qbar} \approx 0$. 

Owing to the large mass scale of the virtual particles that can occur in the
loops, additional diagrams from physics beyond the SM, with heavy particles in
the loops, may contribute. The measurement of \CP violation in these channels
and the  comparison with the $B$-to-charmonium reference value is therefore a sensitive
probe for physics beyond the SM. A discrepancy between $S_{b \to
s q\qbar}$ and \stwob can provide an indication of new physics.  If
the SM and new physics contributions are both significant, the deviations of the \CP
asymmetries from \stwob become mode dependent, because they depend on the relative
size and phase of the contributing amplitudes, which are determined by the quantum
numbers of the new physics and by strong interactions.

The important question is how well can we  bound the contribution of the second,
CKM-suppressed, term to the $b \to s q\qbar$ transition in Eq.~(\ref{eq:ut_loop})?
This term has a different weak phase than the dominant first term, so its impact
on $S_{b \to s q\qbar}$ depends on both its magnitude and relative strong phase.
Naive factorization suggests that for $q = s$ the $\lambda^2$ suppression of the
second term is likely to hold because it would require
an enhancement of rescattering effects to upset this.  However, for $q = u$,
there is a color-suppressed $b \to u$ tree diagram, which has a different weak
(and possibly strong) phase than the leading $\lambda^2$ penguin amplitude.  For
$q = d$, any light neutral meson formed  from $d \dbar$ also has a $u
\ubar$ component, and there is ``tree pollution'' again. The \Bz decays to
$\piz\KS$ and $\omega\KS$ belong to  this category. The mesons $\etapr$ and
$f_0(980)$ have significant $s\sbar$ components, which may reduce the tree
pollution.  Neglecting rescattering, the three-body final state $\Kz\Kzb\Kz$
(reconstructed as $\KS\KS\KS$) has no tree pollution, whereas $\Bz\to\Kp\Km\Kz$
(excluding $\phi\Kz$) does. 

Recently QCD factorization~\cite{Buchalla:2005us,Beneke:2005pu} and
SCET~\cite{Williamson:2006hb} was used to calculate the deviations from $\stwob$
in some of the two-body penguin modes. It was found that the deviations are the
smallest (\,$\lsim 0.02$) for $\phi \Kz$ and $\eta' \Kz$.  This is fortunate
because these are also the modes in which the experimental errors are the
smallest.  The SM shifts enhance $-\eta_f S_f$ (except for $\rho\KS$)
using~\cite{Buchalla:2005us,Beneke:2005pu}, while suppress $S_{\eta'K_S}$
using~\cite{Williamson:2006hb}. $\SU(3)$ flavor symmetry has also been used to
bound the SM-induced deviations from
$\stwob$~\cite{Grossman:2003qp,Gronau:2003kx}.  Owing to the lack of information
on strong phases and the weak experimental bounds on some $b\to d q\qbar$
mediated rates, the resulting bounds tend to be weak. An exception is
$\pi^0\KS$, where $\SU(3)$ relates the relevant amplitudes to $\pi^0\pi^0$ and
$K^+K^-$~\cite{Gronau:2003kx}.  The theoretical understanding of factorization
in three-body decays does not yet allow accurate bounds on $-\eta_f S_f - \stwob$
to be computed.

\begin{figure}[t!]
  \centerline{\includegraphics[width=9cm]{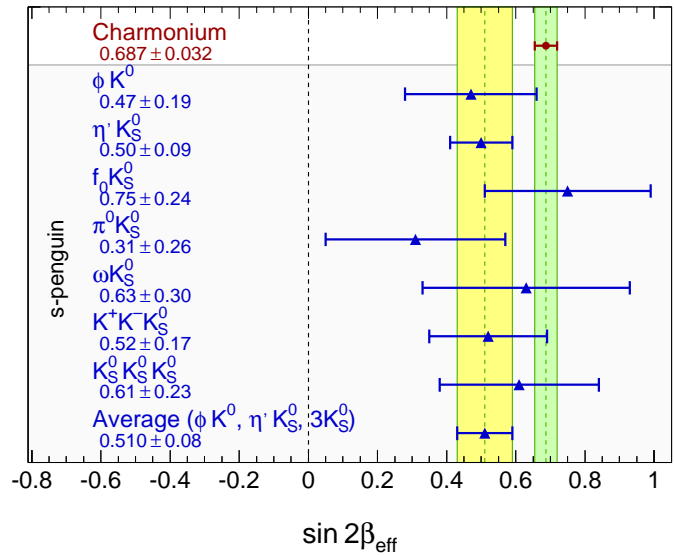}}
  \vspace{-0.3cm}
  \caption[.]{\label{fig:spenguin}\em
	Comparison of world average $\stwob_{\rm (eff)}$ results from
	penguin-dominated decays, and the charmonium reference
	value~\cite{sin2bpenguin,babar_stb,belle_stb,hfag2005}. The light shaded
	band shows the average of the results from  $\Bz\to\phi\Kz,\,
	\etapr\Kz,\, \KS\KS\KS$, considered to be theoretically cleaner than the
	other modes. The result for $\Bz\to\piz\piz\KS$ is not shown, because it
	has a much larger error.}
\end{figure}

There has been considerable excitement about these measurements in the past few
years, which has been somewhat damped by recent updates from the experiments. 
The results for the measurement of $\stwob_{\rm eff}$ (which is equal to 
$-\eta_f S_f$ for the final states with unique \CP content) from the
various penguin modes are compiled in Fig.~\ref{fig:spenguin}. If one restricts
the modes to those with the potentially smallest theoretical uncertainties, \ie,
the final states $\phi\Kz$, $\eta^\prime\Kz$, and $\Kz\Kzb\Kz$, and attempts to
average the effective $\stwob$ results, $\langle\stwob_{\rm
eff}\rangle=0.50 \pm0.08$, which is within $2.2\sigma$ reach of the charmonium
reference value. Beginning of 2005, this $s$-penguin average was $0.40\pm0.09$,
and because the charmonium result was larger at that time, the discrepancy between
the $\stwob$ numbers was at the $3.2\sigma$ level, which explains the popularity
of the results. Better statistics are required to clarify the
situation.\footnote
{
	It is an interesting outcome of the QCD factorization calculations that 
	the SM deviations from $\stwob$ should be positive in all modes, except 
	for $\rho^0\Kz$~\cite{Buchalla:2005us, Beneke:2005pu}. Taking the
	expected deviations into account would enhance the significance of the
	current discrepancy between the  charmonium-based and the penguin-based
	results.
}

Another interesting measurement in the penguin sector is the time-dependent \CP
asymmetry in $B\to \KS\pi^0(K^{*0})\gamma$. \babar and Belle have measured 
it exclusively and inclusively, with the averages
$S_{K^*\gamma}=-0.13\pm0.32$ and $S_{K_S\piz\gamma}=0.00\pm0.28$~\cite{kstrgam,hfag2005}.
Although the $B\to X_s\gamma$ rate
is correctly predicted by the SM at the 10\% level, that measurement sums over
the rates to left- and right-handed photons, and their ratio is also sensitive
to new physics.   In the SM, $b$-quarks mainly decay to $s\gamma_L$ and 
$\bbar$-quarks to $\sbar \gamma_R$, so their interference is suppressed, 
proportional to $r_{f_s} = A(\Bzb\to X_{f_s}\gamma_R) / A(\Bzb\to
X_{f_s}\gamma_L)$.   If only the
electromagnetic penguin operator, $O_7 \sim \sbar \, \sigma^{\mu\nu} F_{\mu\nu}
(m_b P_R + m_s P_L) b$ contributed to the rate, it would give $S_{K^*\gamma} =
-2(m_s/m_b)\stwob$~\cite{Atwood:1997zr}. This also holds in the nonresonant 
$B\to \KS\pi^0\gamma$ case~\cite{Atwood:2004jj}. Grinstein 
\ea~\cite{Grinstein:2004uu} recently realized that
four-quark operators contribute to $r$ that are not suppressed by
$m_s/m_b$.  The numerically dominant term is due
to the matrix element of $O_2 = (\cbar \, \gamma^\mu P_L b) (\sbar \, \gamma_\mu
P_L c)$, and its contribution to the inclusive rate can be calculated reliably,
$\Gamma(\Bzb\to X_s\gamma_R) / \Gamma(\Bzb\to X_s\gamma_L) \approx
0.01$~\cite{Grinstein:2004uu}.  This suggests that for most final states, on
average, $r \sim 0.1$ should be expected. A SCET analysis of the exclusive decay
proved the power suppression of the amplitude ratio, $A(\Bzb\to
\Kzb{}^*\gamma_R) / A(\Bzb\to \Kzb{}^*\gamma_L) = {\cal O}[(C_2/ 3C_7)\, (\lqcd/
m_b)] \sim 0.1$~\cite{Grinstein:2004uu}, but the uncertainties are sizable.

\subsection{The $D$ System}

The $D$ meson system is complementary to $K$ and $B$ mesons, because it is the
only neutral-meson system in which mixing and rare decays are generated by
down-type quarks in the SM (or up-type squarks in SUSY).   Therefore,  flavor
and \CP violation are suppressed by both the GIM mechanism and by the Cabibbo
angle.  As a result, \CP violation in $D$ decays, rare $D$ decays, and $\Dz\Dzb$
mixing are predicted to be small in the SM and have not yet been observed. There
are well-motivated new physics scenarios, such as those based on quark-squark
alignment~\cite{Nir:1993mx}, that predict $\Dz\Dzb$ mixing to be of order
$\lambda^2$, so that if the current bounds improve by a factor of a few, these
models become fine tuned.

The most interesting experimental hint for a possible $\Dz\Dzb$ mixing signal
so far is the lifetime difference between the \CP-even and \CP-odd
states~\cite{Yabsley:2003rn}
\beq
y_{\CP} = {\Gamma(\CP \mbox{-even}) - \Gamma(\CP \mbox{-odd}) \over
  \Gamma(\CP \mbox{-even}) + \Gamma(\CP \mbox{-odd})} = (0.9 \pm 0.4)\,\%\,.
\eeq
Unfortunately, this central value alone cannot not be interpreted as a sign of
new physics, because of possible long-distance hadronic
effects~\cite{Falk:2001hx}.  At the present level of sensitivity, \CP violation
in mixing or decays, or enhanced rare decays, such as $D\to \pi\ell^+\ell^-$
would be the only clean signal of new physics in the $D$ sector. Because the 
\B-factories are also charm factories (however, without coherent production 
of $\Dz\Dzb$ pairs), it is possible to observe such unambiguous signals of 
new physics every time the analyzed data sample increases, independent of 
the tight constraints in the $K$ and $B$ sectors.

\subsection{The $B_s$ System}

Although all present results show agreement with the SM,
it is still possible that the new physics flavor structure is quite different 
from that in the SM, with ${\cal O}(1)$ effects in sectors of flavor 
physics that have not yet been probed with good precision. As mentioned above,
many well-motivated models can lead to large new physics effects between 
the second and the third generations, while leaving the flavor-changing 
transitions between the other families unaffected. The $\Bs$ sector, which is 
currently being investigated at the Tevatron (\cf\  Sec.~\ref{b_oscillation})
and will soon be studied by the LHC experiments, is suited to test such 
scenarios.\footnote
{
	During the Summer of 2005, the Belle experiment performed a three-day 
	engineering run comprising a five-point energy scan between 
	10.825\gev and 10.905\gev, to locate the \FiveS peak, and 
	data taking at the \FiveS (10.869\gev). A data sample corresponding
	to an integrated luminosity of 1.9\invfb was collected during
	that period~\cite{belleups5s}. \FiveS resonant events decay 
	into pairs of neutral and charged $B^{(*)}$ mesons as well as neutral
	$B_s^{(*)}$ mesons with opposite flavors. Belle found a fraction
	of events with $B_s^{(*)}$ pair production over the full $b\bbar$
	production (including the continuum) at the \FiveS of
	$(16.4\pm1.4\pm4.1)\%$. Among these, $\FiveS\to\B_s^*\Bbar_s^*$
	events dominate. All inclusive and exclusive measurements
	performed are found in agreement with earlier results on the 
	\FiveS, with less statistics, obtained at the CLEO 
	experiment~\cite{cleoups5s}.
}

Compared with the \Bd mesons, the properties of the \Bs are still less well known,
although the measurement of $\dms$~\cite{d0bsbsbar,cdfbsbsbar} strongly
indicates that $\BszBszb$ mixing is SM-like.  Although the uncertainty of $\dms$ is
already at the few percent level, we have no significant information yet on the 
$\Bs\Bsb$ mixing phase, $\beta_s$, and the width difference, $\Delta\Gamma_s$, is
poorly known. Many extensions of the SM predicted large modifications of $\dms$
owing to contributions from new particles to \Bs oscillations.  Such models, or
such parameter regions of models are now ruled out. The LHC
experiments will determine $\stwobs$ through the time-dependent measurement of
mixing-induced  \CP violation in the decay $\Bs\to J/\psi\phi$ (and similar
final states).\footnote
{ 
	As all pseudoscalar decays to a pair of vector mesons,
	such as $\Bz\to J/\psi K^{*0}$ or $\Bz\to \rho^+\rho^-$, the 
	final state in $\Bs\to J/\psi\phi$ is an admixture of \CP-even 
	and \CP-odd components.  To avoid  a dilution of the \CP-asymmetry 
	measurement, they must be  separated by an angular analysis, which 
	complicates the measurement. 
} 
The expected precision of
better than $0.02$ for the combined LHC experiments will barely be
enough for a significant measurement if $S_{\Bs\to J/\psi\phi}$ is as predicted
by the SM. However, for the purpose of the search for new physics, 
which could still occur at the ${\cal O}(1)$
level, only the absolute error on $\stwobs$ matters.  Because this phase is
$\lambda^2$ suppressed, it can be predicted accurately (in terms of its absolute
error) by the global CKM fit, giving $\stwobs = 0.0365 \pm
0.0021$~\cite{Laplace:2002ik,ckmfitterFPCP06}.  Hence this represents the first
precision measurement, sensitive to a new physics phase, that can be confronted
with a {\em precision prediction}. It could become the most sensitive test of
the KM theory for \CP violation.

The class of rare decays that proceed through penguin diagrams, such as  $\B\to
K^*\gamma$ and $\B\to K^*\ell^+\ell^-$, will be augmented by the $\Bs$  system
through (among others) the measurement of the decay $\Bs\to\mup\mun$  at the
LHC. The best current limit on the branching fraction is $8\times10^{-8}$ at
90\%~\CL~\cite{bsmumutevatron}, which is well above the  SM model expectation of
$3.4\times10^{-9}$ (whose dominant theoretical  uncertainty is due to 
$f_{B_s}^2$)~\cite{burasbsmumu}. Expectations for the  LHC experiments are hard to
estimate owing to the difficulty in fully simulating sufficient background
statistics, but all experiments expect to see a SM signal within one
year. The decay $\Bd\to\mup\mun$ will be a further  challenge due to its even
lower branching fraction in the SM. It should be observed by all experiments
after several years of data taking. A precision measurement of
$\BR(\Bs \to \mup\mun)/$ $\BR(\Bd \to \mup\mun)$, which has a clean relation to
$\dms/\dmd$, will be beyond the scope of the LHC.

\subsection{Model-Independent Constraints on New Physics in $B$ Oscillation}

\begin{figure}[t!]
\centerline{	\includegraphics[width=8cm]{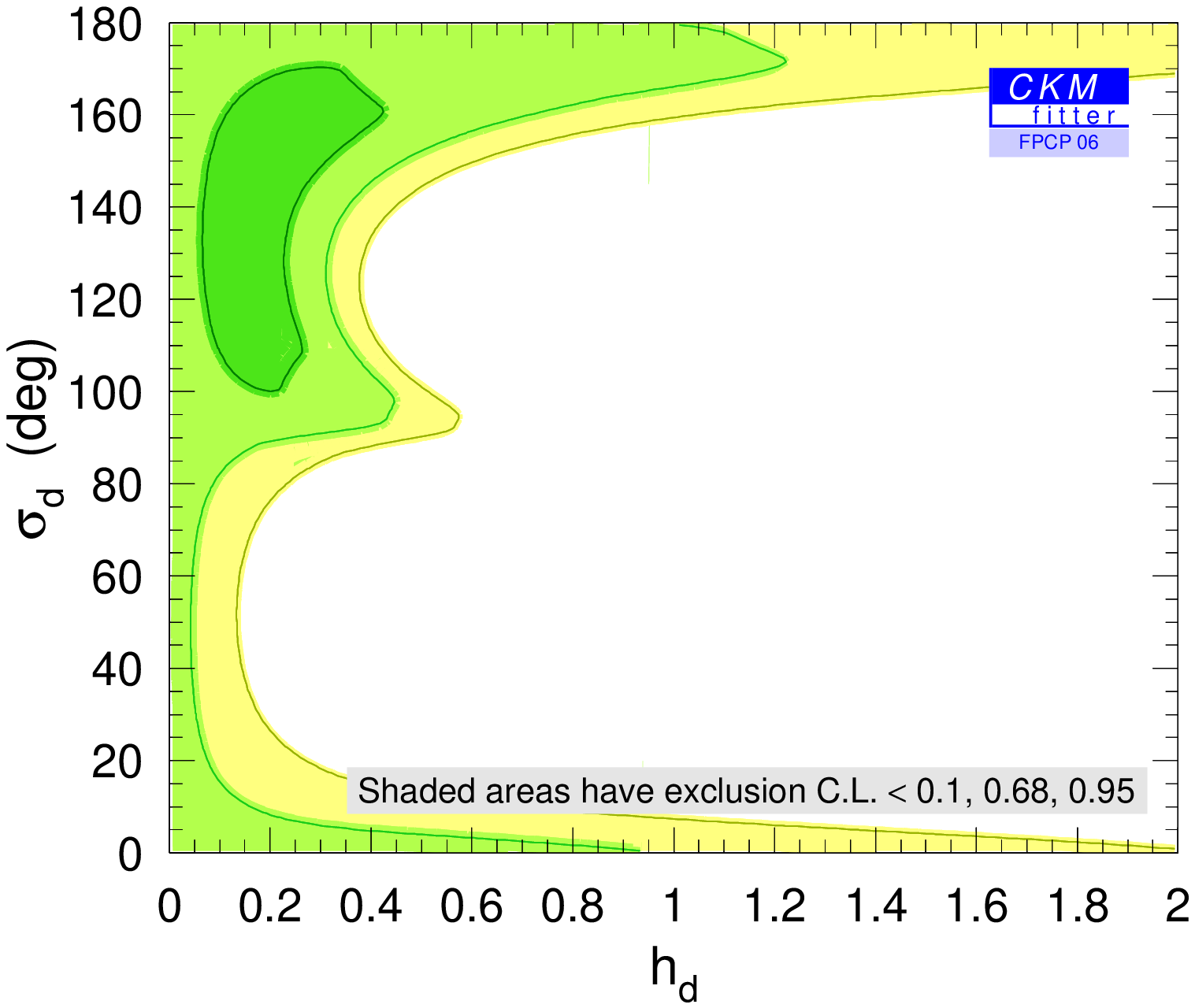}
            	\includegraphics[width=8cm]{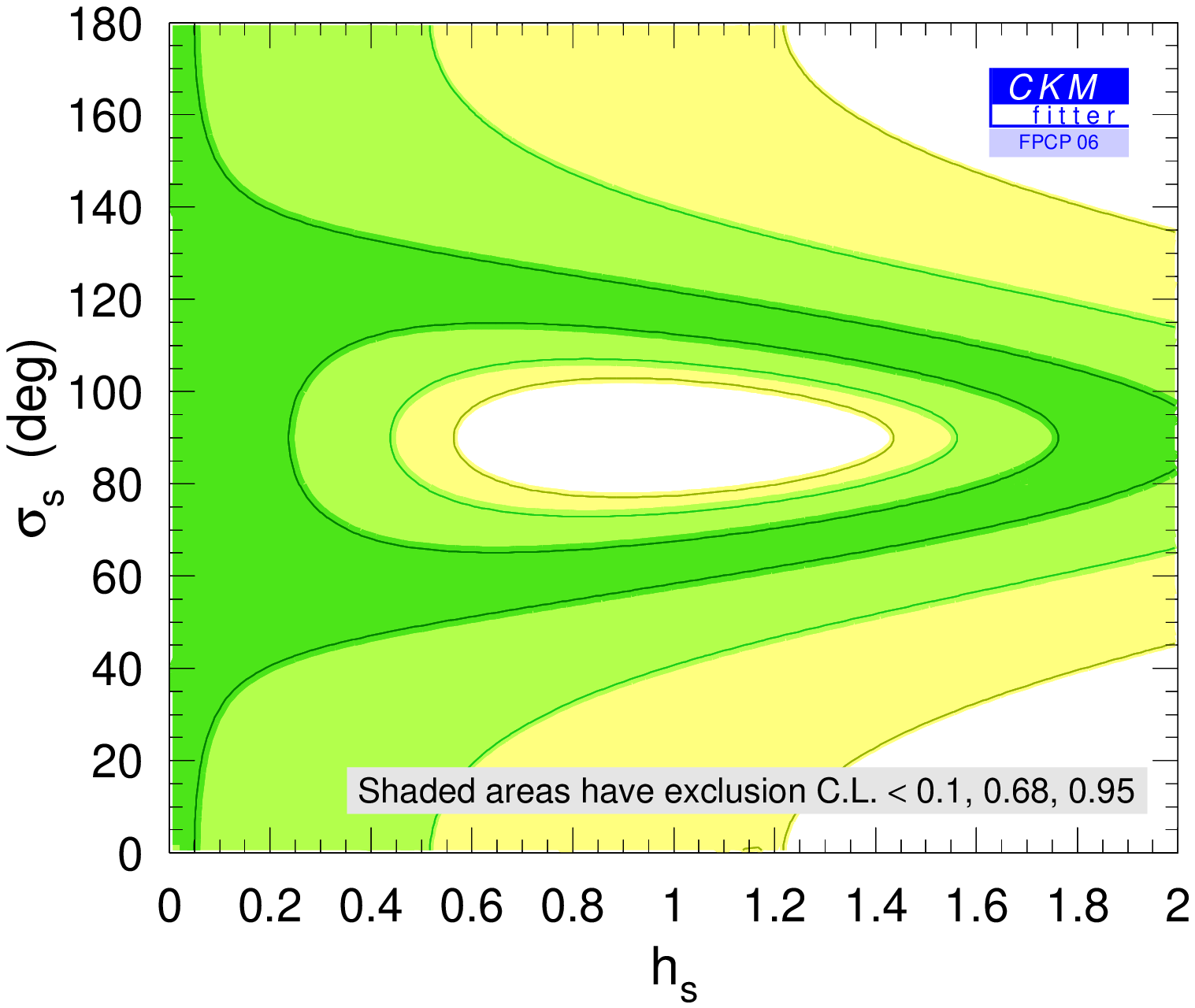}}
\caption{\em Confidence levels of the $h_d,\sigma_d$ (left) and 
	$h_s,\sigma_s$ (right) new physics parameters. The
	dark, medium, and light shaded areas indicate $10\%$, $68\%$ and 
	$95\%$~\CL allowed regions, respectively. 
	The SM corresponds to $h_d=h_s=0$.
	}
\label{fig:hdsd}
\end{figure}

In a large class of models the dominant new physics effects are new 
contributions to the $\BBz$ and $\Bs\Bsb$ mixing amplitudes~\cite{Soares:1992xi},
which can be parameterized as
\beq\label{npmod}
M_{12} = M_{12}^{\rm SM}\, (1 + h_{d,s} e^{2i\sigma_{d,s}}) .
\eeq
The allowed regions for $\sigma_d$ vs.\ $h_d$ and $\sigma_s$ vs.\ $h_s$  are
shown in Fig.~\ref{fig:hdsd} (left and right plots, respectively).  
Numerically, one finds $h_d=0.23\,^{+0.57}_{-0.23}$, corresponding to the
$95\%$~\CL upper limit $h_d<1.7$~\cite{ckmfitter2004}. The constraints on the
new physics parameters are reasonably tight because  $\rhoeta$ has been
determined from (effectively) tree-level measurements independent of new physics
in $\BBz$ mixing (before the $\alpha$ and $\gamma$  measurements most of the
plotted region was 
allowed)~\cite{ckmfitter2004,Ligeti:2004ak,Bona:2005eu,Botella:2005fc}. In
$\Bz\Bzb$ mixing, the ${\rm Im}\, e^{2i\sigma_d} > 0$ region is more strongly
constrained than  the ${\rm Im}\, e^{2i\sigma_d} < 0$ region due to the mild
tension between $\sin2\beta$ and $|V_{ub}|$.  Although the constraints are
significant, new physics with a generic weak phase may still contribute to the
$B^0\Bzb$ mixing amplitude of the order of $30\%$ of the SM.  Because
Eq.~(\ref{npmod}) has two new parameters that need to be determined from the
data, measurements irrelevant for the SM CKM fit, such as the \CP asymmetries in
semileptonic $B_d^0$ and $B_s^0$ decays become important for these
constraints~\cite{Laplace:2002ik,Ligeti:2006pm,Blanke:2006ig}; the former is
measured as\footnote
{
  	To avoid contamination from $\Bs$ mesons, only the
	measurements performed at the $\FourS$ are included in this
	average~\cite{niretal}. In the SM $A_{\rm SL}^s \ll A_{\rm SL}^d$, so
	one may average all data to bound $|q/p|-1$, quoted on
	p.~\pageref{ut_beta_cc}.
} 
$A_{\rm SL}^d =-(0.5 \pm 5.5) \times10^{-3}$~\cite{ASL}.  In $\Bs\Bsb$ mixing,
even after the measurement of $\Delta m_s$, new physics and the SM may still
give comparable contributions to the mixing amplitude, since we have no
information on the mixing phase yet~\cite{Agashe:2005hk,Ligeti:2006pm}. This
will change as soon as there is data on the \CP asymmetry in $\Bs\to
J/\psi\phi$; even a measurement with an uncertainty as large as
$\delta(S_{B_s\to \psi\phi}) \sim 0.1$ would bring the constraints on $h_s,
\sigma_s$ to a level comparable to those on $h_d,
\sigma_d$~\cite{Ligeti:2006pm}.  At present, the lifetime difference between the
\CP-even and odd $\Bs$ states gives a useful additional 
constraint~\cite{niretal,Bona:2006sa,Ligeti:2006pm} (not included in
Fig.~\ref{fig:hdsd}), however, this is largely because the central value
disfavors any deviation from the SM.  After a nominal year of LHC data this 
constraint will probably be less important due to theoretical uncertainties.  
Similar results for the constraints in $\KzKzb$ mixing can be 
found in Refs.~\cite{Agashe:2005hk,Bona:2005eu}.

\section{Conclusions and Perspectives}

The \B-factories and the Tevatron experiments have provided a spectacular 
quantitative confirmation of the three-generation CKM picture. Even more
interesting than  the improved determination of the CKM elements is the
emergence of redundant measurements that overconstrain the CKM matrix. The
measurements of \CP asymmetries, mixing, and semileptonic and rare decays allow 
us to severely constrict the magnitudes and phases of possible new physics
contributions to flavor-changing interactions in a variety of neutral- and
charged-current processes.  We can meaningfully constrain flavor models with
more parameters than the SM; in particular, within the last two years, the
comparison between tree- and loop-dominated measurements significantly bounds 
new physics in $\B\Bb$ mixing.  This illustrates that it is the multitude of
overconstraining measurements and their correlations (and not any single
measurement) that carry the most compelling information. 
Nevertheless, as can be seen in Fig.~\ref{fig:hdsd}, new physics in $\BBz$ and
in $\Bs\Bsb$ mixing may still be comparable to the SM contribution.

Having analyzed these results, one may ask where flavor physics should go 
from here. The sensitivity to short-distance physics will not be limited
by hadronic physics in many interesting processes for a long time to come.  
The existing measurements could have shown deviations from the SM, and if 
there are new particles at the TeV scale, new flavor physics could show 
up any time as the measurements improve.  If new physics is seen in flavor 
physics, we will want to study it in as many different processes as possible.  
If new physics is not seen in flavor physics, it is still interesting to 
achieve the experimentally and theoretically best possible limits.  If
new particles are observed at the LHC, the measurements from the flavor 
sector, as well as flavor-diagonal constraints from electric dipole moments, 
will strongly confine the degrees of freedom of the underlying new physics.

The large number of impressive new results speak for themselves, so that it 
is easy to summarize the main lessons we have learned so far:

\newcommand{\myit}{\item[$-$]}

\begin{itemize}
 
\myit 	$\stwob = 0.687 \pm 0.032$ implies that the $B$ and $K$ 
	constraints are consistent, and the KM phase is the dominant 
	source of \CP violation in flavor-changing processes at the 
	electroweak scale.

\myit 	The difference between the measurements of $\stwob$ from \B decays to
	charmonium plus $\Kz$ and from the penguin-dominated $\etapr\Kz$ mode
	(or the average of $\phi\Kz$, $\etapr\Kz$ and $\KS\KS\KS$) is 
	approximately $2\sigma$ at present. More data are required 
	to make this comparison conclusive.

\myit 	The measurements of $\alpha = (100^{+15}_{-8})^\circ$ and 
	$\gamma = (62^{+35}_{-25})^\circ$ are in agreement with 
	the SM expectations, and begin to improve the constraint on $\rhobar$, 
	$\etabar$. They also provide significant bounds on new physics in 
	$\Bz\Bzb$ mixing.

\myit 	The CDF measurement of $\dms=17.31\,^{+0.33}_{-0.18}\pm0.07\invps$
	(in agreement with the SM expectation, $21.7\,^{+5.9}_{-4.2}$)
	together with the results of the other experiments lift the 
	evidence for $\Bs\Bsb$ oscillation to the $4\sigma$ level.

\myit 	The determinations of $|V_{cb}| = (41.5 \pm 0.7)\times 10^{-3}$, 
	$\ov m_b(\ov m_b) = 4.18\pm0.04\,\GeV$, and $\ov m_{c}(\ov m_c) =
	1.22\pm0.06\,\GeV$  from inclusive semileptonic \B decays reached
	unprecedented  precision and robustness, as all hadronic inputs were
	simultaneously determined from data.

\myit	Strong experimental evidence for the leptonic decay $\Bp\to\taup\nut$ 
	has been found, with a world-average branching fraction of 
	$(10.4\,^{+3.0}_{-2.7})\times10^{-5}$, in
	excellent agreement with the SM.

\myit 	$A_{K^+ \pi^-} = -0.115 \pm 0.018$ implies large direct \CP violation
	disproving ``$B$-superweak" models. It also signifies that sizable 
	strong phases occur in this decay. Many more decays exhibiting
	direct \CP violation will emerge from the increasing data samples.

\myit 	The list is much longer, \eg, improvements in the determination of
	$|V_{ub}|$, observation of $B\to X_s\ell^+\ell^-$ and $B\to
	D^{0(*)}\pi^0$ decays, and a zoology of new $D_s$ and $c\cbar $ states.

\end{itemize}

The next few years promise similarly interesting results (in
arbitrary order):

\begin{itemize}

\myit 	One of the most important outstanding questions is the comparison
	of $\stwob$ from tree and penguin decays. Three times more data 
	than currently used in the analyses could be accumulated until
	data taking of the first generation \B-factories ends.

\myit 	A measurement of $\stwobs$ with mixing-induced \CP-violation analysis
	of $\Bs\to\J/\psi\phi$  (and similar) decays. If $\stwobs$ can be
	measured at the 0.02 level  (absolute error) at the LHC, it will
	constitute the most sensitive probe so far for new \CP-violating phases. 
	Together with other SM-like observations in the  $K$ and $B$
	sectors, it would suggest that the flavor sector of any new physics at
	the TeV scale approximates minimal flavor violation.

\myit 	Improvements in the determination of $\alpha$ and $\gamma$ can be 
	expected. The angle $\gamma$ will also be measured by LHCb.
	This will further constrain the 
	right side of the UT, serving as the SM reference
	for the ratio $\dmd/\dms$.  Reduction of
	LQCD errors is also necessary for this comparison.

\myit 	The rigor of the determination of $|V_{ub}|$ may approach that of
	$|V_{cb}|$, and its error will also be reduced.  It determines the side
	of the UT opposite to $\beta$, so any progress directly improves
	the accuracy of CKM tests (the error  with continuum methods reach
	an asymptote at approximately $5\%$).

\myit   The experimental precision approaching the limits from the theoretical
	uncertainties in $B\to X_s\gamma$ and $B\to X_s\ell^+\ell^-$ (for
	branching fractions,  \CP-violating asymmetries and 
	forward-backward asymmetries) will impact model building.  
	The (continuum) theory is most precise for inclusive decays, which, 
	however, cannot be measured at the LHC.

\myit 	Improving the precision of the current  
	$S_{K^*\gamma} = -0.13 \pm 0.32$ measurement
	will be important to constrain certain extensions of the SM.

\myit 	Improving the precision of the current  
	$A_{\rm SL} = -2 (|q/p|-1) = -(0.5 \pm 5.5) \times 10^{-3}$ measurement
	is also important, in addition to a measurement of 
	$A_{\rm SL}^s$ in $\Bs\Bsb$ oscillation.

\myit 	Future studies should firmly establish and precisely measure the 
	$B\to \rho\gamma$ and $B\to\tau\nu$ rates, and try to measure 
	$B\to X_d\gamma$ inclusively.  They should also search
	for $B_{d,s}\to \ell^+\ell^-$ and other rare or forbidden decays.

\myit 	Similar to the ``new'' $c\sbar $ and $c\cbar $
	states discovered by the \B-factories, new physics could also be
	discovered  in the charm sector.  With increasing sensitivity, one may
	find clear signs of new physics in $D\to \pi\ell^+\ell^-$ or through
	\CP violation in $\Dz\Dzb$ mixing.

\end{itemize}\vspace*{-8pt}

\subsubsection*{Acknowledgments}

We are indebted to Stephane T'Jampens and the CKMfitter Group for providing many
of the plots and results given in this review. We thank Max Baak, Pat Burchat,
J\'er\^ome Charles, Roger Forty, Michael Luke, Stephane Monteil, Arnaud Robert 
and Frank Tackmann for helpful comments on the manuscript. The work of Z.L.\ was 
supported in part by the U.S.\ Department of Energy Contract number 
DE-AC02-05CH11231.

\addcontentsline{toc}{section}{References}



\end{document}